\documentclass[prb,showpacs,showkeys,preprintnumbers,superscriptaddress,a4paper,twocolumn,superscriptaddress,nofootinbib,longbibliography]{revtex4-1}
\usepackage[T1]{fontenc}
\usepackage[utf8]{inputenc}
\usepackage{graphicx}
\usepackage{amsmath,amssymb}
\usepackage{amstext}		
\usepackage{blindtext}
\usepackage{enumitem}
\usepackage{psfrag}
\usepackage{xcolor}
\usepackage{epsfig}
\usepackage[colorlinks,citecolor=blue,
    filecolor=blue,
    linkcolor=blue,
    urlcolor=blue]{hyperref}

   \let\c=\chi

\def\beq{\begin{equation}}
\def\eeq{\end{equation}}
\def\bea{\begin{eqnarray}}
\def\eea{\end{eqnarray}}
\def\ba{\begin{array}}
\def\ea{\end{array}}

\DeclareUnicodeCharacter{03B2}{\ensuremath{\beta}}
\begin{document}
\title{ A primer on Kitaev Model:\\ Basic aspects, material realization, and recent experiments}
\author{Saptarshi Mandal }\email{saptarshi@iopb.res.in}
\affiliation{Institute of Physics, P.O.: Sainik School, Bhubaneswar 751005, Odisha, India.}
\affiliation{Homi Bhabha National Institute, Mumbai - 400 094, Maharashtra, India}

\begin{abstract}
This  elementary review article is aimed to the beginning graduate students interested to know basic aspects of Kitaev model. We begin with a very lucid introduction of Kitaev model and present its exact solution, Hilbert space structure, fractionalisation, spin-spin correlation function and topological degeneracy in an elementary way. We then discuss the recent proposal of realizing Kitaev interaction
in certain materials. Finally we present some recent experiments done on these materials, mainly magnetization, susceptibility, specific heat and thermal Hall effect to elucidate the recent status  of material realization of coveted Kitaev spin-liquid phase. We end with a brief  discussion on other theoretical works on Kitaev model from different many-body aspects.   
 
\end{abstract}

\date{\today}

\pacs{
75.10.Jm 	
}
\maketitle
\section{Introduction}

\indent
\indent
{\textbf{Phases of magnetic system:}} Study of magnetic behavior  and different magnetic phases and the reason of its origin constitute one of the primary area of investigation to condensed matter community. From very early time one is used to know about different magnetic responses such as diamagnetism, ferromagnetism, antiferromagnetism\cite{neel-1971} and paramagnetism, metamagnetism\cite{cullity-2008,coey-2010,salamon-2001} which are considered as the starting point toward the vast field of magnetism. As one delves further we come across many types of other magnetic phases  with more complex noncollinear or noncoplanar orders\cite{boni-2009,Yu2010,gingras-2001,Cheong2022} (such as spiral or helical phases), disordered phases such as spin-glass phase \cite{Binder1977,cannella-1974,levy-1977,Mydosh_2015}  as a result of competing exchange interactions or specific lattice structure(yielding geometrical frustration) or spin-orbit interaction\cite{cao-2021}.  Whether most of the above  phases can be described with  classical spins in consideration, there exist many other  phases which are truly quantum mechanical with no classical analogue. Such phases are quantum dimer phase\cite{kivelson-1988,fradkin-2004,kivelson-2012}, valence bond crystal phase\cite{ueda-1996,zheng-1999,sorella-2000,Iqbal_2012}, resonating valence bond phase\cite{moessner-2001,sorella-2001-rvb}, quantum paramagnet\cite{atanupsg,atanufisher} phases where quantum fluctuations play a dominant role and mostly found for spin-1/2 systems. The reason of appearing a particular phase and its stability as well as the low-energy excitation of this phase constitute a genuine theoretical challenge \cite{Sachdev2008,ulrich-2004,Keeling2008,Laughlin2019}. As far as the diverse magnetic  phases are considered, we are being continuously surprised  with discovery of new phases  theoretically or experimentally such as altermagnetism\cite{tomas-2022,Cheong2024,bai2024} and spin liquid\cite{broholm-2020,lucy-2021,savary-review} etc.  Also the experimental detection of these phases is not straightforward and currently intense speculation is underway for unambiguous experimental confirmation of many unusual properties of quantum spin liquid phase\cite{,Hart2021,Banerjee2023,ref1,refael-2024,Wulferding_2020,saikat-orbit-mag}. In this review we discuss certain aspects of spin liquid phases through a paradigmatic model known as Kitaev model\cite{kitaev-2006}. Below we try to provide  some basic ideas of spin liquid phases, their general features, material perspective and other  theoretical aspects.  \\
\indent

{\bf{Frustration: A route to exotic magnetic phases:}}

Given a spin Hamiltonian characterized by exchange coupling between any two spins, one may or may not simultaneously minimize each interaction pairwise in classical sense. This leads to a situation  known as frustration contrary to the non-frustrated phases like simple ferromagnetic(FM) or antiferromagnetic(AFM) phase where all pairs of interactions can be minimized simultaneously. The origin of frustration could be geometrical\cite{Mila2011,Diep2013} (such as AFM interactions in a triangle) as well as specific choices of the competing exchange coupling(such as a diagonal interactions in  square lattice). In the context of Kitaev model the frustration arises due to competing exchange interaction where  a given spin can not simultaneously minimize all the exchange interactions with neighboring sites. This frustration leads to more complex classical phases such as noncollinear ground states\cite{boni-2009,Yu2010,gingras-2001,Cheong2022} , spin glass phases\cite{Binder1977,cannella-1974,levy-1977,Mydosh_2015}etc. The frustration  facilitates strong quantum fluctuations  leading to various quantum magnetic phases mentioned earlier. In spin liquid phase, the quantum fluctuation is strongly pronounced and it prevents interacting magnetic or spin moments to be ordered classically or even quantum mechanically paired up with neighboring spins such as in quantum dimer phases or valence bond crystal phase \cite{Bal,Motome_2020}. The study of spin liquid phase can be traced back to the physical idea of resonating valence bond(RVB) states by Pauling \cite{pauling1953theory} and its mathematical representation by Anderson\cite{ANDERSON1973153} in certain antiferromagnetic Heisenberg model. Since then spin liquid phases are being extensively investigated on triangular, kagome and pyrochlore lattices \cite{Bal,Zhou-2017}.  The spin liquid phase is attributed  with certain key properties which establish it as an unique phase  of matter with possible applications in quantum technology such as quantum computations\cite{nayak-2008,semeghini-2021}. Below we briefly describe its salient characteristics. \\
\indent

{\bf Degeneracy, Quantum fluctuations and Entanglement:}
One of the manifestation of frustration in minimizing the exchange interactions is the macroscopic degeneracy of the classical ground states\cite{lacroix-2011,ramirez-1994,Mendels_2007,moessnercanada-2001,Diep-2004}. This leads to extensive ground state entropy as well. The entropy of frustrated spin systems has attracted a lot of interest, for example in triangular Ising antiferromagnets\cite{wannier-1950}, in square ice \cite{lieb_1967} and in spin ice\cite{Ramirez1999} systems. Due to this large degeneracy of the ground state manifold, even at low temperature, the spins do not freeze or order and instead  continue to fluctuate among the degenerate states\cite{lieb_1967,Ramirez1999,rudian-2019,arthur-2006}. However such fluctuations are correlated in the sense that there are certain patterns and relations among the fluctuations of different spins and this separates it from the ordinary paramagnetic phase\cite{lacroix-2011,Diep-2004,arthur-2006}.  We note that the fluctuation we consider here is quantum fluctuation not the thermal one which plays dominant role in case of ordinary paramagnet. The origin of quantum fluctuation arises in two ways. Firstly it has an inherent noncummutivity among generators of spin-angular moment and secondly the competing exchange coupling can further enhance it by creating a time-dependent local magnetic field which depends on the instantaneous fluctuating states of neighboring spins. Hence the name spin liquid draws its origin in comparison with ordinary liquid where the movement of molecules or particles are highly correlated yet having no order\cite{savary-review,nakatsuji-2006,Machida2010}.  For system with larger spin, the fluctuations is of classical nature whereas for low spin system such as spin-1/2, the fluctuations are of quantum mechanical and often is comparable to the  size of spin itself \cite{savary-review,Kimura2013,molavian-2007,nakatsuji-2006,sym14081716,Correggi2015,tan-2022}. Very often these fluctuations are phase coherent yielding a macroscopic superposition of quantum states. This macroscopic superposition of  quantum states  remains one of the defining characteristics of spin liquid phase. This superposition leads to quantum entanglement\cite{Giampaolo_2018,illuminati-2013,illuminati-2011,laurita-2015,Tang2023,ravi-2009}  in technical term. This entanglement could be long-range and of the system size. The physical manifestation of this long-range entanglement is that two spins could be far away from each other but their motions are correlated in a definite manner\cite{lozano-2022,lu-2023,thibaut-2019,LACROIX20093038,fabio-2024}. This long-range entanglement separates it from the idea of ordinary classical liquid. For an quantitative estimation of various entanglement measures and its relation with quantum spin liquid phase can be found here \cite{Grover_2013,broholm-2020,pretko-2016,melko_2013}.\\
\indent
{\bf Fractionalization and topological order :} Fractionalization is a novel idea where an elementary excitation is fractionalized or seperated into more than one excitation  whose quantum numbers are different than the original one\cite{laughlin-1983,arovas-1984,wen-1990,hastings-2004,Nachtergaele2007}. These fractionalized excitations are distinct quantum states with well defined quantum numbers.  Once fractionalized, these independent excitations  can be moved apart at arbitrary distances without any energy cost. The most commonly available example is to consider an one dimensional antiferromagnetic Heisenberg chain with a valence bond state where alternate bonds are singlet\cite{Khomskii_2010,LIEB1961407}. Excitation is created  by breaking a spin-singlet with spin 1 excitation which can be separated into two spin-1/2 excitations and moved apart without any energy cost\cite{ghosh-majumder,ghosh-majumder-2,Jordan1928}.  The property of this fractionalization is attributed to certain characteristics of ground state itself and the underlying  long-range entanglement makes it possible to move it without further energy cost\cite{kitaev-2003,saptarshi-2014}. Similarly for the RVB state discussed previously, the elementary excitations are charge neutral spinon with spin  1/2 or a holon with charge `$e$' without any spin\cite{kivelson_1987,laughlin_1988,kalmeyer_1987,martson_1988}. This is an example where the quantum numbers of electron are fractionalized and attributed to two separate  excitations. One of the key observations of the fractionalization is that it results from the non-local entanglement between degrees of freedom which allows only part of the quantum number of the original particles to show up and this non-local entanglement invariably leads to topological order of the states\cite{broholm-2020,read-1989,wen-1991,senthil-2000,oshikawa-2006,wen-1990-prb,wen-vacuum-1989}.  Further the exchange statistics of these excitations could be different than the original fermionic or bosonic excitations. The spin liquid phase realizes all these exotic phenomenon where fractionalization, topological order and exchange statistics are intimately tied up with each other \cite{broholm-2020,read-1989,wen-1991,senthil-2000,oshikawa-2006}.  It may be noted that topological orders could be realized in other quantum magnetic phases as well\cite{fradkin-2004-annals} and long-range entanglement remains key element. \\
\indent
{\bf Models and materials  for spin liquid:} There are several theoretical models where classical ground states are highly frustrated  and these models can serve as a good starting point to look for spin liquids\cite{Bal}. These includes AFM Heisenberg model on lattices with triangular unit such as two and three dimensional triangular lattice, kagome and pyrochlore lattice etc. There exist various models where interactions are bond dependent and known as compass models \cite{nussinov-2015}.  These are also good  starting point to study quantum spin liquid theoretically . As far as the materials are concerned there exist several materials such as two-dimensional organic salt, Herbertsmithite
\cite{broholm-2020}, the spin ice materials\cite{gingras-2001,gingras2009spinice} and Kitaev materials \cite{Takagi2019,Motome_2020,Ma-2018} etc. In this beginners review, we discuss some aspects of Kitaev model in the context of spin liquid physics. This model is remarkable in the sense that it is an exactly solvable model where various aspects of  spin liquid physics such as fractionalization\cite{smandal-prl}, topological degeneracy\cite{mandaljpa,kells-degeneracy} etc are realized exactly and its representation is simple in terms of underlying mathematics. Though the model was originally proposed in view of possible realization of fault-tolerant quantum computations\cite{kitaev-2003,nayak-2008,bonderson-2008,sdsarma-2006,freedman-kitaev}, the model attracted a lot of interest from condensed matter community and is recognized as one of the paradigmatic model in its true sense. Remarkably, soon after its proposal, a lot of material candidates  have been proposed with possible Kitaev interactions  \cite{Takagi2019,Motome_2020,Ma-2018}  in it fueling the excitement of the research community further.

{\bf Contents of the review: } 
 
 This review is primarily meant as a basic introduction to Kitaev model and related development aiming toward enthusiastic readers willing to have some exposition to this field. The review mainly contains three parts. In the beginning, we discuss the Kitaev model and some of its exact aspects such as exact solution through Majorana fermionization\cite{kitaev-2006}, the issue of extended Hilbert space structure and its relation to physical solution,  exact calculation of two-spin correlation function, fractionalization of spins\cite{smandal-prl} and topological degeneracy\cite{mandaljpa} in the thermodynamic limit. In the next we outline how in certain materials the direction dependent interactions as contained in Kitaev model is realized through some selection rules and symmetry consideration\cite{jackeli}. We note that in such materials Kitaev model is realized along with traces of other interactions. These includes Heisenberg as well as Gamma interaction\cite{wang-2021,Rau-2014}. The presence of these additional interactions  do not allow the certain spin liquid aspects of Kitaev model such as fractionalization and short-range spin correlations to be  realized exactly. Thus the question of material realization of Kitaev spin liquid phase are being debated with different views at present. Lastly we review some of the important experiments\cite{johnson-2015,kubota-2015,kataoka-jpsj-2020,Kasahara2018,Do2017,Widmann2019} which were performed on these materials and we try to give an overview  what are the phases  obtained  in these experiments. We argue that some of the phases may actually  correspond to the coveted spin-liquid phase as prescribed in   Kitaev model. We end with a discussion and conclusion summarizing other few aspects of Kitaev model which may be taken by enthusiastic readers for future study.

\section{ Kitaev model}
\label{section-1}
  
\begin{figure}
\psfrag{e}{$\vec{a}_1$}
\psfrag{d}{$\vec{a}_2$}
\includegraphics[width=0.85\linewidth]{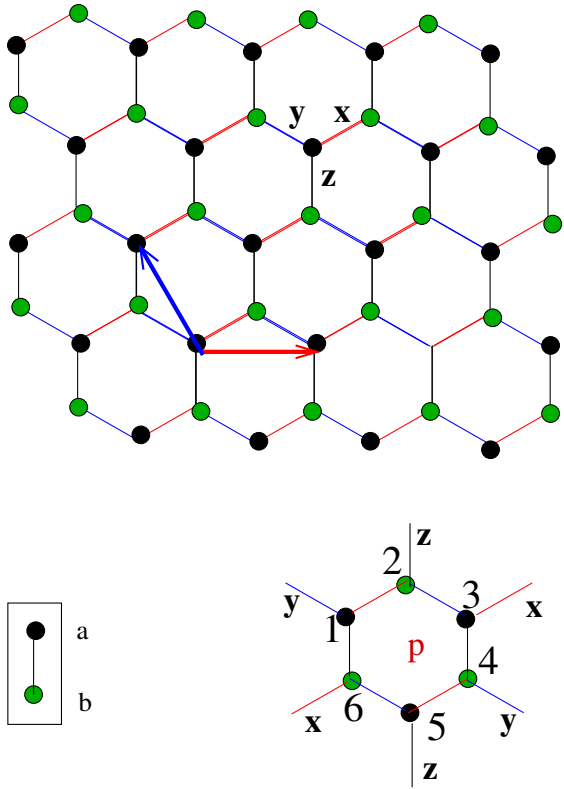}
\caption{\label{fig1} In the left we have shown a cartoon picture of honeycomb lattice. Black and green filled circles denote the ``a'' and ``b'' sublattices of the honeycomb lattice respectively. The green, black and red bonds  contain
$x$-type, $y$-type and $z$-type interactions respectively. `$\vec{a}_1$' and `$\vec{a}_2$' are the two basis vectors. In the lower right corner an elementary  plaquette  is shown with a particular enumeration of sites which has been used to define
the conserved quantity $B_p$ in Eq.~\ref{bpeq}. }
\end{figure}

Let us introduce  the  Hamiltonian for the Kitaev model\cite{kitaev-2006} as follows. The model is defined on a honeycomb lattice and at each site of this lattice we consider  a spin-1/2 particle.  The spin-1/2 particles interact with each other in a specific way.  The origin of such interaction is spin-orbit coupling and will be discussed later, for more details one can have  look at the recent progresses \cite{winter}. To understand that we first note that for honeycomb lattice, there are three different kind  of bonds that can be grouped together according to their alignment  as shown in Fig.~\ref{fig1}. There are  vertical bonds which we label as $z$-bond. There are  bonds with  positive slope and we call it $x$-bond and the bonds which are having negative slope are termed as $y$-bonds. Now it is clear that a given site is connected to three other sites by these three different types of bonds. The spin-spin interaction between two neighboring spins depend on the kind of bonds they are connected with. For example, a given spin located at site `$i$' interacts with another spin at  site $i+\delta_x$ (connected by a x-type bond) with $S^x_i S^x_{i + \delta_x}$, where as the same spin interact with its neighboring site located at  $i + \delta_y$ and $i + \delta_z$ with $S^y_i S^y_{i + \delta_y}$ and $S^z_i S^z_{i + \delta_z}$ respectively. In the above $S^{\alpha}_i, ~\alpha=x,y,z$ denotes the  spin angular moment at a given site `$i$'. We know that honeycomb lattice has two sublattices namely `a' and `b' sublattices as denoted by the green and black filled circles in Fig. \ref{fig1}. Without loss of generality we may assume that `$i$' belongs to `a' sublattice. In this case we have the folowing expressions for $\delta_{\alpha}$. $\delta_x= \frac{\sqrt{3}}{2} \hat{x} + \frac{1}{2} \hat{y},\delta_y= -\frac{\sqrt{3}}{2} \hat{x} + \frac{1}{2} \hat{y}, \delta_z= - \hat{y} $. Here we consider the distance between two nearest neighbours to be unity. The lattice translation vectors $\vec{a}_1$ and $\vec{a}_2$ are obtained as $\vec{a}_1= \sqrt{3} \hat{x}, ~\vec{a}_2= - \frac{\sqrt{3}}{2} \hat{x} + \frac{3}{2} \hat{y}$ as shown in Fig. \ref{fig1}. With this in mind the Hamiltonian of the model can be written as follows, 

\begin{eqnarray}
H= \sum^{N_x}_{\langle ij\rangle_x=1} J_x S^x_i S^x_j + \sum^{N_y}_{\langle ij\rangle_y=1} J_y S^y_i S^y_j + \sum^{N_z}_{\langle ij\rangle_z=1} J_z S^z_i S^z_j .~~~
\label{ham0}
\end{eqnarray}

Here $J_x, J_y, {\rm~and}~ J_z$ denote exchange interactions on $x, y, z$ type of bonds respectively. $\langle ij\rangle_{\alpha}$ (with $\alpha =x,y,z$) denotes a pair of sites connected by $\alpha$-type of bond. Each summation in the above equation runs over all bonds of a particular type and hence no double counting arises. Here $N_{\alpha}$ is the total number of $\alpha$-type bonds. In thermodynamic limit one have $N_x=N_y=N_z=N/3$ where $N$ is total number of bonds in the system. Note that enumeration of a type of bond by integers starting from 1 to $N_{\alpha}$ is arbitrary and does not cause any physical effect in Eq.~\ref{ham0}. In the Eq.~\ref{ham0} the coupling parameter $J_x, J_y, J_z$ are taken positive.  However the analysis we  follow and the detail properties that we  discuss remains unchanged for any combinations of signs for  the coupling parameter. This is another remarkable characteristics of Kitaev model, where unlike any other model such as Ising model  or Heisenberg model  where the  many  spin configuration depends on the sings of coupling parameter. One important aspect of this  peculiarity  is that the  frustration of Kitaev model at classical level  does not depend on the sign of exchange interaction. To appreciate that we can think about $J_x=0, J_y=0$ limit, and we find that either a ferromagnetic (FM) or antiferromagnetic (AFM )alignment of spins on a given $z$-bond can classically satisfy a $z-z$ interaction depending on whether the $J_z$ coupling is negative or positive. However if we now turn on $x$ or $y$ type of exchange couplings, the interactions on $x$ or $y$ bonds remains unsatisfied. This peculiarity makes the Kitaev interactions very special and certain aspect that  follows below remains true for any combinations of coupling strength and it can even be disordered. For more on classical solution the reader is requested to have a look at previous work examining the classical solutions of Kitaev model~\cite{sen-shankar}. 

\subsection{Exact Solution of Kitaev model}
\label{section-2-0}
The Hamiltonian as given in Eq.~\ref{ham0} represents an interacting  quantum spin system and the exact solutions i.e obtaining ground state wave function or some or all eigenfunctions and eigenvalues remains the primary quest for us \cite{sutherland-book,korepin-book,march-2016,Takahashi_1999,Lieb-Wu-1968,essler-1992}. Thus, we first focus on obtaining  the eigenvalues  and eigenfunctions of Eq.~\ref{ham0}. Once we have that, we can calculate any physical quantity of our interest in principle.  The individual states of a given spin-1/2 particle can be represented by the two eigenstates of the Pauli matrices.  If there are `$N$' such spin-1/2 particles, the total  Hilbert space dimension becomes $2^N$. The Hamiltonian in Eq.~\ref{ham0} can be represented  suitably in $2^N \times 2^N$ dimensional hermitian matrix. To obtain the eigenvalues and eigenfunctions one has to diagonalize this matrix which is known as procedure of exact diagonalization. Another alternative way to solve a spin Hamiltonian is to use suitable auxiliary variable to express the spins \cite{tsvelik}  preserving their mutual commutation algebra. One may use either  fermionc or bosonic field operators to express the spin-1/2 angular momentum algebra which is known as fermionization or bosonization  respectively, depending on the auxiliary field operators that are used. For one-dimensional system such fermionization procedure yields useful starting point \cite{Jordan1928,pfeuty,LIEB1961407,bethe}.\\
\indent
It is known that for exactly solvable model presence of conserved quantities play a crucial role. One of the  key feature of this model (Eq.~\ref{ham0}) is the presence of large number of such conserved quantities. For each plaquette,  one can define an operator  which commutes with the Hamiltonian.  We call this  plaquette operator as $B_p$ where the subscript `$p$' stands for the plaquette index. Here by plaquette we mean the smallest closed loop in honeycomb lattice formed by six bonds. Plaquette operators defined on different plaquettes commute among themselves. With reference to the Fig.~\ref{fig1} $B_p$ is defined as,
\begin{equation}
\label{bpeq}
 B_{p}=\sigma^{y}_{1}\sigma^{z}_{2}\sigma^{x}_{x}\sigma^{y}_{4}\sigma^{z}_{5}\sigma^{x}_{6}. 
\end{equation}
 Thus for each plaquette  `$p$' we can define  a $ B_{p}$. It can be easily checked that,
\begin{equation}
[B_p,H]=0 \, , \,\,\,\, [B_p,B_q]=0, ~p \ne q .
\label{conbpch1}
\end{equation}
In the above $p,q$ indicate different plaquette indices. This implies  that $B_p$'s are  conserved quantities for this model.  It is easy to verify that $B^2_p=1$  which implies that eigenvalues of $B_{p}$ are $\pm 1$.  We will see later that these conserved quantities play a significant role in the dynamics of Kitaev model. We now present the formal solution of this spin model as obtained by  Kitaev himself \cite{kitaev-2006}. He showed that this  spin model can be solved exactly using a fermionization procedure which expresses the spin-1/2 operators in terms of Majorana fermions. In the next  we elaborate on this.

\subsection{\label{fso2.1}Fermionization of spin-1/2 operators}

One of the efficient way to analysis an interacting spin system is to use auxiliary fermions to express the spin operators. In  one dimensional system, Jordan-Wigner fermionization \cite{Jordan1928,pfeuty,LIEB1961407,bethe}is one such procedure with numerous applications. Here we outline another fermionization procedure which was used by Kitaev  himself. 
At every site we introduce two Dirac fermions $d_1$ and $d_2$ and construct the following Majorana  operators, $c=(d_{1}+d^{\dagger}_{1}), c^{x}=\frac{1}{i}(d_{1}-d^{\dagger}_{1}),~ c^{y}=(d_{2}+d^{\dagger}_{2}),~c^{z}=\frac{1}{i}(d_{2}-d^{\dagger}_{2})$. One can easily verify that they satisfy  anticommutation relation $[c_{\alpha}, c_{\beta}]=2 \delta_{\alpha,\beta}$, they are self-conjugate ($c_{\alpha}=c^{\dagger}_{\alpha}$) and when multiplied with itself gives unity ($c^{2}_{\alpha}=1$). Here $\delta_{\alpha,\beta}$ are Kronecker delta function. One can then define the spin-1/2 Pauli matrices by forming suitable bilinear of this Majorana fermions as follows. $\sigma^{x}=ic^{x}c,~\sigma^{y}=ic^{y}c,~\sigma^{z}=ic^{z}c$. It is easy to verify that above defined Pauli operators  satisfy usual commutation relation. However one notices that the constraints over Pauli spin algebra namely  $ \sigma^x \sigma^y \sigma^z=  i$ is also to be satisfied. Given the definition of $\sigma$ operators, one may verify that $\sigma^x \sigma^y \sigma^z= i (2 d^{\dagger}_1 d_1-1) (2d^{\dagger}_2 d_2-1)$. Thus it is evident that $\sigma^x \sigma^y \sigma^z= i$ is true only for even particle sector. To implement this condition, it is useful to define $D=-i \sigma_x \sigma_y \sigma_z$. Thus in even (odd) particle  sector $D=1(-1)$. Now if one defines a projection operator $P_i=\frac{1}{2} (1+ D_i)$ and constructs a total projection operator $\mathcal{P}= \prod_i P_i$ and applies on the wave function obtained one recovers the physical solution. This issue will be discussed again in next section in more detail. 

\subsection{\label{qdham1}Quadratic Hamiltonian}
To proceed we  write the Hamiltonian in terms of fermionic operators as discussed above. After inserting the relations $\sigma^{\alpha}_i= i c^{\alpha}_i c_i,~~\alpha=x,y,z$ in  Eq.~\ref{ham0} one obtains,

\begin{eqnarray}
\label{2}
 H &=&\sum^{N_x}_{\langle ij \rangle_x =1}J_{x}(ic^{x}_{i,a}c^{x}_{j,b})ic_{i,a}c_{j,b} \nonumber \\
 &+&
\sum^{N_y}_{\langle ij \rangle_y=1}J_{y}(ic^{y}_{i,a}c^{y}_{j,b})ic_{i,a}c_{j,b} \nonumber \\
&+& \sum^{N_z}_{\langle ij \rangle_z=1}J_{z}(ic^{z}_{i,a}c^{z}_{j,b})ic_{i,a}c_{j,b}.
\end{eqnarray}
Here the summation follows same meaning as explained in Eq. \ref{ham0}. Additionally we use the sublattice indices and without loss of generality assign index `$a$' to the site `$i$' and `$b$' to the site `$j$'. We observe that each terms in the above Hamiltonian is quartic in  Majorana fermion operators.  Generally such quartic Hamiltonian is quite difficult to solve. However it can be easily noted that  operators in the parenthesis of each term of the above Hamiltonian  commute with the Hamiltonian and commute among themselves.  This means they follow the following commutation relation which the interested readers are requested to check before proceeding further.
\begin{eqnarray}
[ic^{\alpha}_{i,a}c^{\alpha}_{j,b}, ic^{\beta}_{i,a}c^{\beta}_{j,b} ]=0, [ic^{\alpha}_{i,a}c^{\alpha}_{j,b}, ic_{i,a}c_{j,b} ]=0 .
\end{eqnarray}

In the above equation $\alpha, \beta=x,y,z$ and $(i, j)$ denotes a pair of nearest-neighbor  sites joined by a $\alpha$-type bond. It means that they are conserved quantities as far as this fermionized Hamiltonian is concerned. This fact makes Eq.~\ref{2} to be effectively quadratic in Majorana fermions. Let us call, $ic^{x}_{i,a}c^{x}_{j,b}=u^{x}_{i,j}$ for the $x$-bond. Similarly we define $ u^{y}_{i,j}$ and $ u^{z}_{i,j}$ on $y$ and $z$ bonds respectively. It is obvious that, $u^{\alpha}_{i,j}=-u^{\alpha}_{j,i}$ and its eigenvalues can take value $\pm 1$ given the fact $(u^{\alpha}_{i,j})^2=1$.  Here we follow the convention of keeping the indices of the site belonging to the `$a$' sublattice first and then for `$b$' sublattice in the expression of $u_{i,j}$(for brevity  explicit mention of $\alpha$ in the expression of $u^{\alpha}_{ij}$ will be avoided in some cases). We note that in thermodynamic limit, in honeycomb lattice,  the number of sites are equal to the twice of the  number of a given type of bonds. This is because a given type of bonds connect all sites and two sites are connected by a given bond. Thus considering there are three different kind of bonds we have $3N/2$ number of bonds if there are $N$ sites in honeycomb lattice system in thermodynamic limit. Using the above replacement for the conserved quantities on each bond, the Hamiltonian takes the following form,
\begin{eqnarray}
H &=&\sum^{N_x}_{\langle ij \rangle_x=1} J_{x}u^{x}_{i,j}ic_{i,a}c_{j,b}+ \sum^{N_y}_{ \langle ij \rangle_y}
J_{y}u^{y}_{i,j} ic_{i,a}c_{j,b}\nonumber \\
&&  +\sum^{N_z}_{ \langle ij \rangle_z=1}J_{z} u^{z}_{i,j}ic_{i,a}c_{j,b} = \Phi^{\dagger}  \mathcal{H}([u]) \Phi ,
\label{3}
\end{eqnarray}
where $ \Phi^{\dagger}=(c_1,c_2,...,c_N)$ is an one dimensional row matrix of length `$N$' and $\mathcal{H}[u]$ is a $N \times N$ matrix which depends on the values of $u_{ij}$ on each bond. Now we see that above Hamiltonian describes a tight binding Majorana fermion hopping interactions but the hopping  amplitudes are  coupled with  conserved operator or fields $ u^{\alpha}_{i,j}$ on each bonds. As we mentioned that these operators have eigenvalues $\pm 1$. These $u^{\alpha}_{i,j}$ are called $Z_2$ gauge fields because of the eigenvalue to be $\pm 1$. Depending upon the values of these $Z_2$  gauge fields the eigenvalues of the system will change. Physically one way to visualize is that initially we had quartic Majorana fermion interaction of the form $ic^{\alpha}_{i,a} c^{\alpha}_{j,b} i c_{i,a} c_{j,b} $. This process can happen in many ways. For example this can be thought  of as two Majorana fermions hopping process happening simultaneously such as $c^{\alpha}_{i,a} \rightarrow c^{\alpha}_{j,b},~c_{i,a} \rightarrow c_{j,b}$ or $c^{\alpha}_{\alpha,i,a} \rightarrow c_{j,b} ,~c_{i,a} \rightarrow c^{\alpha}_{j,b}$ or similar processes. Alternatively it can be thought of  as two on site interactions happening simultaneously  yielding an energy cost. This on site interaction involves the Majorana fermions at a given site only. All possible equivalent explanation exists for such four body interactions and they all correspond to one physical process.  However the fact that the combination $ic^{\alpha}_{i,a} \rightarrow c^{\alpha}_{j,b} $ is a conserved quantity signify that once  eigenvalue of $ic^{\alpha}_{i,a} \rightarrow c^{\alpha}_{j,b}$ is fixed to either +1 or -1 (like an initial condition) it remains same and does not change with time. Lets assume that the eigenvalue is fixed at +1(or -1) then the hopping process $ c_{i,a} \rightarrow c_{j,b}$ happens with a phase +1( or -1). This also signify that the process $c^{\alpha}_{i,a} \rightarrow c_{j,b} ,~c_{i,a} \rightarrow c^{\alpha}_{j,b}$ or other possible processes are not allowed by the system. Further in Eq.~\ref{3}, if we transform $c_{i,a} \rightarrow \lambda_i c_{i,a} $, we observe that to keep the Hamiltonian invariant under such transformation $u^{\alpha}_{i,j}$ must change according to $u^{\alpha}_{i,j} \rightarrow \lambda_i u^{\alpha}_{i,j} \lambda_i$, where the allowed value of $\lambda_i$ is again $\pm 1$. Thus the Hamiltonian given in Eq.~\ref{3} has a underlying $Z_2$ symmetry \cite{wen}. \\
\indent
   It is to be noted that the new conserved quantities, $u_{i,j}$, are absent in the original spin Hamiltonian. In the original spin Hamiltonian as given in Eq.~\ref{ham0}, there is no conserved quantity associated with a bond. Earlier we found that there was only one type of conserved quantity associated with each  plaquette only and it is given in Eq.~\ref{bpeq}. Also it is easy to check that in terms of $u_{i,j}$ $B_p$ is given as $B_{p}=\prod_{(j,k)\epsilon {\rm boundary(p)} } u^{\alpha}_{j,k} $ i.e    $B_p$ is a product of six  emergent conserved quantities ($u_{i,j}$'s) (after fermionization of the spin Hamiltonian) defined on each bond. Initially our physical degrees were the spins represented by a product of two Majorana fermions. The bond conserved quantity $u_{i,j}$ is also product of two Majorana fermions but each Majorana fermion is taken from the two spins attached at the end of a particular bond. And one can verify that a $u_{ij}$ can not be expressed in terms of spins. It is the product of $u_{ij}$ over the bonds of a  plaquette which can be expressed as a product of six spin operators and called $B_p$. One can easily verify that the square of $B_p$ is one indicating that the eigenvalue of the operator $B_p$ is $\pm 1$. We have also seen that eigenvalue of the bond-conserved quantity $u_{ij}$ is also $\pm 1$. As $B_p$ is expressed as a product of six $u_{ij}$ we can understand that for a given eigenvalue of $B_p$ there are many choices for the eigenvalues of the participating $u_{ij}$'s. Among the total $2^{6}$ configurations that six $u_{ij}$ provides, half of them  yields $B_p=1$ and other half yields $-1$. For a given  eigenvalue of $B_p$ say +1, among the various combinations, we will observe that a given $u_{ij}$ changes its eigenvalue from +1 to -1 though value of $B_p$ is fixed to 1. For this reason $u_{ij}$s are called gauge fields analogous to magnetic vector potential and $B_p$'s are physical observable analogous to magnetic field. Physically $u_{ij}$ can not be measured and its  expectation value or outcome in experiment will be zero because of gauge averaging over many combinations that one $u_{ij}$ takes. However eigenvalue of $B_p$ is a physical observable. Thus we  use the phrase that $B_p$ is gauge invariant and $u_{ij}$ are not. \\
\noindent

\begin{center}
{\bf Hilbert space structure, extended vs physical}
\end{center}

\begin{center}
\begin{figure}[ht!]
\psfrag{a}{$2^{N}$}
\includegraphics[width=0.8\linewidth]{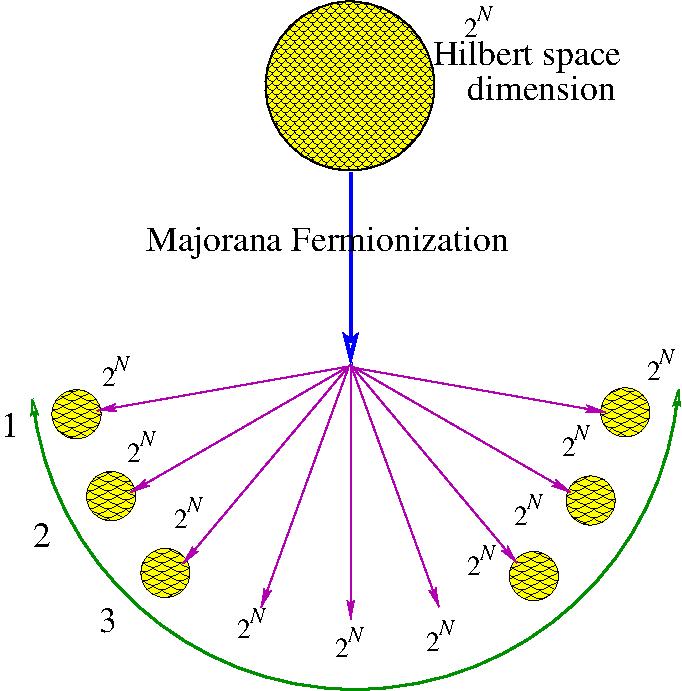}
\caption{\label{hilbert}The big circle at the top denotes the original Hilbert space dimension of $2^{N}$. At the lower  $2^{N}$ copies of the original Hilbert space is shown. This happens due to Majorana fermionization using four Majorana fermions at a given site. The description of Kitaev model in each of these $2^{N}$ copies are identical up to an gauge transformations which would connect one gauge copy to another. }
\end{figure}
\end{center}

 Before proceeding further we elaborate on the Hamiltonian represented in  Eq.~\ref{3}.  This will illustrate the inherent mathematical structure clearly. Usually this mathematical structure is not easily comprehensible in model Hamiltonian of frustrated magnetism. However   in Kitaev model such structure becomes manifestly invariants and this motivates us to explain in detail the consequences of mapping of original spin problem into a Majorana fermion hopping problem coupled with $Z_2$ conserved gauge field. Such  description may  happen for other system approximately and we believe that a thorough understanding of it in the context of Kitaev model will help the reader to extend their imagination easily to explore the intricacies of other complex system or models of frustrated magnetism. Lets consider a  finite Honeycomb lattice such that it has $N_1$ dimer or $z$-bond in the $\vec{a}_1$ direction and $N_2$ dimer or $z$-bonds in $\vec{a}_2$ direction as shown in Fig. 1. Thus we have a system of $N$ spins with $N=2 N_1 N_2$. Because there are two spins in a given $z$-bonds. Total number of  bonds of a particular type say $x,~y,~z$ are equal and they are $N_1 N_2$. Thus we have a total number of bonds $N_b=3 N_1 N_2=3N/2$. The original spin Hamiltonian of $\mathcal{N}$ spins is defined in $2^{\mathcal{N}}=M$ dimensional Hilbert space. This means that we are to solve a $M \times M$ matrix with $M$ real eigenvalues and corresponding eigenstates. Now while we  employed a fermionization procedure where at each site there are two fermions implying that  each site is associated with a Hilbert space dimension of four yielding total Hilbert space of $4^{\mathcal{N}}= (2^2)^N=2^N 2^N=M_1=2^M$ which means we have now enlarged Hilbert space yielding $M_1$ number of eigenstates/eigenvalue which is much more than the original Hilbert space. Thus though the original problem in spin-space was dimension $2^N$, after fermionization now we have a problem of dimension $2^N \times 2^N$. This situation is depicted in Fig.~\ref{hilbert}. 

\begin{figure}
\psfrag{a}{$N \times N$}
\includegraphics[width=0.9\linewidth]{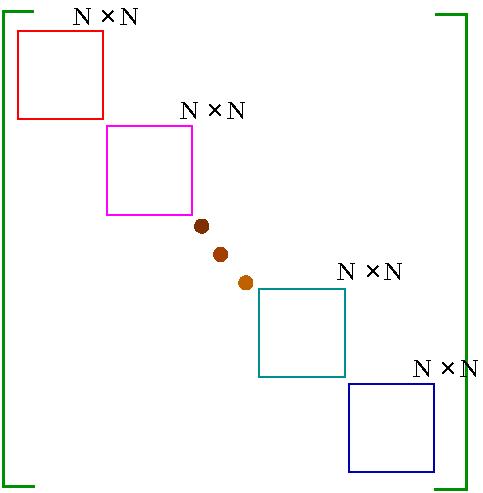}
\caption{\label{matrix} For a system of `$N$' sites we have a matrix of dimension $N \times N$ for a given realization of $u_{ij}$ on each link. Such matrix is shown by the small colored matrix. For a given $N$ sites,
there are $3N/2$ bonds yielding a total possibility of $2^{3N/2}$ such configuration. Thus the dimension of the outer big matrix is $2^{3N/2} \times 2^{3N/2}$. For more detail on the counting on the Majorana fermion and its relation to complex fermions see text below. }
\end{figure}

    Let us try to find another connection which  explains the relation between the eigenstates of the original spin problem given by Eq.~\ref{ham0} or Eq.~\ref{3} and Majorana fermion hopping problem given by Eq.~\ref{2}. In the spin space we must have $M=2^N$ number of eigenstates and eigenvalues. How do we obtain that from Eq.~\ref{3} and what is the consequences of enlargement of Hilbert space dimension. This is explained here.  For a given distribution of gauge fields on the bonds of honeycomb lattice, we have a matrix  of dimension $N$ which has $N$ eigenvalues and $N$ single particle Majorana fermion eigenstates.  If one diagonalizes $\mathcal{H}[u]$ in Eq.~\ref{3}, we obtain $\lambda[u]_i, i=1,N$ with eigenvectors $|\lambda[u] \rangle_i= \sum_i \gamma[u]_i c_i$. However these Majorana fermions are to be regrouped to yield $N/2$ complex fermions which have a definite occupation number representation~\cite{kitaev-2006,bruus}.  From this $N/2$ single particle eigenstates one can obtain $2^{N/2}$ many-body states by constructing eigenstates having arbitrary particle number. The energy eigenvalues for such many-body states can be written as $\sum_i \lambda[u]_i$. One can easily note that there are $2^{N/2}$ such many body eigenstate which can be obtained by $\sum_m { \frac{N}{2} \choose m}$. However we remember that such description is true for a given distribution of conserved quantities on every bond. This situation is depicted in Fig.~\ref{matrix} where the dimension of the outer big matrix is $\frac{N}{2} \times \frac{3N}{2}= \frac{3N^2}{4}$. Inside this big matrix the smaller diagonal matrix represent a certain distribution of  conserved gauge fields and the dimension of this smaller matrix is $N$ yielding $N/2$  single particle  fermionic eigenstate which in turn yield $2^{N/2}$ many-body states.  Now we already mentioned there are in total $N_b=3N/2$ number of bonds yielding in total $2^{3N/2}$ such combinations. This means, with reference to Fig.~\ref{matrix}, we have  $2^{3N/2}$ such small diagonal matrices. Each of this smaller matrix yields  eigenstates of $2^{N/2}$ numbers. Thus the total number of eigenstates is $2^{N/2}  2^{3N/2}= 2^N 2^N $ which matches with our earlier counting. \\
\noindent
Earlier we mentioned that $u_{ij}$'s are not gauge invariant quantities and they are not physically observable. The gauge invariant physically observables are the plaquette conserved quantity $B_p$. There are many combinations of $u_{ij}$ which yields a configuration of $B_p$ for each plaquette. And it is important to note that the energy eigenvalues only depend on the distribution of $B_p$ not $u_{ij}$. If two configurations of $u_{ij}$ yield same distributions of $B_p$, they  have identical eigenvalues. In reference to Fig.~\ref{matrix}, this means that there are many gauge equivalent smaller matrices which give identical distributions of $B_p$ and have identical energy eigenvalues.

\vspace{0.5cm}
\begin{center}
\Large{Lieb Theorem }
\end{center}

\begin{figure}
\psfrag{A}{A}
\psfrag{B}{B}
\psfrag{C}{C}
\psfrag{D}{D}
\psfrag{+}{+}
\psfrag{-}{$-$}
\psfrag{1}{1}
\psfrag{2}{2}
\psfrag{3}{3}
\psfrag{4}{4}
\psfrag{5}{5}
\psfrag{6}{6}
\psfrag{7}{7}
\psfrag{8}{8}
\includegraphics[width=0.99\linewidth]{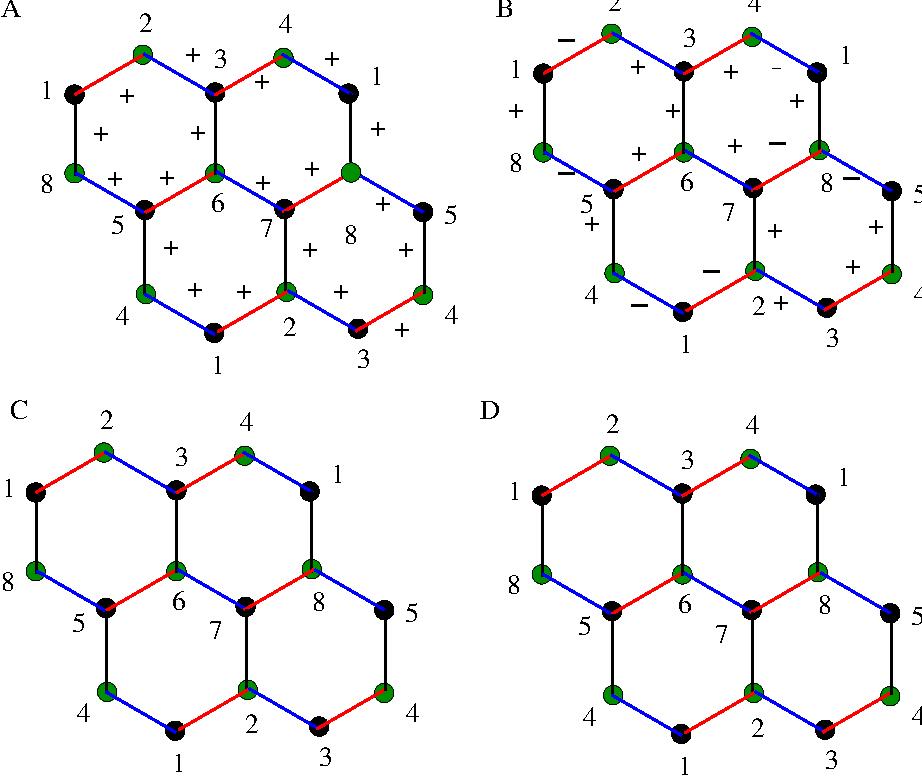}
\caption{\label{fourchoice}In the panel A and B, two choices of conserved $Z_2$ gauge fields are shown to yield identical distribution of $B_p$. For C and D two equivalent $Z_2$ gauge field choices can be found to yield $B_p=-1$ for all the plaquettes.}
\end{figure}

We observe from Eq.~\ref{3} that Hamiltonian is functional of configurations of conserved quantity $u_{ij}$ defined on each bond. Such Hamiltonian can be represented by block diagonal form in the basis of eigenstates of $u_{ij}$ as represented by the small square matrix of $N \times N$ dimension in Fig.~\ref{matrix}. Each sub-block refers to a certain distribution of $u_{ij}$s and as explained before   each block corresponds to a certain distribution of gauge invariant conserved quantity $B_p$ for each plaquette. There are many distinct configurations of $u_{ij}$ which yields  an unique configuration of $B_p$. We also know that  eigenvalue of $B_p$ is $\pm 1$.In Fig. \ref{fourchoice}, A and B, we have shown
two inequivalent gauge field distributions for uniform $B_p=1$ for each plaquette. One can similarly find two different distributions of gauge fields with $B_p=-1$ for each plaquette. Now the important question is the following. Does   ground states obtained from each sub-block in Fig.~\ref{matrix} yields same energy or does it depend on the distribution of $u_{ij}$ or  on the distributions of $B_p$. As the $u_{ij}$ is not a gauge invariant quantity and physically not observable, the energy is not directly dependent on the distribution of $u_{ij}$s rather it depends on the distribution of $B_p$s. The next question is which distribution of $B_p$ yields the absolute minima. From a very remarkable theorem \cite{lieb-1994} by E. Lieb we know that uniform configuration of $B_p=1$ for each  plaquette yields the global minima. This is obtained by fixing $u_{i,j}=1$ for every link (there are many other configurations which yields $B_p=1$, however each of them would yield identical ground state energy). This has also been confirmed by Kitaev numerically \cite{kitaev-2006}. For the uniform choices of $u_{i,j}$s (which corresponds to global minima ) we can easily diagonalize the Hamiltonian and get the ground state wave function.

\begin{center}
\Large{Physical wave function and projection operator}
\end{center}

With the discussion of the foregoing paragraph, let us call the wave function obtained by diagonalizing  Eq.~\ref{3} with uniform configuration of $u_{ij}=1$ (yielding $B_p=1$ for each  plaquette) as $|\psi\rangle_{ext}$. We have deliberately added the subscript `$ext$'  to remind the fact that the above wave function is obtained  in the extended Hilbert space. Whenever we make an operation which takes us from an Hilbert space $H^{>}$ with more number of states to another Hilbert space $H^{<}$ with less number of states such that some states are excluded in $H^{<}$, we need a projection operator. If any states $\Psi^{>}$ belongs to $H^{>}$, the corresponding states in $H^{<}$ after projection is obtained as
$\Psi^{<}= P \Psi^{>}$ where $P$ is the projection operator. The  projection operator  removes the unphysical states and keep only the physical states in the expansion of $\Psi^{>}= \sum_i \gamma_i  |i \rangle $ where $\gamma_{i}$ is complex coefficient and $|i \rangle$ is the normalized basis vector belonging to $H^{>}$. Actually $| i \rangle$ can be decomposed into two groups $| i_< \rangle$ and $| i_> \rangle$ such that $ | i_< \rangle$  constitutes the normalized  basis  vector of $H^{<}$. The job of $P$ is to remove or annihilate the states $| i_> \rangle$ such that $P \Psi^{>}$ involves only the states belonging to $H^{<}$. 
\begin{equation} 
\label{4}
|\psi\rangle_{\rm phy}=\hat{P}|\psi\rangle_{ {\rm ext}}
\end{equation}
\indent 

Now the complex fermions $c_i, c^{x}_{i,a},c^{y}_{i,a},c^{z}_{i,a}$ that have been used to define the Majorana fermionization of spin operators have the states $|00 \rangle, ~|10 \rangle= c_1^{\dagger}|00 \rangle , ~|01 \rangle= c_2^{\dagger}|00 \rangle,~ |11 \rangle =c^{\dagger}_1c_2^{\dagger}|00 \rangle$. It is clear that the original spin  states are needed  to be mapped to these four states and there are  enlargement of states. To understand the mapping let us recall that $D=\sigma_x \sigma_y \sigma_z=i$ is an identity which must be hold true for any states within the physical Hilbert space. If we calculate $D$ according to the definition  given before one finds $D=i (1-2 d^{\dagger}_1 d_1) (1-2 d^{\dagger}_2 d_2)$. Now we see that $D=i$ holds true only for the states $|00\rangle$ and $|11\rangle$. Thus any physical states should  have the general representation $\Psi_{ph}= a_{00}|00\rangle  + a_{11} |11\rangle$. But while working on the extended Hilbert space we  encounter states $\Psi_{ext}= a_{00}|00\rangle  + a_{11} |11\rangle +a_{01}|01\rangle  + a_{10} |10\rangle$. How do we get rid of the unphysical states $|01 \rangle$ and $|10 \rangle$. Note that the operator $P=(1+D)/2$ acting on this unphysical states yields zero and keeps the physical states as it is. Thus it is straightforward to check that $P \Psi_{ext}= \Psi_{ph}$. Now this is for  a given site and we call this projection
operator $P_i= (1+ D_i)/2$. The above procedure can be extended to all the sites of entire system and the total projection operator is defined as,  
\begin{equation}
\label{5}
\hat{\mathcal{P}}=\prod_{i \epsilon {\rm all~~sites}}\frac{(1+D_i)}{2} .
\end{equation}

Now we discuss a little more about specific property of this projection operator and its practical implementation to obtain physical wave function as discussed in \onlinecite{daniel-2011}. From Eq. \ref{3}, after a canonical transformation $\mathcal{Q}^{\mu}$, one obtains  new Majorana operators as follows,
\begin{eqnarray}
(b_1, b_2,...,b_{2N-1}, b_{2N})= (c_1,....., c_{2N}) \mathcal{Q}^{\mu},
\end{eqnarray}
where  in $\mathcal{Q}^{\mu}$, the index $\mu$ refers to the fact $\mathcal{Q}$ depends on the gauge field distributions $[\mu_i]$. The above transformation renders Eq. \ref{3} as $H_{\mu}= \frac{i/2} \sum_m b_{2m-1}b_{2m}$ where $\epsilon_m$s are the positive eigenvalues of $2i \mathcal{H}[u]$. One can define the new fermionic operator as $\alpha_m= \frac{1}{2} (b_{2m-1} + i b_{2m})$ such that  the Hamiltonian becomes diagonal in the occupation number of $\alpha$ operator with $H= \sum_{m} \epsilon_{m} (n_m-1/2)$ with the ground state energy $E_0= -\frac{1}{2}$. Now it is instructive to look at the projection operator $\mathcal{P}$ in more details as explained below,
\begin{eqnarray}
\mathcal{P}=\prod^{2N}_{i=1} \left( \frac{1 + D_i}{2}  \right)= \frac{1}{2^{N}} \sum_{\{ i \}} \prod_{i \in \{ i\}} D_i .
\end{eqnarray}

In the above,  summation includes all possible subsets of indices $\{ i\}$. One can easily very that $\mathcal{P}$ can be written as sum of terms such that for a given product of $\mathcal{L}_i=\prod_{i \in \{ i\}} D_i$, there exists its conjugate  $\mathcal{L}_j=\prod_{j\in \{ j\}} D_j$  so that  $\mathcal{L}_i \mathcal{L}_j= \prod^{2N}_i D_i$. This allows us to write,
\begin{eqnarray}
\mathcal{P}&&= \left( \frac{1}{2^{2N-1}} \sum^{\prime}_{\{ i\}} \prod _{i \in \{i\}} D_i  \right) \cdot \left(\frac{1+ \prod^{2N}_{i} D_i }{2}\right), \nonumber \\
&& =\mathcal{S} \cdot \mathcal{P}_0.
\label{eqsp0}
\end{eqnarray}
In the above we use $\mathcal{S}= \left( \frac{1}{2^{2N-1}} \sum^{\prime}_{\{ i\}} \prod _{i \in \{i\}} D_i  \right)$ and $\mathcal{P}_0= \left(\frac{1+ \prod^{2N}_{i} D_i }{2}\right)$
In Eq. \ref{eqsp0} $\mathcal{S}$ includes all possible summations such that its conjugates are not considered.  Now we note that $\prod^{2N}_{i} D_i$ in $\mathcal{P}_0$ contains all possible $c_i, b^x_i, b^y_i, b^z_i$. It is evident that $b^{\alpha}_i$'s can be group together to yield $u^{\alpha}_i$. However this rearrangement of $b^{\alpha}_i$ with its partner to form $u^{\alpha_{ij}}$  brings a phase that depends on particular definitions of $u^{\alpha}_{ij}$ in a given lattice. Apart from this we  also have $ \prod c_i$. Considering all these, as shown \cite{daniel-2011}, $\mathcal{P}_0$ can be written as,
\begin{eqnarray}
	\label{parity}
2 \mathcal{P}_0 = 1 + (-1)^{\theta} {\rm det} (\mathcal{Q}^{\mu}) \hat{\pi} \prod_{\langle i,j \rangle} u_{i,j} ,
\end{eqnarray} 
where $\hat{\pi}= \prod^{N} _{m}(1-2n_m)$ and denotes the parity of the fermions. The factor $\theta$ is an integer which depends on the definition of $u_{i,j}$ on a given lattice. The usefulness of the expression given in Eq.~\ref{parity} is that the second term is either $+1$ or $-1$ depending on the parity of the state and  thus makes it easy to implement the projection operator. In short this analysis at once tells that either the odd or even particle sector only contributes in physical wave function. We would like to mention that this parity operator is useful even in the case of implementation of Jordan-Wigner fermionization (without enlarging the Hilbert space dimension) to map the spin operators by appropriate complex fermions as confirmed by recent studies\cite{pervez2023decipheringI,pervez2023decipheringII}. 

\subsection{The ground state}
\label{grst}
We have already argued that it is the uniform configuration of $B_p=1$ which contains the global minima of the spectrum. Here we consider the choice $u_{ij}$ = 1 for each bond which is one of
the realizations of $B_p=1$ for each plaquette. After doing that the  Majorana fermion
hopping Hamiltonian given in Eq.~\ref{ham0} reduces to a translationally  invariant Hamiltonian facilitating easy solution using Fourier transformations. The translational invariant Hamiltonian  is given by,

\begin{eqnarray}
\label{ham}
 H &&=\sum^{N_x}_{\langle ij \rangle_x=1}J_{x}ic_{i,a}c_{j,b}+
\sum^{N_y}_{ \langle ij \rangle_y=1}J_{y}ic_{i,a}c_{j,b} + \nonumber \\
 &&~\sum^{N_z}_{ \langle ij \rangle_z=1}J_{z}ic_{i,a}c_{j,b}.
\end{eqnarray}
To diagonalize the above Hamiltonian, it is necessary to  consider a system of definite size.
Here we consider a lattice with $M$ and $N$ unit cells in the directions of $\vec{a}_1$ and $\vec{a}_2$ respectively as shown in Fig.~\ref{fig1}.  Unit cell is defined incorporating two sites connected by a vertical (z-type) bond. The general expression of the position of any unit cell is obtained by the vector $\vec{R}_{mn}= m \vec{a}_1 + n \vec{a}_2$ where we consider the mid point of any reference unit cell as the origin. It is easy to find that a pair of sites appearing in the first term of Eq. \ref{ham}  can be represented by $ J_x i c_{\vec{R}_{mm,a}} c_{\vec{R}_{mn,b} + \vec{a}_1 + \vec{a_2}}  $. Similarly the second and third terms are represented by $ J_y i c_{\vec{R}_{mn,a}} c_{\vec{R}_{mm,b} + + \vec{a_2}}  $ and $ J_x i c_{\vec{R}_{mn,a}} c_{\vec{R}_{mm,b}}  $ respectively. The summation now runs over ``$m$'' and ``$n$'' ranging from 1 to $M$ and $N$ respectively.  The momentum space vectors are obtained as ${\vec{k}}=\frac{p}{M}{\vec{b}_1} + \frac{q}{N}{\vec {b}_2}$, where ${\vec b_{1,2}} $ are the reciprocal lattice vectors  given by,
\begin{equation}
\label{revec}
{\vec{b_{1}}}= \frac{4\pi}{3}(\frac{\sqrt{3}}{2} \textbf{e}_{x} +\frac{1}{2} \textbf{e}_{y}) \,\,;\,\, {\vec{b_{2}}} = \frac{4\pi}{3}  \textbf{e}_{y} .
\end{equation}
 Here `$p$' and `$q$' varies from $-M/2$ to $M/2$ and $-N/2$ to $N/2$ respectively. The above discussion defines the Brillouin zone. To solve the above Hamiltonian  given in Eq. \ref{ham} we may define the  Fourier transformation for the Majorana fermions $c_{i,a}$ and $c_{j,b}$ appropriately. To ensure that one needs to  map the site index $i,a$ ($j,b$) to $\vec{R}_{mn,a}$ and $\vec{R}_{mn,b}$ appropriately  as discussed before. Henceforth for easy representation we use $c_{mn,s}$ for $c_{\vec{R}_{mn,s}}$ with $s=a,b$. With this discussion we define the Fourier transformation for the $c_{mn,s}$ operators as,
 \begin{equation}
 	c_{mn,s}= \sum_{k}\frac{1}{\sqrt{MN}} e^{i\vec{k}.\vec{R}_{mn}} c_{k,s}.
 \end{equation} 

We  notice  that the property $c^{\dagger}_i=c_i$  implies $c_{k}=c^{\dagger}_{-k}$. After performing the Fourier transformation, we get the Hamiltonian in k-space as follows,
\begin{eqnarray}
\label{k-hamch2}
H&=& \sum_{k \epsilon {\rm HBZ}} {(c^{\dagger}_{k,a}   c^{\dagger}_{k,b})} \left( \begin{array}{rr} 
0& if^{*}_{k}\\
-if_{k}&0 \end{array} \right) \left( \begin{array}{rr} 
c_{k,a}\\
c_{k,b} \end{array} \right).
\end{eqnarray}
In the above equation `HBZ' stands for half Brillouin zone. Note that for the condition $c_k=c^{\dagger}_{-k}$, all the Majorana modes are not independent. There is a specific way a Majorana fermion  with positive momentum is related with a Majorana fermion with a negative momentum with the same magnitude. Equivalently  the $c_k=c^{\dagger}_{-k}$  implies  that annihilating a Majorana fermion with momentum $k$ is same as creating a Majorana fermion with momentum $-k$. The spectral function $f_{k}$ is given by,
\begin{equation}
\label{spectral}
f_{k}= J_{z} + J_{x} e^{-ik_{1}} + J_{y} e^{-ik_2}.
\end{equation}
In above expression `$k_{1}$' and `$k_2$' are the components of $\vec{k}$ along `$x$' bond and `$y$' bond respectively. They are given by,

\begin{eqnarray}
k_1&=&\vec{k}.\hat{n}_1 \,\,;\,\, k_2=\vec{k}.\hat{n}_2, \\
\hat{n}_1&=& \frac{1}{2}\hat{e}_x +\frac{\sqrt{3}}{2} \hat{e}_y \,\,;\,\,\hat{n}_2= \frac{-1}{2}\hat{e}_x +\frac{\sqrt{3}}{2} \hat{e}_y.
\end{eqnarray}
Here $\hat{n}_1$ and $\hat{n}_2$ are the unit vectors along the `$x$' and `$y$' bond respectively. The Hamiltonian given in Eq.~\ref{k-hamch2}  can be diagonalized easily with the following unitary  transformation given below,
\begin{eqnarray}
\label{kham1}
\left( \begin{array}{r} c_{k,a}\\c_{k,b} \end{array} \right) = \frac{1}{\sqrt{2}} \left( \begin{array}{rr} v_{k} &-v_{k} \\
1&1\end{array} \right)  \left( \begin{array}{r} \eta_{k} \\ \xi_{k} \end{array} \right),
\end{eqnarray}

with $v_{k}=i f^{*}_{k}/ |f_{k}|$. The diagonalised Hamiltonian is given by,
\begin{eqnarray}
\label{h-diag}
H= \sum_{k} E_{k} (\eta^{\dagger}_{k} \eta_{k} - \xi^{\dagger}_{k} \xi_{k}),
\end{eqnarray}
where $E_{k}=|f_{k}|$ is the quasiparticle energy associated with new field operators $\eta_k$ and $\xi_k$. The ground  state is obtained by filling up all the negative energy states of  $\xi_k$ quasiparticles and can be written as,

\begin{equation}
\label{gr-st}
|G \rangle = \Pi_{k, {\rm HBZ}} \xi^{\dagger}_{k} |0 \rangle,
\end{equation}

where $ |0 \rangle$ represents the vacuum state such that $\eta |0 \rangle = \xi_{k} |0 \rangle =0$. Here the summation is over the half first Brillouin zone as explained before.  To find whether the spectrum is gapless or not we solve for  $E_k=0$ which implies $f_k=0$. It turns out that the, $f_k=0$ has solutions if and only if $|J_x| ,|J_y|, |J_z|$ satisfy the following triangle inequalities,

\begin{eqnarray}
\label{phasecon}
&&|J_x| \le |J_y| +|J_z| , \,\,\, |J_y| \le |J_x| +|J_z| , \nonumber \\ 
&& |J_z| \le |J_x| +|J_y|.
\end{eqnarray}
\begin{center}
\begin{figure}[h!]
\psfrag{x}[lb][lb][1]{$J_x=1, J_y=J_z=0$}
\psfrag{y}[lb][lb][1]{$J_y=1, J_x=J_z=0$}
\psfrag{z}[lb][lb][1]{$J_z=1, J_x=J_y=0$}
\psfrag{x}[lb][lb][1]{$J_x=0$}
\psfrag{y}[lb][lb][1]{$J_y=0$}
\psfrag{z}[lb][lb][1]{$J_z=0$}
\includegraphics[width=0.8\linewidth]{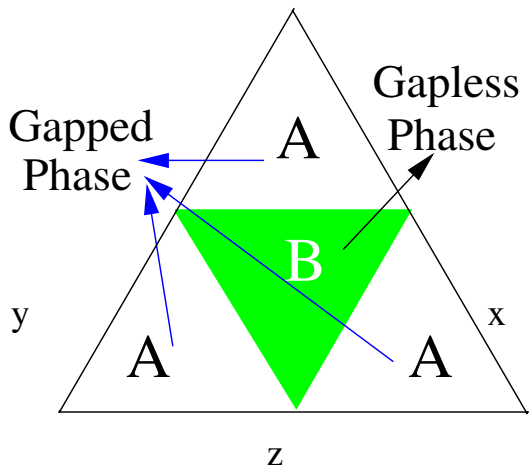}
\caption{\label{phased}Phase diagram for Kitaev model in the parameter space. A point in the above triangle describes relative magnitudes of $J_x, J_y, J_z$. Three sides of the triangle describe $J_x=0, J_y=0$ and $J_z=0$ as given in the figure. The region `A' is gapped and the region `B' is gapless. The gapless region acquires a gap in the presence of magnetic field. }
\end{figure}
\end{center}
The above inequalities as given in Eq.~\ref{phasecon} can be represented as a point inside an equilateral triangle which has been shown in Fig.~\ref{phased}.

In Fig.~\ref{phasegapless} and Fig.~\ref{phasegapped}, we plotted how the spectrum looks like for gapless and gapped phase respectively.
\begin{center}
\begin{figure}
\psfrag{d}{-2}
\psfrag{z}{$E_k$}
\psfrag{b}{2}
\psfrag{x}{$k_x$}
\psfrag{y}{$k_y$}
\psfrag{c}{-5}
\psfrag{a}{0}
\psfrag{e}{5}
\includegraphics[width=0.99\linewidth]{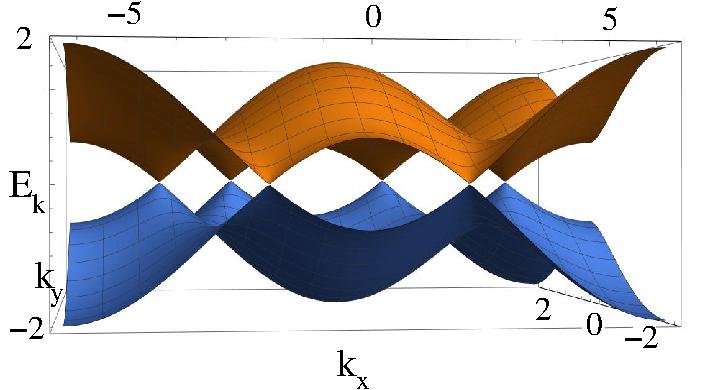}
\caption{\label{phasegapless} Spectrum $E_{k}$ as defined in Eq.~\ref{h-diag} has been plotted  for $J_x=J_y=J_z=1$. We observe that there are six points in the Brillouine zone where $E_k$ vanishes resulting a gapless spectrum. The dispersion near this gapless points are also linear.}
\end{figure}

\begin{figure}
\psfrag{d}{-4}
\psfrag{z}{$E_k$}
\psfrag{b}{2}
\psfrag{x}{$k_x$}
\psfrag{y}{$k_y$}
\psfrag{c}{-2}
\psfrag{a}{0}
\psfrag{e}{4}
\psfrag{f}{-5}
\psfrag{g}{5}
\includegraphics[width=1.\linewidth]{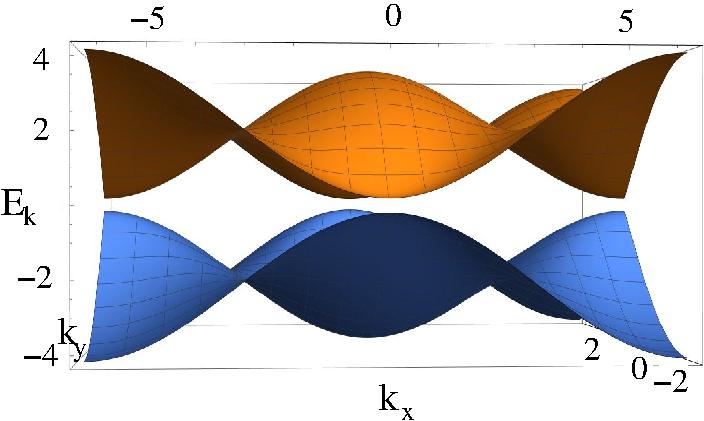}
\caption{\label{phasegapped}In the above $E_{k}$ has been plotted for $J_x=J_y=1,~ J_z=2.2$. We note that there is a gap between  valence band and conduction band. The spectrum near the minimum for conduction band is quadratic.}
\end{figure}
\end{center}

If the inequalities are strict, there are exactly two solutions: ${\bf k}=\pm {\bf q_*}$, one in each HBZ . The region defined by inequalities in Eq.~\ref{phasecon} is the shaded region  B in Fig.~\ref{phased}; this phase is gapless. The region marked by A is gapped. The low energy excitations are different in these two phases. In the gapless phase the low-energy excitations are the Majorana fermions but in the gapped phases the low-energy excitations correspond to the vertex excitations which corresponds to the excitation of $B_p$s, the conserved quantities. In the presence of magnetic field the phase B acquires a gap. These two regions are topologically distinct as indicated by spectral Chern number which is zero for phase A and one for the phase B  \cite{kitaev-2006}. We have argued that as the projection operator $\hat{P}$ commutes with the Hamiltonian, the solution obtained in the extended Hilbert space is exact. One can indeed show  that there  exist a non zero projection. But this method of solving gives eigenstates of the Hamiltonian as well as the eigenstates  of the conserved quantities in terms of Majorana fermions whose occupation number is not well defined. In the next section, we would extend the Majorana fermionization in  an useful way such that  eigenfunctions of the Hamiltonian as well as $u^{\alpha}_{ij}$  is represented by the usual  occupation number representation of complex fermions \cite{bruus}.

\section{Elementary correlation functions}
\label{sec-order}

We have seen that the initial spin Hamiltonian  given in Eq.~\ref{ham0} is reduced to an effective  quadratic fermionic Hamiltonian as given in Eq.~\ref{2}. The original physical object was spin-1/2  magnetic moment. For a two spins belonging to two different sites, spin angular momentum  commutes with each other if we express them by suitable representation. For spin-1/2 particle, the Pauli matrices are a faithful representations. However, the effective fermionic Hamiltonian consists of Majorana fermions  which anticummute irrespective of the site they belong to. This conversion of effective degrees of freedom is  known as emergent degrees of freedom due to interactions. However there is a definite connection between them.  The  eigenstates either obtained by diagonalizing the  original spin Hamiltonian or the effective Majorana Hamiltonian has one to one correspondence and they have identical eigenvalue spectrum. In  dealing with physical systems, generally one is interested in the ground state wave function at low temperature.  Now we must ask  what characterizes the ground states, what  are the properties of the ground states of different phases, that will differentiate one phase from another. Various order parameters are used to determine a certain phase and distinguish from other. The magnetization or spin-spin correlation function between the effective degrees of freedom is one of such  measures. In many occasion such correlation functions can only be calculated approximately. However, remarkably, Kitaev model renders us to compute the correlation functions exactly \cite{smandal-prl}. Mathematically the correlation between two observables or operators $\hat{O}_1$ and $\hat{O}_2$ is expressed as $<\hat{O}_1 \hat{O}_2>$ where $<...>$ denotes ground state expectation value at zero temperature \cite{goodstein1994}. At finite temperature it implies a thermal average. Physically it means what is joint probability that if the operator $\hat{O}_1$ takes value $O_1$ and the operator $\hat{O}_2$ takes value $O_2$. Apart from correlation function magnetization is also used to characterize phases of magnetic and interacting spin systems. Magnetization is defined as the average value of magnetic moment in a system and at zero temperature for quantum mechanical system it is defined as $< \vec{M} >$ where $\vec{M}= \frac{1}{N} \sum_i \vec{m}_i$ and  the angular bracket implies expectation with respect to ground state. Here $\vec{m}$ is the magnetic moment at a given site. Now we  follow an exact calculation of magnetization and two-spin correlation function to find out  what kind of order is exhibited by the ground state wave function. To do that we first discuss an extension of Kitaev's Majorana fermionization such that the mathematical steps to calculate the spin-spin correlation function or magnetization is easily understood.

\subsection{\label{bond-ferm}Bond fermion formalism}
   We have  seen in  Sec.~\ref{fso2.1} that  two complex fermions yield four Majorana fermions as  a single complex fermion can be rewritten into two Majorana fermions. Now to facilitate the easy computation of spin-spin correlations we  invert the above procedure by regrouping two different Majorana fermions to define a complex fermion. We have seen that at every bond there has been one conserved quantity named $u^{\alpha}_{i,j}$  made out of the Majorana fermion $c^{\alpha}_{i,a}$ and $ c^{\alpha}_{j,b}$. Here `$i$' and `$j$' denotes the two sites of a bond, `$a$' and `$b$' denotes sublattice indices and `$\alpha$' denotes a specific bond($\alpha=x,y,z$). We  regroup these two Majorana fermions to define a complex fermion named $ \chi_{\langle ij \rangle_{\alpha}}$ which lives on the bond joining sites `$i$' and `$j$'. We call this procedure as bond fermion formalism. From now on we follow the convention that the site `$i$' in the bond  $\langle ij\rangle_{\alpha}$  belongs to `$a$' sublattice and the site `$j$'  belongs to `$b$' sublattice. Also from now on we do  not mention the sublattice index `$a$' and `$b$' explicitly. We define complex fermions on each bond as,
\begin{eqnarray}
\label{chiadef1}
\chi_{\langle ij\rangle_{\alpha}}&=&\frac{1}{2}\left(c_i^{\alpha}+ic_j^{\alpha}\right),\\
\label{chiadef2}
\chi^\dagger_{\langle ij\rangle_{\alpha}}&=&\frac{1}{2}\left(c_i^{\alpha}-ic_j^{\alpha}\right).
\end{eqnarray}
\begin{center}
\begin{figure}[h!]
\center{\includegraphics[width=0.45\textwidth]{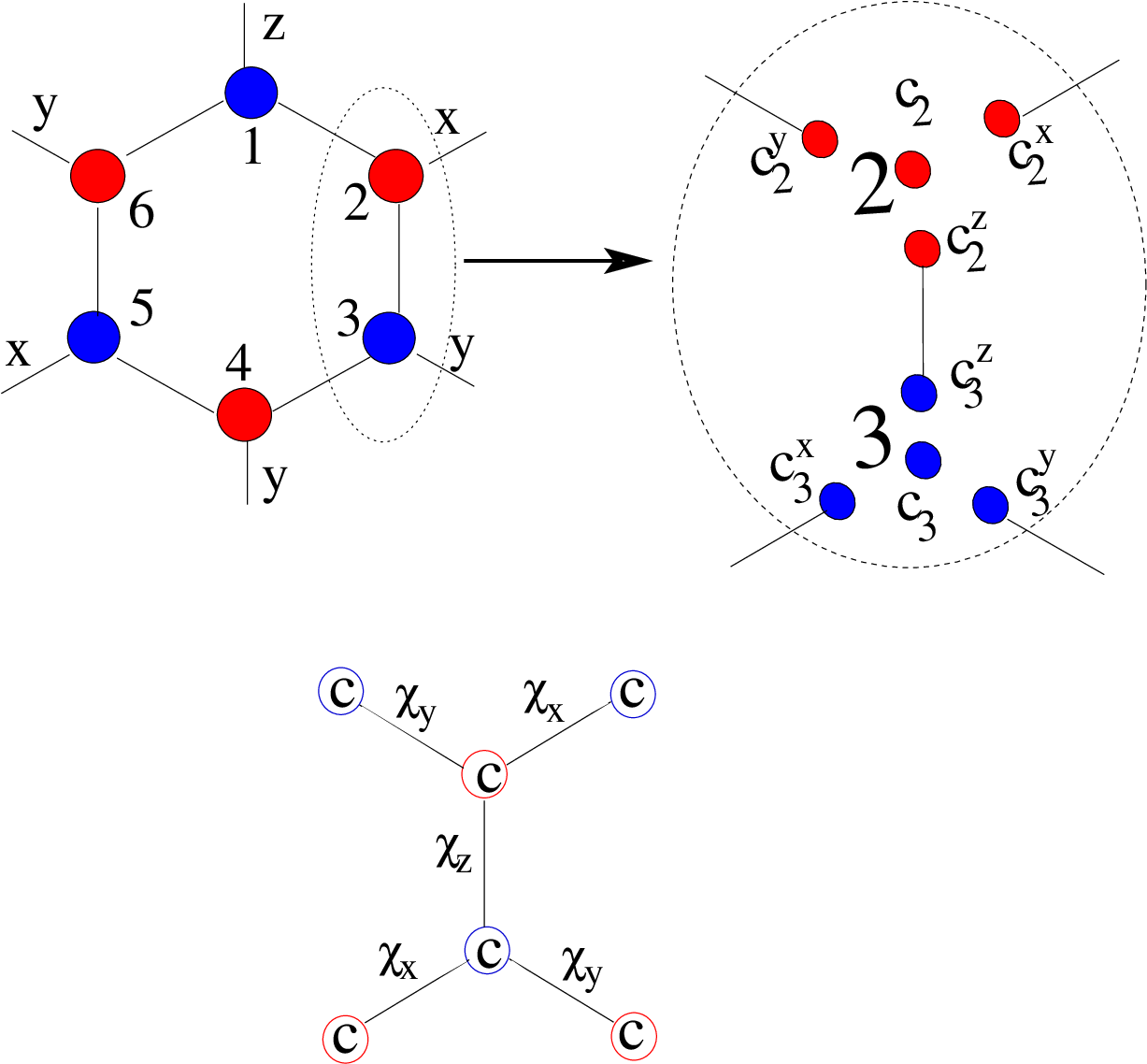}}
\caption{\label{bndfm}Elementary hexagon and `bond fermion' construction is shown schematically. A z-bond comprising site 2 and 3 is enlarged and each site is replaced by four filled in circles which represent four Majorana fermions $\left(c,c^{x},c^{y},c^{z}\right)$  representing a given spin at a site. The central Majorana fermion remains associated with the site only. However each other blue(or red) Majorana  pairs up with another red(or blue) Majorana on the other end and forms a complex fermion on the bond. This is schematically shown in the lower panel where bond fermion $\chi_{\langle23\rangle}$ is formed due to pairing of $c^z_2$ and $c^z_3$ according to Eq. \ref{chi23}. }
\end{figure}
\end{center}

For example with reference to the Fig.~\ref{bndfm}, for the $z$-bond joining site 2 and  3, and for the $y$-bond joining site 1 and  2, we define,

\begin{eqnarray}
\chi_{\langle23\rangle_z}&&=(c^{z}_2 +ic^{z}_3),
\label{chi23}  \\
\chi_{\langle12\rangle_y}&&=(c^{y}_1 +ic^{y}_2).
\label{chi12}
\end{eqnarray}
Then it follows that for the site `2' and `3' the $\sigma^{z}$ operator becomes,
\begin{eqnarray}
\sigma^{z}_2 &&= ic_2(\chi_{\langle23\rangle_z} + \chi^{\dagger}_{\langle23\rangle_z}), \\
\sigma^{z}_3 &&= c_2(\chi_{\langle23\rangle_z} - \chi^{\dagger}_{\langle23\rangle_z}).
\end{eqnarray}

Below we write the  result of this refermionization for a bond of type `$\alpha$' joining site `$i$' and `$j$',
\begin{eqnarray}
\label{chif}
&&\chi_{\langle ij\rangle_{\alpha}}=\frac{1}{2}\left(c_i^{\alpha}+ic_j^{\alpha}\right), \\
&&\sigma_i^{\alpha}=ic_i\left(\chi_{\langle ij\rangle_{\alpha}}
+\chi^\dagger_{\langle ij\rangle_{\alpha}}\right), \\
&&\sigma_j^{\alpha}= ic_j\left(\chi_{\langle ij\rangle_{\alpha}}
-\chi^\dagger_{\langle ij\rangle_{\alpha}}\right).
\end{eqnarray}
It is clear that three components of a spin operator at a given site gets connected to three  different $\chi$ fermions defined on the three different bonds emanating from it. The bond variables are related to the number operators of these fermions,
${\hat u}_{\langle ij\rangle_{\alpha}}\equiv ic_i^{\alpha}c_j^{\alpha} = 
2\chi^\dagger_{\langle ij\rangle_{\alpha}}\chi_{\langle ij\rangle_{\alpha}}-1$. Thus the effective picture is understood easily from the Fig.~\ref{bndfm}.  We identify  a $\chi$ fermion on every bond whose occupation number can be zero or one. This occupation number determines the value of $u_{\langle ij \rangle}$ on that bond. But these fermions are conserved and  serve as an effective $Z_2$ gauge field for hopping `$c$' fermions. As $\chi$ fermions are conserved, all eigenstates can therefore be chosen to have a definite $\chi$ fermion occupation number. The Hamiltonian is then block diagonal in occupation number representation, each block corresponding to a distinct set of $\chi$ fermion occupation numbers for every bonds. Thus all eigenstates in the extended Hilbert  space take the following factorized form,
\begin{eqnarray}
\label{es1}
\vert{ \Psi}\rangle =  \vert {\cal M}_{\cal G};{\cal G }\rangle & \equiv &
\vert {{\cal M}_{\cal G}}\rangle
\vert{\cal G }\rangle,\\
\label{es2}
{\rm with}~~ \chi^\dagger_{\langle ij\rangle_{\alpha}}\chi_{\langle ij\rangle_{\alpha}}
\vert{\cal G} \rangle
&=&n_{\langle ij\rangle_{\alpha}}
\vert \cal G \rangle ,
\end{eqnarray}
where $n_{\langle ij\rangle_{\alpha}}=(u_{\langle ij\rangle_{\alpha}}+1)/2$ and 
$\vert{\cal M}_{{\cal G}}\rangle$ is a many body eigenstate in the matter 
sector determined by `$c$' fermions, corresponding to a given $Z_2$ field configuration determined by $ \vert {\cal G} \rangle $. \\
\indent
\begin{figure}[h!]
\psfrag{a}[cb][cb]{$|\tilde{\psi} \rangle$}
\psfrag{c}[cb][cb]{$\sigma^{z}_{i} $}
\psfrag{b}[cb][cb]{$|\psi^{\prime} \rangle$}
\hspace{1.5cm}\includegraphics[width=0.8\linewidth]{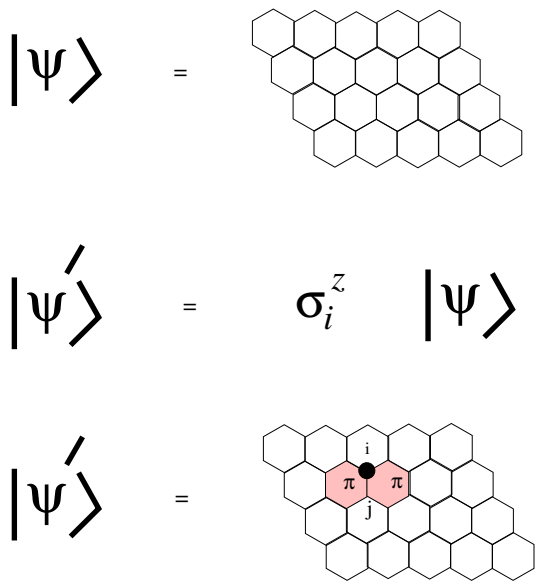}
\caption{\label{spflx}How a spin fractionalizes into two static $\pi$ fluxes and a dynamic Majorana fermion is shown. $ |\Psi \rangle$ is a state with zero flux. We apply $\sigma^{z}_{i} $ where site `$i$' is connected with site `$j$'. As a result we get a state $|\Psi^{\prime} \rangle$ with two static $\pi$ fluxes (represented by pink colored plaquettes) at the plaquette sharing bond $\langle ij \rangle$ and a dynamic Majorana fermion represented by black filled circle.}
\end{figure}
\indent
Now we discuss the results of the above transformation and find that it brings an immediate visualization of the spin operator. We observe  form Eq.~\ref{chif} that the effect of $\sigma^{\alpha}_i$ on any eigenstate becomes very clear. When the spin operator acts on any eigenstate, in addition to adding a Majorana fermion at site `$i$', it changes  the bond fermion number from zero to one and vice versa  (equivalently, $u_{\langle ij \rangle_{\alpha}} \rightarrow - ~ u_{\langle ij \rangle_{\alpha}}$), at the bond $\langle ij \rangle_{\alpha}$. The end result is that one $\pi$ flux  is added to each of the two plaquettes that are shared by the bond 
$\langle ij \rangle a$ (Fig. ~\ref{spflx}). Mathematically the action of any spin operator on any eigenstate can thus be written as, 
\begin{eqnarray}
\label{pinot}
\sigma_i^{\alpha} = ic_i\left(\chi_{\langle ij\rangle_{\alpha}}
+\chi^\dagger_{\langle ij\rangle_{\alpha}}\right) ~~\rightarrow ~~ ic_i~
{\hat\pi}_{1\langle ij \rangle_{\alpha}}~{\hat\pi}_{2\langle ij \rangle_{\alpha}},
\end{eqnarray}
with ${\hat \pi}_{1\langle ij \rangle_{\alpha}}$ and  ${\hat \pi}_{2\langle ij \rangle_{\alpha}}$ defined as operator that creates additional $\pi$ fluxes to two adjacent plaquettes shared by the bond $\langle ij \rangle_{\alpha}$ (Fig.~\ref{spflx}). Now it is easy to understand the action of one more spin which is connected with the previous ones. It yields ${\hat \pi}_{1\langle ij \rangle_{\alpha}}^2 = 1 $, since adding two $\pi$ fluxes is equivalent to adding (modulo $2\pi$) zero flux. This signifies that while action of single spin operator creates the gauge fermion occupation number to change (either decrease or increase by one),  gauge fermion occupation number can be brought back to  initial values by action of  the same spin or a different  neighboring spin. Only criteria is that the spin angular component of both the spins have to be same and this is determined by the nature of the bond they are connected with. It is now straightforward to understand that two states with different flux configurations have vanishing overlap as they belong to different distribution of $\chi$ fermion occupation numbers. Mathematically this implies that,
\begin{eqnarray}
\langle \mathcal{G} | \mathcal{G}^{\prime} \rangle = \delta_{n_{\mathcal{G}}, n_{\mathcal{G}^{\prime}}},
\label{nnpr}
\end{eqnarray}
where $ n_{\mathcal{G}}$ and $n_{\mathcal{G}^{\prime}} $ represent the distribution of $\chi$ fermions for the state $| \mathcal{G} \rangle$ and $| \mathcal{G}^{\prime} \rangle$ respectively. This observation is extremely helpful to compute spin-spin correlations exactly. Not only two-spin correlation function, magnetization can also be calculated exactly. Apart from two-spin correlation functions other multi- spin correlations can be calculated with straightforward generalizations of this fact that for any  spin-spin correlation to be non-zero the first necessary condition is that the simultaneous action of all the spin operators on the ground state must not change the flux configurations or equivalently must not alter the gauge fermion  occupation number of the bonds. This fact can be extended to all the eigenstates of the Kitaev model as well which is remarkable  for Kitaev model. \\
\subsection{Magnetization}
\label{sec-mag}
\begin{figure}
\includegraphics[width=0.99\linewidth]{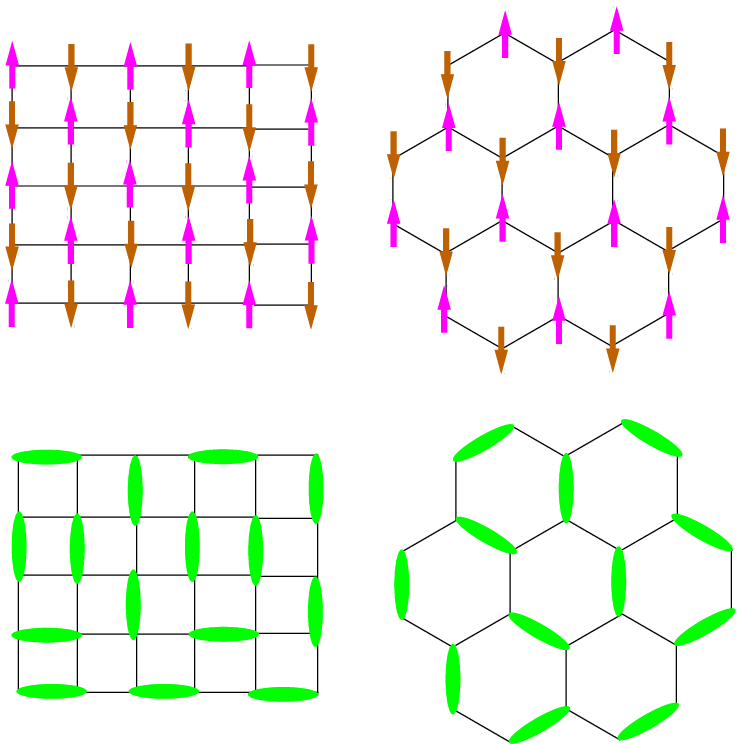}
\caption{\label{spliq}In the top panel we have shown antiferromagnetic spin configurations in square lattice and in honeycomb lattice. In the below we have shown a dimer state  in square and honeycomb lattice where the green ellipses represent a dimer. The particular arrangement of dimers to cover the every site of the lattice is not unique and there exist alternative arrangement as well. The alternative arrangements often are topologically different in the sense that one can not obtain one arrangement from the other by local operation of  destroying few singlet bonds and creating new ones. One often needs a global operation in achieving that\cite{sumathi-rao}.}
\end{figure}

Magnetization $\vec{\mathcal{M}}$ is an important physical quantity which is easily measurable and  can be controlled experimentally as well. A particular component of  magnetization  is defined as  below,

\begin{eqnarray}
\label{mageq}
\mathcal{M}^{\alpha}&&= \frac{1}{\mathcal{N}}  \sum^{\mathcal{N}}_{i=1}  \langle M^{\alpha}_i \rangle . 
\end{eqnarray}
Here $M_{\alpha}$ denotes the $\alpha$ component of total magnetic moment $\vec{M}$ at a given site `$i$'.
In the above $ \langle M^{\alpha}_{i} \rangle $ denotes the expectation value with respect to ground state. For ferromagnetic  phase it can be found that at least one component of $\langle M^{\alpha}_{i} \rangle  $ is non-zero at every site and it is identical for every site. On the other hand for antiferromagnetic phase $\langle M^{\alpha}_{i} \rangle$ is also non-zero at every site however their values are opposite at different sites and makes some pattern depending on the underlying structure yielding average magnetization zero. We note that though we have used $\vec{M}$ to define the magnetization, for our case it can also be defined in terms if spin-angular moment $\vec{S}$. For spin-1/2 $\vec{S}= \frac{\hbar}{2} \vec{\sigma}$ where $\sigma^{\alpha}$ denotes $\alpha$ component of Pauli matrices. In the following we shall omit $\frac{\hbar}{2}$  for brevity. In Fig.~\ref{spliq} we have shown such ferromagnetic and antiferromagnetic structure in square and honeycomb lattice. We can see easily that for antiferromagnetic state $ \mathcal{M}_{\alpha}$ is zero though for all the sites having
red spins are of opposite magnetization than that of blue spins. At a given site the average value of spin momentum is not zero. In the lower panel we have shown a different   state where  each shaded green region defined on a pair of sites refer the following singlet state $|s \rangle = \frac{1}{\sqrt{2}} (| \uparrow \downarrow \rangle - |\downarrow \uparrow \rangle )$. For such a state average value of $\sigma^{\alpha}_i=0$ at any site \cite{moessner-2001,philip,alain}. This state is fundamentally different than the antiferromagnetic state. For the antiferromagnetic state at a given site $\sigma^{\alpha}_i$ is not zero. Now let us see what is the value of $\langle \sigma^{\alpha}_i \rangle =0 $ for the  ground state of the Kitaev model. From the definition of spin operator as expressed in Eq.~\ref{pinot} we see that action of a spin on the ground state is to create two additional flux in the ground state as shown in Fig.~\ref{spflx}, in addition it also adds a Majorana fermion to the ground state. Mathematically this is expressed as 
\begin{equation}
\label{mageq2}
\sigma_i^{\alpha}\vert{\cal G}\rangle\vert{\cal M}_{\cal G}\rangle=
c_i\vert{\cal G}^{i\alpha}\rangle\vert{\cal M}_{\cal G}\rangle.
\end{equation}

Now as the state of different flux configurations are mutually orthogonal due to  different occupation number  of the conserved $\chi$ fermions, we obtain,

\begin{eqnarray}
&&\langle {\cal G} \vert \langle {\cal M}_{\cal G} \vert \sigma_i^{\alpha} \vert{\cal G}\rangle \vert{\cal M}_{\cal G}\rangle = \langle {\cal G} \vert \langle {\cal M}_{\cal G} \vert  c_i \vert{\cal G}^{i \alpha}\rangle \vert{\cal M}_{\cal G}\rangle =0 ,~~~~~~
\end{eqnarray}

because $ \langle {\cal G} \vert{\cal G}^{i\alpha}\rangle =0 $ following Eq.~\ref{nnpr}. Thus we see that the magnetization is zero for Kitaev model. It is zero for every site unlike the AFM state where the magnetization at a given site is not zero but when averaged over the system it is zero. For spin singlet state also the magnetization is zero at a given site.   However there is an important difference between the AFM state, and the spin singlet state shown in Fig.~\ref{spliq}. For AFM state $\langle \sigma^z_{i} \sigma^z_j \rangle = \pm 1$ depending on weather the site `$i$' and `$j$' are both   aligned in the same direction or in opposite direction. Note that it does not depend on the distance between the two sites. It is said that the state has a long-range ordered state.  For singlet state $\langle \sigma^z_i \sigma^z_j \rangle$ is not zero if the site `$i$' and `$j$' both belong to same singlet otherwise it is zero.  Thus there is no long-range correlation in the singlet  state.  However for the singlet state $\langle \sigma^{\alpha}_i \rangle$ is zero which  bears similarity with Kitaev model. Vanishing magnetization at any site individually is one of primary signature of spin liquid. It can happen that at a given site  any residual component of spin such as $\sigma^x_i$ is non-zero but other two component is zero. However there are important differences between the singlet state and Kitaev model which will be established once we calculate the two-spin correlation function. With the knowledge of few magnetic phases as discussed above, we now move on to calculate the spin-spin correlation function of Kitaev model which  would establish its spin liquid nature. 
It is straightforward to check that the  spin singlet product state is an unique state with no long-range correlation but short-range nearest neighbor correlation exist with $\langle \sigma^{\alpha}_i \sigma^{\alpha}_j  \rangle$ non-zero if `$i$'
and `$j$' are nearest neighbor and they belong to a given dimer.

\subsection{Two-spin correlations, fractionalization, de-confinement}
In the previous section we have shown explicitly that any component of magnetization at a given site vanishes. This points out spin rotational symmetry of the eigenstates
to be preserved. This can be compared with the vanishing average position or orientation of a molecule or atom in ordinary liquid phase of matter such was water. There is further similarity as well and hence the particular word spin liquid come into exist. In ordinary liquid state density-density  correlation function is short ranged and can be measured experimentally. In spin liquid the spin-spin correlation function is also short range. In many cases it extends up to few sites and  vanishes exponentially with distance.
Here we give an outline of computation of  spin-spin correlation functions in Kitaev model which is valid for any region of the phase diagram. This procedure is also valid for any eigenstate of Kitaev model which is a remarkable feature unlike other models. We show that two-spin correlation function is zero beyond nearest neighbor which indicates that Kitaev model is a spin liquid indeed. The derivation here is given in the extended Hilbert space but can be easily implemented without any difficulties to physical Hilbert space as can be seen using Jordan-Wigner transformations \cite{mandaljpa}. This happens because after the implementation of projection operator to ground state  wave function obtained in extended Hilbert space  one obtains many daughter states differed with respect each other only in  gauge sector by fermion occupation number in certain manner so that action of two spins on any of them still creates orthogonal states and mutual overlap between such states are again vanishes.  The above fact is expressed technically as the following. Because  spin operators are gauge invariant (commute with the projection operator) the  result of computation of spin-spin correlation function in the extended Hilbert space yields correct value.  First to begin with, we consider the two spin dynamical correlation  functions, in an arbitrary eigenstate of the Kitaev Hamiltonian in some fixed gauge field configuration ${\cal G}$,
\begin{equation}
\label{d2scf1}
S_{ij}^{\alpha \beta}(t)= \langle{\cal M}_{\cal G}\vert\langle{\cal G}\vert 
\sigma_i^{\alpha}(t)\sigma_j^{\beta }(0)
\vert{\cal G}\rangle\vert{\cal M}_{\cal G}\rangle.
\end{equation}
 Here $A(t) \equiv e^{iHt} A e^{-iHt}$ is the Heisenberg representation of an 
operator A, keeping $\hbar = 1$.  Physically the  quantity $ S_{ij}^{\alpha \beta}(t)$ in Eq.~\ref{d2scf1} gives the  joint probability amplitude  of finding a spin at `$j$' (at time zero) along  `$\alpha$' axis and of finding another spin at `$i$' to be  along `$\beta$' axis.  As discussed before we write the action of spin operator on any eigenstate as, 
\begin{eqnarray}
\label{sigones1}
\sigma_j^{\beta}(0)\vert{\cal G}\rangle\vert{\cal M}_{\cal G}\rangle&=&
c_j(0)\vert{\cal G}^{i\beta}\rangle\vert{\cal M}_{\cal G}\rangle ,\\
\label{sigones2}
\sigma_i^{\alpha (t)}\vert{\cal G}\rangle\vert{\cal M}_{\cal G}\rangle&=&
e^{i(H-E)t}c_i(0)\vert{\cal G}^{i\alpha}\rangle\vert{\cal M}_{\cal G}\rangle,
\end{eqnarray}
 where, $\vert{\cal G}^{i\alpha (j\beta)}\rangle$ denotes the state with extra
$\pi$ fluxes added to ${\cal G}$ on the two plaquette adjoining the 
bond in $\alpha$ or $\beta$  directions. It means that if $\alpha=x$, then the 
two plaquette excitations are created adjoining the site `$i$' to another  site `$k$' joined
by $x$-bond. Similar explanation goes for the $\beta$ index as well. In the above equation
$E$ is the energy eigenvalue of the eigenstate 
$\vert{\cal G}\rangle\vert{\cal M}_{\cal G}\rangle$. Since the $Z_2$ flux on 
each plaquette is conserved quantity due to the fact that it is determined by the occupation number
of gauge fermions ($\chi$ fermions) on the bonds, it is obvious that the correlation
function in Eq.~\ref{d2scf1} which is the overlap of the two 
states in equations \ref{sigones1}, \ref{sigones2} is zero unless 
the spins are on neighboring site. This indicates that for non-vanishing values of correlation function  we must
have `$j$' as the nearest neighbor site of `$i$'. Also note that the components $\alpha$ and $\beta$ must be identical so
that the effect of $\sigma^{\alpha}_i$ and $\sigma^{\beta}_j$ are such that  if the first one creates (annihilates) a pair of flux the second one  annihilates (creates) it bringing the flux configuration same as before. The above simple observations says that the
dynamical spin-spin correlation has the form,
\begin{eqnarray}
\label{d2scfres1}
S_{ij}^{\alpha \beta}(t)&&=g^{\alpha}_{ij}(t)\delta_{\alpha,\beta},{\rm ~for}~i,j {\rm~~nearest~neighbours}~,~~~~~~ \\
&&= 0 ~~~~~~~~~~~~~~~~~~~{\rm otherwise.}
\end{eqnarray}
 Computation of $g^{\alpha}_{ij}(0)$ is  straightforward 
in any eigenstate $|\cal M_{\cal G}\rangle$. For the ground state 
where conserved $Z_2$ charges are of value unity for all plaquette, we may have an analytical expression of the ground state wave function as given in Eq.~\ref{gr-st}.  For simplicity  we provide the expression for equal-time correlation by having $t=0$ without loss of generality. For correlation function at different time interested readers are requested to consult the original article \cite{smandal-prl}. Using Eq.~\ref{sigones1} and Eq.~\ref{sigones2}, the equal-time two-spin correlation function for the 
bond $\langle ij \rangle_{\alpha}$ reduces to the following simple expectation value :

\begin{eqnarray}
\langle \sigma^{\alpha}_{i}\sigma^{\alpha}_{j}\rangle \equiv
S^{\alpha\alpha}_{\langle ij \rangle_{\alpha}}(0)
= \langle{\cal M}_{\cal G}\vert\langle{\cal G}\vert 
c_i c_j
\vert{\cal G}\rangle\vert{\cal M}_{\cal G}\rangle.
\label{cicj}
\end{eqnarray}

In the above equation we have omitted the time index for $c_i$ and $c_j$ operator for simplicity. One can simplify the expressions in Eq.~\ref{cicj} by using the Fourier transformation of $c_i$ and $c_j$ operators and then employing the unitary transformation for $c_k$  operators as given in Eq.~\ref{kham1}, we obtain the following final expression for the two-spin equal-time correlation function,
\begin{eqnarray}
\nonumber
\langle \sigma^{\alpha}_{i}\sigma^{\alpha}_{j}\rangle \equiv
S^{\alpha\alpha}_{\langle ij \rangle_{\alpha}}(0)
=\frac{\sqrt{3}}{16\pi^{2}}\int_{BZ} \cos\theta(k_{1},k_{2})dk_{1}dk_{2}.
\end{eqnarray}
Where $\cos\theta(k_{1},k_{2})=\frac{\epsilon_{k}}{E_{k}}$, 
$ E_{k}=\sqrt{(\epsilon_{k}^{2}+\Delta^{2}_{k})}$, in the Brillouin zone. 
$\epsilon_{k}=2(J_{x} \cos k_{1} +J_{y}\cos k_{2} +J_{z})$,  
$\Delta_{k}=2(J_{x}\sin k_{1}+J_{y}\sin k_{2})$,
$k_{1}=\textbf{k}.\textbf{n}_1$, $k_{2}=\textbf{k}.\textbf{n}_2$ and 
$\textbf{n}_{1,2}=\frac{1}{2}\textbf{e}_{x} \pm
\frac{\sqrt{3}}{2}{\textbf{e}_{y}}$ are unit vectors along $x$ and $y$ bonds.
At the point, $J_x=J_y=J_z$, we get 
$S^{\alpha\alpha}_{\langle ij \rangle_{\alpha}}(0) = -0.52 $.
\begin{figure}
\includegraphics[width=0.95\linewidth]{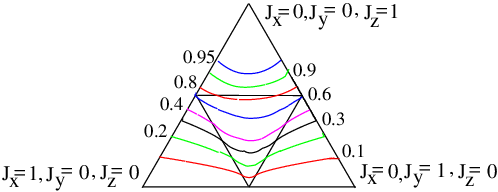}
\caption{\label{zzcore0}Contour plot of nearest-neighbour $z-z$ correlation in the parameter space. We note that $z-z$ correlation increases as magnitude of $J_z$ increases.}
\end{figure}
It is of interest to  examine how this non-vanishing two-spin correlation depends on the
parameters of the Hamiltonian mainly on $J_x, J_y, J_z$. The Fig.~\ref{zzcore0} shows 
how the nearest neighbor $z-z$ correlation varies in the parameter ( $J_x , J_y, J_z$ ) space.
As we have calculated the above two-spin correlation function for two spins connected by a $z$-bond,
we observe that as we increase the magnitude of $J_z$, the value of correlation also increases as shown in Fig. \ref{zzcore1} .
In particular two limiting cases in the parameter values is worth examining. Consider the case
$J_z=0$ where Kitaev model reduces decoupled x-y chains.  Thus  the two spins which  have been joined by an exchange coupling parameter
are now  without any such coupling between them, though the spins maintain their original interactions with
nearest neighbor spins in the respective chains. As there is no interaction between these two spins, we
expect that there will be no correlation between them and from Fig.~\ref{zzcore01}, we do find that
indeed the correlation between them is zero. The other limiting cases is $J_z= \infty$ or $J_x=J_y=0$, in this
limiting case the two spins connected by a `$z$' bonds does not interact with any other spins except the one
connected by $J_z$ bonds. Thus one expects that the correlation will be stronger and reach the saturation value one
as shown in Fig.~\ref{zzcore01}. Another interesting fact to note that there is little change in slope as we increase 
from gapped region to gapless region. \\
\begin{figure}
\label{zzcore}
\includegraphics[width=0.9\linewidth]{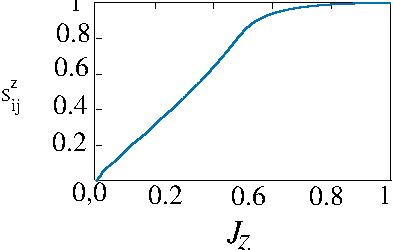}
\caption{\label{zzcore01} In the above we plot the $\langle \sigma_{1z} \sigma_{2z} \rangle (= {\rm s^z_{ij}})$ on a give $z$-bonds.
We have taken $J_x=J_y=0.5$ and vary $J_z$ from 0 to 1. As expected for $J_z=0$ correlation is zero and for large values it saturates to 1.}
\end{figure}
\\
\indent 
\vspace{0.5cm}
\\
\textit{\bf{Fractionalization and Deconfinement}:}  We note that here we have only discussed the derivation of equal-time correlation function which means that in Eq.~\ref{d2scf1} both $\sigma$'s have identical argument  for time. We find that the spin-spin correlation function is short-range and also bond dependent. This property of short-range and bond dependent nature of correlation function remains unchanged for different time correlation function as well. However the derivation and the exact expression  requires a little mathematical digression and for that we do not present that in this article. Instead  we discuss the issue without going into technical details to follow  the ensuing final result. Let us try to understand  the physical consequences of Eq.~\ref{d2scf1}, Eq.~\ref{sigones1} and ,Eq.~\ref{sigones2}. The r.h.s of Eq.~\ref{d2scf1} involves inner product of two parts which are $\sigma_j^{\beta (0)}\vert{\cal G}\rangle\vert{\cal M}_{\cal G}\rangle$ and  $ \langle{\cal M}_{\cal G}\vert\langle{\cal G}\vert \sigma_i^{\alpha}(t)$ which are shown in Eq.~\ref{sigones1} and Eq.~\ref{sigones2}. The meaning of Eq.~\ref{sigones1} is easy to understand which tells what happens  when a given spin acts on an eigenstate. It  creates two fluxes in the gauge sector  $|\mathcal{G} \rangle$ of the eigenstate and also adds a Majorana fermion in the matter sector( $| \mathcal{M}_{\cal G} \rangle$). Let us try to expand the meaning of Eq.~\ref{sigones2} which represents how the action of a given spin on an eigenfunction evolves as function of time. To this end we yield the intermediate steps reaching Eq.~\ref{sigones2} as follows.

\begin{eqnarray}
&& \sigma_i^{\alpha}(t) \vert{\cal G}\rangle\vert{\cal M}_{\cal G}\rangle \nonumber \\
&&= e^{i H t} \sigma_i^{\alpha}(0) e^{-iHt} \vert{\cal G}\rangle\vert{\cal M}_{\cal G} \rangle \nonumber \\
&&= e^{i H t} \sigma_i^{\alpha}(0) e^{-iE t} \vert{\cal G}\rangle\vert{\cal M}_{\cal G} \rangle \nonumber \\
&&= e^{i (H-E)t } \sigma_i^{\alpha}(0) \vert{\cal G}\rangle\vert{\cal M}_{\cal G} \rangle \nonumber \\
&&= e^{i (H-E)t } c_i(0) \vert{\cal G}^{i,\alpha} \rangle \vert{\cal M}_{\cal G} \rangle .
\end{eqnarray}

In the first and second step we have used the  definition of time evolution of an operator and applied eigenstate condition with energy $E$ respectively. The third step is a simple rearrangement  and fourth step implements the effect of spin $\sigma^{\alpha}_i$ which is equivalent to create two fluxes in the $\vert{\cal G}\rangle $ (to have $\vert{\cal G}^{i,\alpha} \rangle $) and also adding a Majorana fermion $c_i$ in the matter sector $\vert{\cal M}_{\cal G} \rangle$. Now we need to consider the effect of the exponential operator $ e^{i (H-E)t }$ on this. Remember that from Eq.~\ref{mageq2} that $c_i(0) \vert{\cal G}^{i,\alpha} \rangle \vert{\cal M}_{\cal G} \rangle $ is not an eigenstate of $H$. The  situation is complicated  for two reason. Firstly $ \vert{\cal G}^{i,\alpha} \rangle $  makes all the conserved quantities '$u$' to be 1 except on a bond connecting site `$i$'. Secondly $ c_i(0) \rangle \vert{\cal M}_{\cal G} \rangle $ is not an eigenstate for such distribution of $u$'s in Eq.~\ref{3}. In general if a state is not an eigenstate of a Hamiltonian then the action of Hamiltonian on it results into superposition of many other eigenstates. Thus the action $ \sigma_i^{\alpha}(t) \vert{\cal G}\rangle\vert{\cal M}_{\cal G}\rangle$ can be mathematically represented as follows,
\begin{eqnarray}
&& \sigma_i^{\alpha}(t) \vert{\cal G}\rangle\vert{\cal M}_{\cal G}\rangle \nonumber \\
&&=  e^{-iE t} \left( \sum^{N}_{j=1} c_j(t) \mathcal{A}_{ij} \vert{\cal M}_{\cal G}\rangle_j \right) \vert{\cal G}^{i,\alpha} \rangle,  
\end{eqnarray}
where $ \vert{\cal M}_{\cal G}\rangle_j$ represents `$j$'th eigenstate in the presence of two fluxes and the index `$j$' runs over all the sites of the system. In each of these $\vert{\cal M}_{\cal G}\rangle_j$, the fluxes are in identical position, they are static but the Majorana fermion added
to each of this eigenstate $\vert{\cal M}_{\cal G}\rangle_j$ are in different position in general. This effect is 
remarkable  and  signifies that effectively a given spin is composed of two objects, a Majorana fermion and  two fluxes. While the fluxes  do not move, the Majorana fermion moves as time evolves and it is not confined in a given region unlike  static flux pairs. This implies that physically a spin is fractionalized into two objects. The fact that Majorana fermion is moving gives rise to a deconfinement phenomenon. In Fig.~\ref{zzcore1}, we depict pictorially the phenomenon of fractionalization and deconfinement. The exact formula for $\sigma_i^{\alpha}(t) \vert{\cal G}\rangle\vert{\cal M}_{\cal G}\rangle $ can be seen in the original manuscript \cite{smandal-prl}.
 \begin{figure}
 	\psfrag{d}{0.2}
 	\psfrag{f}{$J_z$}
 	\psfrag{b}{1}
 	\psfrag{c}{0.8}
 	\psfrag{a}{$0,0$}
 	\psfrag{e}{$\langle \sigma^z_i \sigma^z_j \rangle$}
 	\includegraphics[width=0.7\linewidth]{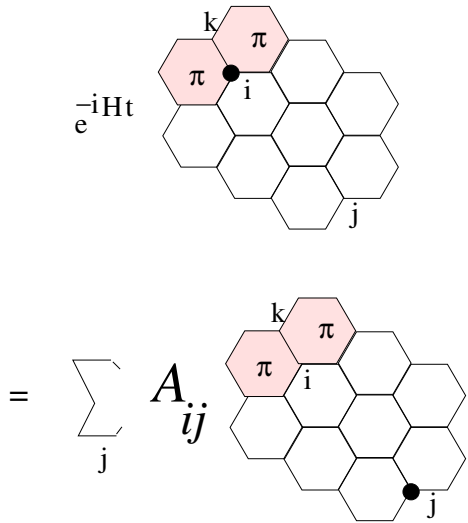}
 	\caption{\label{zzcore1} Consequence of fractionalization of spin into dynamic Majorana fermion and a pair of static fluxes is shown pictorially. Evolution of the state $|\Psi^{\prime} \rangle$ (used in Fig.~\ref{spflx}) results in linear superposition of many states where pair of fluxes do not change their positions but Majorana fermion does. }
 \end{figure}


Now in comparison with the singlet state we discussed before we find a number of differences. The similarity between the two is that in both the cases the two-spin correlation function is short range i.e it exists only for nearest-neighbor sites only. However there are a number of important differences. For example in the singlet state, the two-spin correlation function is non-zero if the two spins belong to a given singlet. However for two neighboring spins belonging to different singlet states, the correlation vanishes. This is the first difference. The second difference is that for Kitaev model the two-spin correlation is bond dependent, only $\sigma^{\alpha}_i \sigma^{\alpha}_j$ is non-zero for $\alpha$-type of bond. However for  a pair of sites belonging to a  given singlet $\langle \sigma^{\alpha}_{i} \sigma^{\alpha}_j \rangle$ exists for $\alpha=x,y,z$, i.e it is isotropic. 

\subsection{Multi-spin Correlation function, topological degeneracy}
\label{multi}
\begin{figure}
\includegraphics[width=0.99\linewidth]{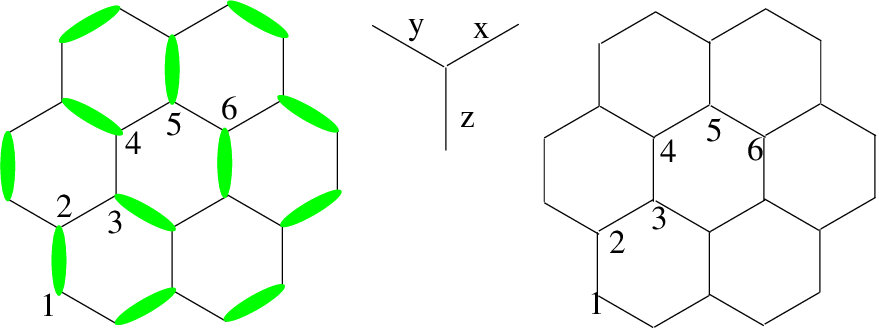}
\caption{\label{mulsp}Multispin correlation function exists for Kitaev model. For the sites 1 to 6, $\langle \sigma^z_1 \sigma^y_2 \sigma^y_3 \sigma^y_4 \sigma^z_5 \sigma^y_6 \rangle$ is non-zero. For detail see text below. }
\end{figure}
We have seen that for Kitaev model magnetization is zero at every site. Also two-spin correlation function is extremely short-range such that it does not extend beyond nearest-neighbor. It is natural to ask  what is the difference between an ordinary paramagnetic or other disordered magnetic state  for which long-range two-spin correlation is also zero. We  now  discuss that. Though the two-spin correlation exists only for nearest neighbor spins there is a underlying long-range correlation among them such that spins at long distances are entangled unlike paramagnetic or other disordered  magnetic state where such long range correlation does not exist. We  also compare it with the singlet state described before and find a very important differences. The existence of long-range multi-spin correlation function depends again on  the creation and annihilation of flux configurations in  $| \mathcal{G} \rangle$ such that the final and the initial flux configurations remains same. We note that this was the main  reason for non-vanishing two-spin correlations. In the right panel of Fig.~\ref{mulsp}, we have drawn a cartoon of Kitaev model where $x,y$ and $z$ type of bonds are shown. Various numerics from 1 to 6 denote many spins for which we wish to show that a non-vanishing multispin correlation function exist.  We notice that for bond $1-2$, $\sigma^z_1 \sigma^z_2$   does not change the flux configurations in $| \mathcal{G} \rangle$, similarly   for bond $2-3$, $\sigma^x_2 \sigma^x_3$ does not change the flux configuration in $| \mathcal{G} \rangle$. Thus for the site $1-2-3$, $\sigma^z_1 \sigma^z_2 \sigma^x_2 \sigma^x_3 \sim \sigma^z_1 \sigma^y_2 \sigma^x_3$ does not change the flux configurations in $| \mathcal{G} \rangle $ and this three-spin  correlation is non-zero. Thus for the spin 1 to 6 the following multi-spin correlation is non-zero,
\begin{equation}
S_{1-6}= \sigma^z_1 \sigma^y_2 \sigma^x_3 \sigma^y_4 \sigma^z_5 \sigma^y_6.
\end{equation}

\begin{figure}
\includegraphics[width=0.6\linewidth]{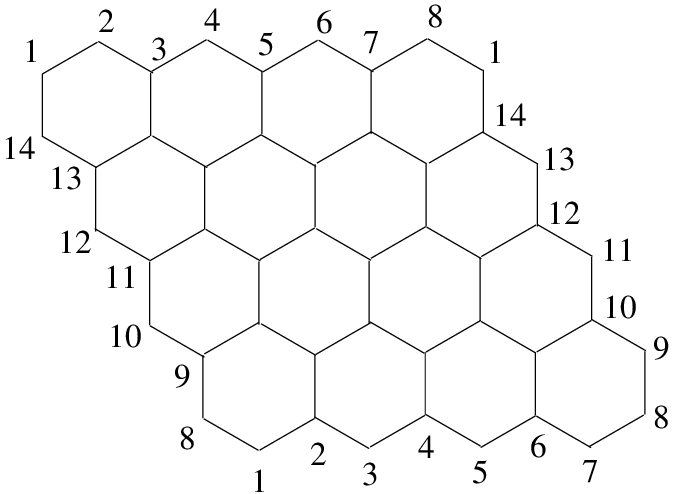}
\caption{\label{torus}A cartoon picture of torus geometry for a $4 \times 4$ unit cells is shown. The numerics show the specific way the boundaries are connected with each other. The above geometry is not unique and  different geometry of lattice exist for torus realization. }
\end{figure}
Thus for any path of arbitrary length there is  a non-vanishing multi-spin correlation associated with it. It shows that all the spins irrespective of the distance between them are really entangled implying their individual state depends on each other in a specific way. 
This is not true for ordinary paramagnetic or disordered spin configurations. Also the above multi-spin correlation is not present in the singlet state as shown in the left panel because there is no correlation exist between spin $2$
and spin $3$. However there could be an alternative singlet covering such that singlet state is formed on bonds $2-3$, $4-5$ etc. If we consider  the superposition of this two states, we might have a multispin correlation non-zero. Thus though Kitaev spin liquid phase  and the spin singlet phase both have short range two-spin correlations,  they are  of different nature. Similarly the long-range multi-spin correlations for both of these phases are of different nature. It may be noted that the existence of short-range spin-spin correlations as well as non-vanishing multi-spin correlation functions can also be established by symmetry considerations~\cite{nussinov-2}. The long-range multi-spin correlations or string operators lead to another novel concept which is
known as topological order. The word topology here is used to denote the fact that the movement of a given spin really depends on each other and every spin is entangled with every other spin in a particular manner. This also says that a spin can not be arbitrarily oriented and it must conform to  certain global pattern or the spin-orientations of other spins. Such global pattern of spin configurations of all the spins might lead some global constraint such that certain patterns are mutually exclusive leading to an existence of long-range  string order parameter which takes distinct values for each different pattern of spin configuration.  We  now explain this fact for Kitaev model defined on torus of honeycomb lattice. Torus means two dimensional lattice such that it is periodically connected in both the direction. In Fig.~\ref{torus}, we have shown such periodic boundary condition applied to a $4 \times 4$ arrangement of unit cells. As can be seen that the lower and upper  sides are connected as well as the left and the right sides. Note  that each one dimensional edges are periodically connected. Now we  evaluate some numbers which the reader are requested to  cross-check for this $4 \times 4$ system having $32$ sites.  For a general $N \times N$ system, there
are total number of sites $M= 2 N^2$. Total number of plaquettes is also $M$. We know that associated with each plaquette there is a conserved quantity called $B_p$. Thus there are in total $M$ number of $B_p$. Now one can check that for torus geometry  once all the $B_p$
are  multiplied together it becomes unity such that,
\begin{eqnarray}
\prod^{M}_{p=1} B_p=1 .
\end{eqnarray} 
This tells that all the $B_p$ is not independent and there are in total $M-1$ independent $B_p$. It can be shown that there is a conserved quantity associated with every closed loop, but they are not independent as they can be constructed by a product of certain $B_p$'s. However there two global loop operator  which extend from one end to another which can not be expressed as a product of $B_p$ and there are two such loop operators which we call $W_1$ and $W_2$. These are analogous to Wilson loop operator in quantum field theory \cite{giles,wilson}. For example take a horizontal zig-zag line constituted out of alternating $x$ and $y$ bonds. the product of $\sigma_z$ for all the sites on that line is a conserved quantity.
\begin{eqnarray}
\label{wone}
W_1 = \prod^{N_{h}}_{i=1} \sigma^z_i .
\end{eqnarray}
In the above $N_h$ denotes the total number of sites in a given $x-y$ chain. Referring to Fig. \ref{torus} the site `$i$' runs over 1 to 8.  Similarly for a $x-z$ chain extending from one end to another, we can construct $W_2$ which is conserved. Thus total number of conserved quantities become $M+1$. Now it can be checked that $W^2_i=1$ implying that eigenvalues of $W_i$ is $\pm 1$. The presence of these two additional independent conserved quantities implies that there will be four different eigenstates of certain identical configurations of $B_p$ but with different eigenvalues of $W_i$. These four eigenfunctions can be represented as,
\begin{eqnarray}
&&|\Psi_1 \rangle= |\mathcal{M}_{\mathcal{G}} \rangle | \mathcal{G}(B_p); W_1=1, W_2=1 \rangle , \\
&&|\Psi_2 \rangle= |\mathcal{M}_{\mathcal{G}} \rangle | \mathcal{G}(B_p); W_1=-1, W_2=1 \rangle ,\\
&&|\Psi_3 \rangle= |\mathcal{M}_{\mathcal{G}} \rangle | \mathcal{G}(B_p); W_1=1, W_2=-1 \rangle ,\\
&&|\Psi_4 \rangle= |\mathcal{M}_{\mathcal{G}} \rangle | \mathcal{G}(B_p); W_1=-1, W_2=-1 \rangle .
\end{eqnarray}
\begin{figure}
\includegraphics[width=0.8\linewidth]{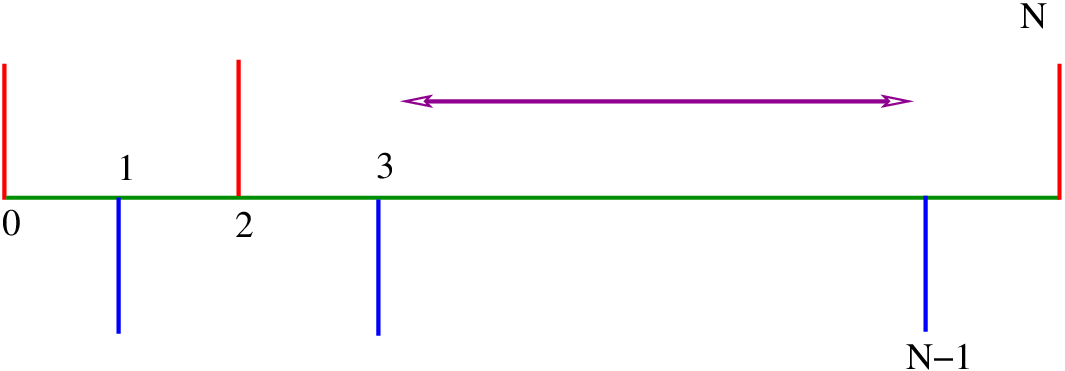}
\caption{\label{topoja} Various red lines represent the spin states with even number of  spin in the $+ z$ direction. The blue lines show odd number of spins in the $- z$ direction. The red states contribute to $W_1=1$ and the blue states  contribute to $W_1= -1$.}
\end{figure}
  These $W_i$  manifests a hidden topological order\cite{sumathi-rao} which we will try to understand by considering how many ways $W_1$ defined in Eq.~\ref{wone} can be made one for one dimensional arrangement of spins. In Fig.~\ref{topoja}, we have shown by red line the state for which even number of spins are aligned in positive z-direction.  For a one dimensional chain of $N$ sites there are $2^{N-1}$ such states.  Similarly by blue line, we represent states with odd number of spins oriented along negative $z$-axis. All the red states in a given $x-y$ line contribute to the  wave function to make $W_1=1$ and such states can not be converted to an eigenfunction with $W_1=-1$ by local flipping of a given spin or a number of spins. For that we need to convert all the red states to the blue one. It is non-local transformation in the Hilbert space and thus topological. However the energy does not depend on the eigenvalues of $W_i$, it only depends on the flux configurations. Thus all the states $| \Psi_i \rangle, i=1,4$ have identical energy in the thermodynamic limit \cite{mandaljpa}. This is a manifestation of topological degeneracy.

\section{Kitaev-Heisenberg model}
\begin{figure}
\includegraphics[width=0.99\linewidth]{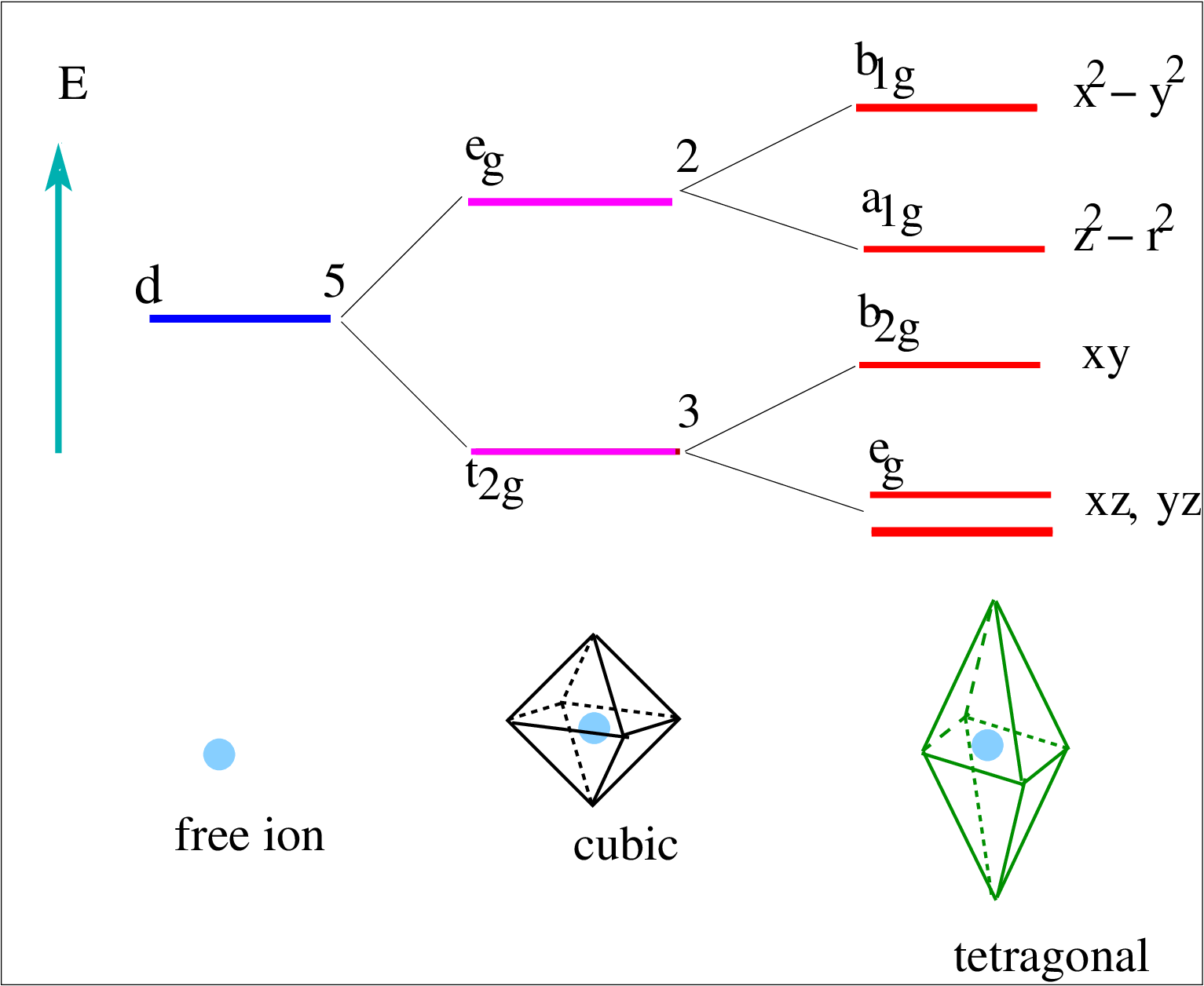}
\caption{For a free ion in Hydrogen-atom like situation, the atomic $d$-levels are five-fold degenerate which is represented by blue line. However in molecular arrangement with cubic symmetry, these five-fold degeneracy is broken into two-fold degenerate high energy $e_g$ levels and three-fold degenerate ground state $t_{2g}$ lavels represented by magenta color. Further in tetragonal arrangement, the levels split further with two-fold degenerate ground state level and three excited states represented by red lines. The arrow in cyan color represent the energy(E) axis in postive direction.}
\label{crystal-field}
\end{figure}

Kitaev model belongs to a class of models known as quantum compass model \cite{nussinov-2015} where the two-spin  interactions on a lattice is highly anisotropic in the  three Cartesian components and depend on the nature of  bonds.  Kitaev model is one of such realization where the two-spin interactions along three different directions of the honeycomb lattice contain different components of spin. As known, the origin of magnetic exchange interaction results from a competition between Coulomb interactions and Pauli exclusion principle governing the wave function to be antisymmetric. The spatial profile of the participating orbitals determine Coulomb interaction energy   and it leads to a certain form of spin-part of the wave function which governs  the effective exchange interaction between the participating electrons. In the absence of any spatial preferences, the $SU2$ symmetric Coulomb interaction yields normally an isotropic Heisenberg interaction $J \vec{S}_i \cdot \vec{S}_j$. However there are other interactions which are very anisotropic\cite{john-1962,Yu2023} such as Ising interaction\cite{wolf_2000,Chibotaru_2015,Lee2016,Wang_2016,Gong2017,Huang2017,cuccoli-2003} and $x-y$ interactions  owing to the crystallographic anisotropy of the structural details\cite{Park_2016,Burch2018,joy-1992,Kim2019}. In general the final form of the  exchange coupling depends on the symmetry of the system\cite{BALTES19672635,attila-rmp-2022} and in many cases spin-orbit(SO) coupling\cite{DZYALOSHINSKY1958241,morya1960} plays a  fundamental role in governing the anisotropic exchange interaction\cite{borisov-2021} as the term $\lambda \vec{l} \cdot \vec{s}$ suggests a directional (or spatial) dependence of the spins $\vec{s}$ on  orbital details yielding $\vec{l}$. However to realize such spin-orbital effect the spin-orbit coupling $\lambda$ must be comparable to the Coulomb interaction energy in a given material. In this respect the heavy-ions with  large spin-orbit interactions  are very relevant\cite{alexander-2021,Li2022,marcin-2019}. For the present  purpose pertaining to Kitaev model material with structure $\rm A_2BO_3$ has been in discussion for decades. Here we give a brief overview of the qualitative and physical reason how the Kitaev interaction is realized in such materials.  \\
\indent
We first begin having a discussion of crystal field splitting in 3d transition metals (TMs) where such strong spin-orbit coupling has been observed. The electrons in a 3d level within a molecular environment loses its five fold degeneracy and depending  on immediate surrounding further degeneracy can be lifted as shown in Fig. \ref{crystal-field}.
Few transition metal ions which contain strong spin-orbit coupling are Ru, Rh, Os and Ir. As mentioned
\cite{jackeli,Jackeli-2009} optical data  obtained for 4p impurities of Ir  in $\rm SrTiO_3$ corresponds to  the SO coupling $\lambda \sim$ 380 meV which is considered a very high value. This high value of spin-orbit interaction is even greater than intersite interactions between two neighboring $t_{2g}$ orbitals or exchange coupling among spins in insulating iridates. This means the  spin-orbit coupling in the atoms contribute in forming a total angular momentum locally. Here we consider a d-5 configuration which yields an effective hole in an otherwise three-fold degenerate $t_{2g}$ level. This is particularly realized in strong octahedral field found in Rh4p and Ir4p. Following\cite{Jackeli-2009}  we outline briefly the superexchange theory in this kind of materials which contains along with Heisenberg interactions other exotic exchange interactions such as quantum compass models.  This paves way for material realizations of otherwise theoretical predictions of fractional statistics or anyonic excitations. To elucidate this it is pertinent to consider the iridium compound $\rm Sr_2IrO_4$ ~\cite{HUANG1994355,GCao-1998,sjmoon-2006,bjkim-2008} which is known as an insulator having weak ferromagnetic  even with large  ferromagnetic moment for individual ions. \\
\indent
We now delve into the situation where strong spin-orbit coupling plays a deciding role in governing certain bond-directed exchange interactions. From Fig. \ref{crystal-field}, we note that under tetragonal (cubic) environment the three-fold degenerate $t_{2g}$ levels are comprised of $xy, xz$ and $yz$  orbitals. Though it belongs $d$-orbitals and hence the orbital moments should be $l=2$, we note that in this degenerate manifold the value of $l_z$ are given by $l_z=0$ for $|xy\rangle$ and $l_z=\pm 1$ for $- \frac{1}{\sqrt{2}} \left(i |xz\rangle \pm |yz\rangle \right) $ states. This implies that the effective angular momentum is 1 instead of 2 \cite{abragam1970}. If we denote $\vec{s}$ as the hole spin operator (as there are 5 electron in three degenerate levels, it can be effectively represented by a hole), the total moment can be expressed as $\vec{M}= 2 \vec{s} - \vec{l}$. To obtain the single-ion Hamiltonian it is necessary to consider the spin-orbit coupling as well as possible tetragonal splitting (see Fig. \ref{crystal-field}) yielding a $\Delta l^2_z$ terms. Thus the single-ion Hamiltonian is given by ${\mathcal{H}}_0= \lambda \vec{l} \cdot \vec{s} + \Delta l^2_z$. Thus in this subspace the states can be represented as $\Psi=\left( |0, \uparrow \rangle,~|0,\downarrow \rangle,~ |1,\uparrow \rangle,~ |1, \downarrow \rangle,~|-1, \uparrow \rangle,~ |-1,\downarrow \rangle \right)$ which yields the following matrix  representation of the single-ion effective Hamiltonian, 
\begin{eqnarray}
 \hat{H} &&= \left(  \begin{array}{cccccc} 0&0&0& \sqrt{2} \lambda&0&0 \\
	0&0&0&0&\sqrt{2} \lambda&0 \\
	0&0& \Delta_+ &0&0&0 \\
	\sqrt{2} \lambda&0&0&  \Delta_{-} &0&0 \\
	0&\sqrt{2}\lambda &0&0& \Delta_{-}&0 \\
	0&0&0&0&0& \Delta_{+} \end{array} \right)~.~~~
\end{eqnarray}
In the above  $\Delta_{\pm}= \Delta \pm 1/2$. The eigenvalues are doubly degenerate  and obtained as $~ \frac{1}{2} (1+ 2 \Delta),~~ -\frac{1}{4} ( \delta \pm \sqrt{\delta^2 +  \gamma^2})$ where $,~~\delta= 1-2 \Delta,~~~ \gamma= 4 \sqrt{2} \lambda$. The ground state energy is obtained as $E_0= - \frac{1}{4} ( \delta + \sqrt{\delta^2 +  \gamma^2}) $ and the degenerate ground state wave functions are obtained as $| \tilde{\uparrow} \rangle = \sin \theta |0, \uparrow \rangle - \cos \theta | + 1, \downarrow \rangle,~~~| \tilde{\downarrow} \rangle = \sin \theta |0, \downarrow \rangle - \cos \theta | + 1, \uparrow \rangle$
$\tan 2\theta =  4 \sqrt{2} \lambda /(1- 2 \Delta) =\gamma/\delta $.

 For our purpose the relevant states are  the degenerate ground state doublets $| \tilde{\uparrow} \rangle$ and $| \tilde{\downarrow} \rangle$ which play the role of effective Kramers degenerate pseudo spin states and it governs the low-energy effective spin-Hamiltonian.  \\
\indent 

\begin{figure}[h!]
\includegraphics[width=0.99\linewidth]{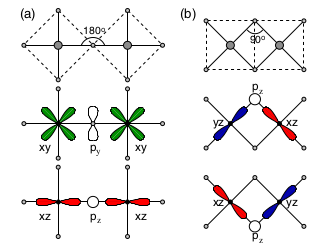}
\caption{In the left upper panel oxygen atom makes a $180^{\circ}$ bond angle with two nearest-neighbor TM ions. In the right panel, an arrangement of $90^{\circ}$ has been shown among two TM ions and oxygen atoms in between. For the former the oxygen atoms connect two identical orbitals such as $d_{xy}$ to $d_{xy}$ or $d_{xz}$ to $d_{xz}$. However for the latter, the oxygen atom connects two orthogonal orbitals such as $d_{yz}$ to $d_{xz}$. Further, for this case there are two ways of hopping from a given TM ion to another.  {\it Permission from the authors to reuse the figure from  Phys. Rev. Lett. {\bf 102}, 017205 (2009)\cite{Jackeli-2009} is gratefully acknowledged.}}
\label{jackeli-angle}
\end{figure}

\begin{figure}[h!]
	\includegraphics[width=0.99\linewidth]{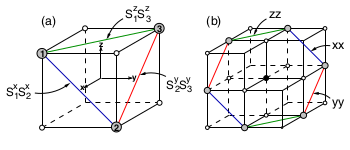}
	\caption{In (a) a structural unit of $ABO_2$ compound has been shown. Oxygen atoms (empty circles) forms $90^{\circ}$ with neighboring TM ions (shown by filled circles). The TM ions forms a layered triangular lattices. The exchange interactions between TM ions are direction dependent as shown by three different colors. In (b) a structural unit of $A_2BO_3$ has been shown. It has layered honeycomb structure and can be obtained from $ABO_2$ by periodic removal of TM ions. {\it Permission from the authors to reuse the figure from  Phys. Rev. Lett. {\bf 102}, 017205 (2009)\cite{Jackeli-2009} is gratefully acknowledged.}}
	\label{jackeli2}
\end{figure}

 In real materials the transition metal(TM) ions are connected via an intermediate oxygen ion which leads to a superexchange mechanism to play the deciding role of effective spin Hamiltonian between these TM ions. The superexchange theory was first proposed by Kramers\cite{kramers1934}  and subsequently developed by  Anderson\cite{anderson-1950} and  many others\cite{goodenough1955,kanamori1959,kondo1959,gilleo1960,yamashita1958}. For a recent review one may look at \onlinecite{Streltsov_2017,magnetochemistry8010006}.
 In superexchange mechanism the relative position of the intermediate oxygen atom plays an important role to determine the final form of effective Hamiltonian. For our interest, we elucidate two such different cases and explain how it may govern different effective Hamiltonian. As shown in Fig. \ref{jackeli-angle} (a) and (b) the oxygen atoms make an angle of $180^{\circ}$ and $90^{\circ}$ with two adjacent TM ions. $\rm LaTiO_3$\cite{giniyat-2005} is an example where the angle is $180^{\circ}$. The important point to note that in this case the two orbitals which are connected by the intermediate oxygen atom are identical i.e they are diagonal in orbital which in this case is $d_{xy}$ type or $d_{xz}$ type, and they are connected via an oxygen $p_y$ and $p_z$ orbital respectively.  This particular bond angle is obtained for a corner sharing tetrahedron. However one may note that such a corner sharing tetrahedron may also give bond angle  other than $180^{\circ}$ such as $\rm \alpha-MnO_2$\cite{mandal2014}. However for $180^{\circ}$, one obtains the effective Hamiltonian as $H_{ij}= J_1 \vec{S} \cdot \vec{S}_j + J_2 (\vec{S}_i \cdot \vec{r}_{ij}) (\vec{r}_{ij} \cdot \vec{S}_j)$. Here $\vec{r}_{ij}$ denotes unit vector which connects the two TM ions and hence the $J_2$ term signifies an anisotropic interaction along with the isotopic Heisenberg interaction denoted by $J_1$.\\
\indent
Now we move on to discuss the case of $90^{\circ}$ angle formed by the intermediate oxygen atoms in between two TM ions as shown in Fig. \ref{jackeli-angle} (b). A crucial difference between the former and the present case is that here two TM ions are connected by two oxygen atoms and hence there are two paths to tunnel from one TM ion to another. Secondly  for this case the oxygen atom connects two orthogonal orbitals such as $d_{yz}$ and $d_{xz}$ via a  $p_z$ orbitals. As pointed out \cite{Jackeli-2009}, the transition amplitude via this two alternative paths in between the two TM ions cancel due to destructive interference. (One way to think is to represent the transition amplitude via the upper  and lower path as $t_u=\langle d_{yz}| T | d_{xz} \rangle, t_d= \langle d_{xz}| T | d_{yz} \rangle $. Now consider $t_d$ and use a rotation $\pi/2$ clockwise along z-axis such as $ y \rightarrow x, x\rightarrow -y$ which renders  $t_d =  -\langle d_{yz}| T | d_{xz} \rangle=-t_u$. Here we have used the fact that $d_{x_1 x_2} \sim x_1 x_2 \Phi{r}$.)
Due to this cancellation, the contribution from this lowest order exchange interaction mechanism which is otherwise isotropic vanishes identically. This zero isotropic exchange interaction coming from the lowest order process comes from the selection rule of angular momentum as well \cite{Takagi2019}. Note that the $l_z$ values $d_{xz}$ or $d_{yz}$ are $\pm 1$ and for Oxygen $p_z$ $l_z=0$, thus it requires a change of orbital angular momentum of magnitude $2$ which is forbidden. Though the lowest order exchange mechanism does not contribute, the higher order mechanism involving the higher multiplet excited states do contribute a finite exchange interaction which is anisotropic. More excitingly, such anisotropic interaction depends on the spatial orientation of the bonds formed by two TM ions. Thus the final form for such interaction is given by   $H^{\gamma}_{ij}= -J S^{\gamma}_i S^{\gamma}_j$ where the bond $ij$ is assumed to be in the  $\alpha \beta$ plane which is perpendicular to the  $\gamma(=x,y,z)$ axis. \\
\indent
Now we discuss few  materials where such anisotropic direction dependent exchange interaction arises. First is to consider layered materials with chemical formula $\rm ABO_2$ where $\rm A$ represents an alkali ions and $\rm B$  a transition metal ions (for example $\rm BaZnO_2, BaCoO_2, BaMnO_2$ \cite{Spitsbergen:a02765} ). In such compounds there exists a triangular unit  as shown in Fig. \ref{jackeli2} (a) where two TM ions are connected by an oxygen atoms forming a $90^{\circ}$ bond. The sides of the triangle formed by three TM ions are shown by green, blue and red where different anisotropic direction dependent exchange interactions exist. Now to obtain Kitaev model, one may periodically remove some magnetic ions and replace it by suitable non-magnetic ions. The resulting structure with honeycomb layers are then formed with $A_2BO_3$ structure as shown in   Fig. \ref{jackeli2} (b) and it has exactly Kitaev like anisotropic interaction. There are many such materials with $\rm A_2BO_3$ structures  such as $\rm Li_2RuO_3$\cite{miura-2007} and $\rm A_2IrO_3$\cite{Takagi2019} where such Kitaev like interactions has been realized.  

It may be mentioned that apart from the Kitaev interaction (which results from indirect hopping between two d-orbitals via an excited $J_{eff}=3/2$ states) there exists remnant Heisenberg interaction originating from direct hopping between two $d$-orbitals and mixture of the above two processes gives another anisotropic term known as $\Gamma$ interaction. Thus the final Hamiltonian looks like ,
\begin{eqnarray}
\mathcal{H}&&= \sum^{y,z}_{\gamma=x} \sum^{N_{\gamma}}_{\langle i,j\rangle_{\gamma}=1} -K s^{\gamma}_i s^{\gamma}_j + \sum^{N}_{\langle ij \rangle = 1}J \vec{S}_i \cdot \vec{S}_j \nonumber \\
&& + \sum^{y,z}_{\gamma=x} \sum^{N_{\gamma}}_{\langle i,j\rangle_{\gamma}=1} \Gamma ( s^{\gamma^{\prime}}_i s^{\gamma^{\prime\prime}}_j + s^{\gamma^{\prime\prime}}_i s^{\gamma^{\prime}}_j ) ~.~~~~~
\end{eqnarray} 
In the above first and second terms denote Kitaev and Heisenberg interaction respectively. The third term denotes the anisotropic $\Gamma$ interaction. Here $N= N_x + N_y + N_z$, total number of bonds and summation for Heisenberg interaction runs over all bonds. The summation in $\Gamma$ interaction follows the convention used in Kitaev interaction. We note that in the last term $\gamma \ne \gamma^{\prime} \ne \gamma^{\prime \prime}$. This implies that on a $x$-type of bond, the two-spin interaction looks like $s^{y}_is^{z}_j + s^{y}_js^{z}_i$. The identical convention follows for $y$ and $z$-type of bonds. We notice that in this scheme the Kitaev interaction is ferromagnetic and other interactions are antiferromagnetic\cite{Liu-2022}.  The above scheme of obtaining the effective Kitaev interaction is based on $4d^5$ or $5d^5$ systems and normally known as the {\it Jackeli-Khaliullin mechanism}. Later there have been other proposals to realize Kitaev interactions and among them we briefly mention the proposal made by Motome et. al \cite{Motome_2020}. They have proposed a mechanism based on $f^1$ system ( ex. $\rm Pr^{4+}$) or high spin $\rm d^7$ spin with $\rm Co^{2+}$ or $\rm Ni^{3+}$. The salient characteristic here is that the Kitaev exchange interaction is obtained as a direct hopping between two $f$-orbitals
 via intermediate oxygen orbitals (where as for $\rm 4d^5$ system this mechanism cancels due to selection rule). As a result the obtained Kitaev interaction is antiferromagnetic one and also larger in magnitude then the other case. For more detailed discussion on this aspect we refer the original reference \cite{Motome_2020,Takagi2019}.

\section{Prelude to experimental work}
\begin{figure}[h!]
	\includegraphics[width=0.99\linewidth]{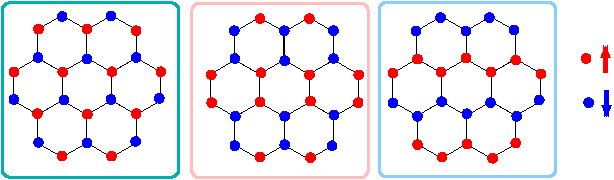}
	\caption{A cartoon picture of Neel order, stripe order and zig-zag order have shown in the left, middle and right panel respectively.}
	\label{afm-zigzag}
\end{figure}

Now we move on to discuss some important experiments done on materials having Kitaev interaction. Among various thermodynamic quantities and transport signatures we consider magnetization, susceptibility, specific heat and thermal Hall effect. For a brief review on various magnetic responses of magnetic materials we refer \onlinecite{Mugiraneza2022} where a detailed discussion is included for varieties of magnetic systems such as ferromagnets, antiferromagnet, ferrimagnet and paramagnets. This would guide to understand the complex magnetic responses of Kitaev materials. To have an understanding of experimental identification of spin liquids on various materials we refer \onlinecite{jinsheng-2019}. This review paper\cite{Takagi2019} gives a good account of theoretical and experimental review of Kitaev quantum spin liquid phase. For our purpose we mainly focus on studies done on $\rm{\alpha-RuCl_3}$. Depending on temperature and other physical conditions $\rm RuCl_{3}$ could be in various crystal structures and these different crystal structures (or polymorphs) are named as $\alpha, \beta, \gamma,$-RuCl$_3$, each having different crystal structures\cite{fletcher-1967}. For $\alpha$-RuCl$_3$ crystal structure contains layered honeycomb structures, while for $\beta$-RuCl$_3$ the structure is similar to zirconium trichloride with space group P6$_3$/mcm\cite{douglas2007structure}. We note that most of the Kitaev materials are shown to posses collinear order of various antiferromagnetic long range order at low temperature due to the presence of other non-Kitaev interactions. However at intermediate temperature and in the presence of magnetic field, signature of revival of Kitaev spin liquid phase has been speculated  by experts in the field. In an earlier study by the author and collaborators\cite{saptarshi-2011}, fate of Kitaev model was investigated in the presence of Ising interaction which destroys the exact solvability of Kitaev model.  There it was found that in the presence of Ising interaction several long-range ordered phases such as Neel AFM  phase and Dimer phase ( alternatively named spin-chain phase)  appear when Kitaev exchange coupling($K$) is small in comparision to Ising interaction ($\rm J_I$), mainly when $\rm K < 0.1 J_I$. In Fig. \ref{afm-zigzag}, we have shown few well known classical long-range ordered ground states that appear at low temperature due to various types of non-Kitaev interactions. The left panel describes the Neel order, the middle panel denotes the stripe order and  the right panel is for zig-zag order.

Further within a mean-field approximation it is found that   when the exchange coupling  for Kitaev interaction becomes more than $10\%$  of Ising interaction, the Kitaev spin liquid phase is recovered. Similar conclusion has been observed in an exact diagonalization study\cite{jackeli} where the spin-chain phase is named as stripe order. Further investigations for various sign and magnitude of other non-Kitaev interactions such as Gamma interaction\cite{Rau-2014} exhibits a variety of classically ordered phases and the fate of this ordered phases into Kitaev spin liquid phase is a question by itself which promises competing many-body physics to play interesting dynamics.\\
\indent

\begin{figure}[h!]
	\includegraphics[width=0.5\linewidth]{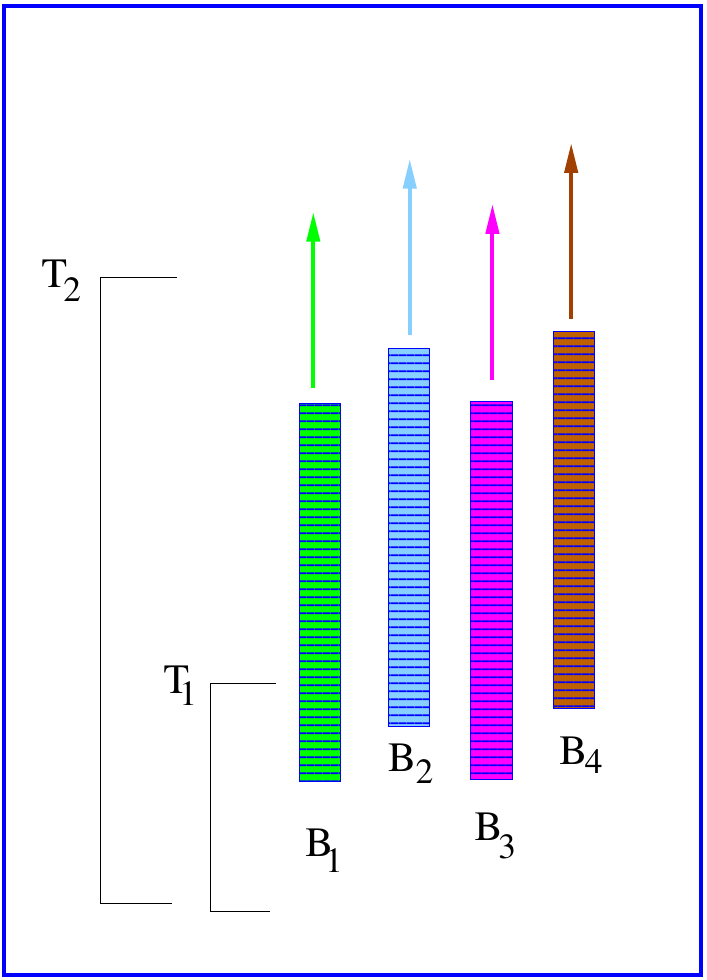}
	\caption{Physical origin of two peak structure of specific heat is shown graphically. A free fermionic spectrum is obtained for each  equivalent flux configurations. The  ground state energy for each distinct flux configurations  are different and correspond to low-energy excitations. For more details see text.}
	\label{spcartoon}
\end{figure}

The presence of other non-Kitaev interactions makes it difficult for the coveted spin-liquid phase to be realized theoretically and equally challenging for experimental detection of various remarkable exact properties that Kitaev model offers. As the spin-fractionalization into a dynamic Majorana fermion and pair of static fluxes is not realized exactly the so called two peak structure in specific heat is not guaranteed to be observed. In Fig. \ref{spcartoon}, we explain the reason of origin of two peak structure in pure Kitaev model. At zero temperature, for each in equivalent flux configurations one obtains a free-fermionic many-body ground state whose energies are different for each background flux configurations, generally. This corresponds to low-energy excitations coming from different flux configurations and should yield a low temperature peak in specific heat. When the temperature is increased the contribution from the excitations from each flux configurations contribute yielding a second peak.
  Similarly for a pure Kitaev model the susceptibility and magnetization must be different than ordinary known behavior such as antiferromagnetic,  ferromagnetic or paramagnetic behavior. 
  As far as the the transport signature of Kitaev materials is concerned, only thermal Hall effect offers a possibility to infer the existence  of Kitaev spin liquid phase. This is so because Kitaev materials are inherently non-metallic and usual electronic transport signatures are not useful. However in some cases, it has been claimed that the specific heat capacity shows a linear dependence on temperature which signifies the fractionalization of spin into dynamic Majorana fermion and static fluxes. It is then understood that the dynamic Majorana fermions are responsible for the linear temperature dependence of specific heat.
    
\indent 
\begin{figure*}[ht!]
	\includegraphics[width=01\linewidth]{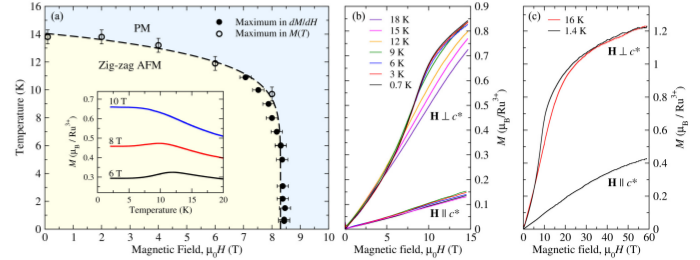}
	\caption{\label{sus-mag1}
		In the left phase diagram for single-crystal $\rm \alpha-RuCl_3$ is plotted in temperature vs  magnetic field (applied in plane). In the middle panel magnetization is plotted against applied magnetic field for up to 15 Tesla for different temperatures. The same has been plotted in right panel for magnetic field up to 60 Tesla. The maximum of $dM/dH$ obtained from middle panel has been used to obtain the phase diagram in left panel (the filled in circles.)	The empty circles denote the maxima in $M-T$ plot shown in the inset of left panel. The dashed line is a guide to the  eye to mark the phase boundary that exists between the zig-zag antiferromagnetic phase (shown by yellow shade) and paramagnetic phase (shown by light blue shade).  {\it Permission from the authors to reuse the figure from  Phys. Rev. B {\bf 92}, 235119 (2015)\cite{johnson-2015} is gratefully acknowledged.}}
\end{figure*}

\subsection{\large{Suscpetibility and magnetization}}
  {\bf {Crystral structure of $\rm{\alpha-RuCl_3}$ :}} We first discuss the experimental investigation done on $\rm{\alpha-RuCl_3}$ in Ref. \onlinecite{johnson-2015}. In the beginning a complete and elaborate discussion was made to report the x-ray diffraction measurements on crystals of  $\alpha-\rm{RuCl_3}$ without any twinning. A single magnetic phase transition has been reported at low temperature which conforms the results obtained from powdered sample. It posses a monoclinic crystal structure having  C2/m space group. The diffraction pattern does not seem to conform the supposed model of trigonal structure of $\rm P 3_1 12 $ which is also argued\cite{kim-2016-dft,wang-weyl-2021} to be very close in energy with $\rm C2/m$ . The crystal structure of  $\rm{\alpha-RuCl_3}$ and that of $\rm{Na_2IrO_3}$ (and $\rm{\alpha-Li_2IrO_3}$ Ref. \onlinecite{skchoi2012prl,OMALLEY20081803}) having  honeycomb  stacked layers are seen to be iso-structural. $\rm Na_2IrO_3$ has monoclinic crystal structure with space group $\rm C2/c$ where $\rm Ir$ ions form a honeycomb layer. (On the otherhand $\rm Li_2IrO_3$ appears in various polymorphs. The primary one is $\rm \alpha-Li_2IrO_3$ having honeycomb layers of $\rm Li$ ions. The secondary is called $\rm \beta-Li_2IrO_3$ where the polymorph crystallizes in three dimensional hyperhoneycomb lattice. Lastly there exists $\rm \gamma-Li_2IrO_3$ where a 3D stripy honeycomb structure is seen. Interestingly all the three polymorphs are tricoordinated and candidate Kitaev spin liquid material\cite{Takagi2019,yogesh-first-prl,Glamazda2016}.)  Further the magnetic propagation vector obtained from powder neutron diffraction pattern establishes the zig-zag magnetic ordering. The closely related stripe ordering  does not conform to the theoretical  diffraction data. Having established the magnetic ordering, its stability under external field has been investigated where field has been applied in the plane of honeycomb layers. Furthermore first-principle calculation were followed to  calculate the electronic band structure and stability of crystal structure to determine the magnetic ground state of $\rm Ru^{3+}$ ions.\\
  
\indent
{\textbf{Possible antiferromagnetic phase, magnetization plot:}} In Fig.  \ref{sus-mag1} (a) the phase diagram in $T-M$ plane(temperature(T) vs magnetic field (M)) has been plotted. In Fig. \ref{sus-mag1}(b) and (c), magnetizations (M) at different temperatures  are plotted against external magnetic field. The study used pulsed magnetic field and no differences were found in the rising and falling part of the field pulses within experimental accuracy which rules out existence of any hysteresis. From Fig.  \ref{sus-mag1}(b), for the field applied perpendicular to the honeycomb layer, one finds that near magnetic field $8$ T, there is sharp departure from the straight line behavior which decreases as one increases the temperature. This refers to onset of phase transition and the phase diagram thus obtained is shown in Fig. \ref{sus-mag1} (a) where the black dots refer maximum in  $dM/dH$ and empty dots correspond maximum of  $M(T)$. In the inset of Fig. \ref{sus-mag1} (a), additional $M(H,T)$ plots against temperature are shown. This  clearly shows a maxima at $8$T  and it disappears for larger magnetic field. The phase diagram  shown in Fig.  \ref{sus-mag1} (a) refers to continuous phase boundary which encloses a single antiferromagnetic phase at low temperature and here  magnetic field  is applied in the honeycomb layers. The little hump found in magnetization plot in the inset of Fig. \ref{sus-mag1}(a) is also found in numerical studies in Kitaev clusters \cite{pervez2023decipheringI,pervez2023decipheringII}. This  also shows that at large magnetic field the hump is absent and one obtains smooth transitions. Another fact of interest is that a comparison of data for  $\rm H \parallel c^*$ and  $\rm H \perp c^*$ shows that the former is 5 or 6 times smaller then the later. This establishes that the features which is observed only exists if the field is applied along the plane. These differences of response for different direction of magnetic field also refers to strong anisotropy of $\rm Ru$ g-factor. For comparison with other  magnetization studies we refer \onlinecite{majumder-2015}.\\
\indent

\begin{figure}[h!]
\includegraphics[width=0.9\linewidth]{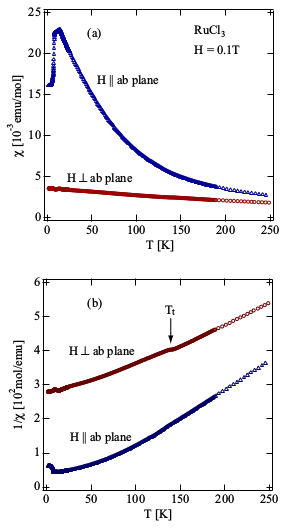}
\caption{\label{susmag2} In the upper panel magnetic susceptibilities $\chi=M/H$ of $\rm \alpha-RuCl_3$ has been shown against temperature for magnetic field parallel and perpendicular to the `ab' plane as indicated. The magnetic field is kept at 0.1 T.  In panel (b) inverse susceptibilities have been shown. It reveals minor discontinuity shown by arrow and it indicates structural phase transitions.  {\it Permission from the authors to reuse the figure from  Phys. Rev. B {\bf 91}, 094422 (2015)\cite{kubota-2015} is gratefully acknowledged.}}
\end{figure}

Now we discuss the experimental results obtained in \onlinecite{kubota-2015}. Here the article presented magnetization, susceptibilities and specific heat measurements  on  $\rm \alpha-RuCl_3$. In the following we first discuss the susceptibility to connect with the previously  discussed magnetization. Like before, the magnetic susceptibilities are  highly anisotropic  conforming  $\rm Ru^{3+}$ having a state of low-spin. The anisotropy also suggests decreased  entropy. Remarkably magnetic ordering is seen to occur in many steps which are attributed to competing exchange interactions. Vertical Bridgman technique was used to prepare the single crystal of $\rm RuCl_3$ from a melt. SQUID magnetometer(Quantum Design MPMS XL) were used for measuring magnetization. The range of temperature used are in between 1.8 K and 100 K and magnetic field up to 7 T has been used. Both in  and out of plane magnetic field have been used. Also a high magnetic field upto 57.5 T has been used for temperature   4.2 K and 1.3 K which uses induction method along with a multilayer pulse magnet. Fig. \ref{susmag2} (a)  show  the temperature dependence of the magnetic susceptibilities  and  Fig. \ref{susmag2} (b) shows the same for   inverse susceptibilities. Both were  measured for a magnetic field H parallel and perpendicular to the `ab' plane.  \\
\indent
 
 {\bf{Anomalies in susceptibilities:}} From Fig. \ref{susmag2}, we observe that the susceptibilities perpendicular to `ab' plane is much lesser than that parallel to the field. As mentioned earlier this strong anisotropy refers to anisotropic `g' factor which has been discussed in detail in Ref. \onlinecite{kubota-2015}.  Both the susceptibilities shows some anomaly between the temperature 15 K and 10 K. Apart from that the inverse susceptibilities perpendicular to the plane shows a little discontinuity near 150 K.  The performed x-ray crystal analysis could not reveal any structural phase transition associated at this temperature. However in 
 $\rm CrCl_3$ which is closely related to $\rm RuCl_3$, a structural phase transition is observed at temperature $T_t \simeq 240$ K \cite{morosin-1964}. The transition takes the monoclinic structure  $\rm C2/m$ to the trigonal structure $\rm R\bar{3}$.  Thus we can assume that the low temperature structure of $\rm RuCl_3$ is  $\rm R\bar{3}$.\\
 \indent
\begin{figure}[h!]
\includegraphics[width=0.9\linewidth]{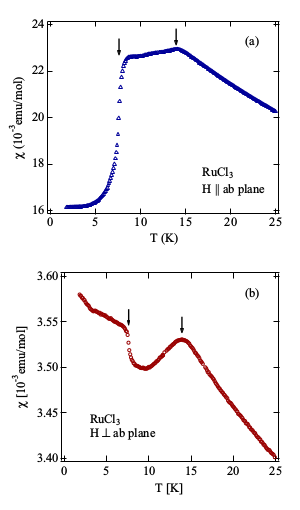}
\caption{\label{susmag3} Zoomed in susceptibilities have been shown up to temperature 25 K. External magnetic field has been kept at 0.1 T as in Fig. \ref{susmag2}. The arrows denote  magnetic anomalies and it is present for both the cases i.e for magnetic field parallel and perpendicular to `ab' plane. {\it Permission from the authors to reuse the figure from  Phys. Rev. B {\bf 91}, 094422 (2015)\cite{kubota-2015} is gratefully acknowledged.}.}
\end{figure}
\begin{figure}[h!]
\includegraphics[width=0.9\linewidth]{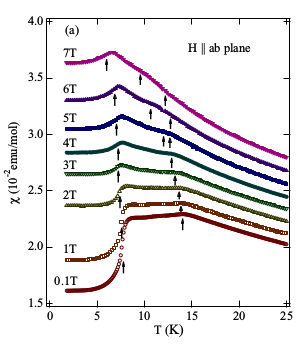}
\caption{\label{susmag4} Magnetic susceptibilities of  $\rm \alpha-RuCl_3$  have been plotted against temperature up to 25 K for a series of  external magnetic fields applied parallel to `ab' plane. The arrows indicate  specific heat anomalies at that temperature. For a better visualization the susceptibility data are shifted upward by multiples of $2\times 10^{-3}$ emu/mol. {\it Permission from the authors to reuse the figure from  Phys. Rev. B {\bf 91}, 094422 (2015)\cite{kubota-2015} is gratefully acknowledged.}}
\end{figure}
Fig. \ref{susmag3} zooms in the low temperature behavior of susceptibilities which indicate the presence of magnetic phase transitions at  T = 13.9 and 7.6 K, supported by the data from both the susceptibilities. These phase transitions are confirmed by specific heat data as well. These phase transitions resemble a ferromagnetic saturation from the paramagnetic region at $\rm T= 13.9$ K for $\rm H \parallel \rm ab$ and then a AFM transition at $\rm T=7.6$ K.  The out of plane susceptibilities shows non-monotonous changes in between $7.6 <  \rm T < 13.9 $, being finally increased as one lowers the temperature further.  This behavior is similar to that reported in Sears et al. \cite{sears-2015} and Majumder et al. \cite{majumder-2015}. However, we note that according to \onlinecite{kubota-2015}, in the case of  $\rm H \perp ab$, the magnetic susceptibilities for some $\rm \alpha-RuCl_3$ samples decrease at 7.6 K with decreasing temperature in contrast to what has been shown in Fig. \ref{susmag3}(b). Finally Fig. \ref{susmag4}(a) represents how the susceptibilities  change in the presence of increasing magnetic field $\rm H \parallel \rm ab$. The motivation is to induce a different phase closer to spin liquid than the observed AFM phases. The behavior as shown in Fig. \ref{susmag4}(a) shows additional peaks (shown by arrows) in specific heat measurements.\\
\indent
{\bf{Studies on small clusters: }} We end the discussions on magnetization and susceptibility  with a brief review of an interesting work on $\rm Os_xCl_3$ where it is claimed that a Kitaev-like system is realized in the form of nano-domains. This material is synthesized by heating $\rm O_s$ metal at $500^{\circ} \rm C$ under $\rm Cl_2$ gas in a quartz
ampule and has a  layered $\rm CdCl_2$   structure with the triangular lattice partially occupied by $\rm O_s$ ions on average. The triangular lattice is composed of nano-domains with a honeycomb arrangement of $\rm O_s$ ions, as confirmed  by electron microscopy and Raman scattering experiments. Kitaev model being an example where the bond-dependent short range correlation function does holds even at smallest possible clusters this study offers a direct  realization of such aspect. In Fig. \ref{susmag6} we present the results obtained in the study \onlinecite{kataoka-jpsj-2020}. 
\begin{figure}[ht!]
\includegraphics[width=0.8\linewidth]{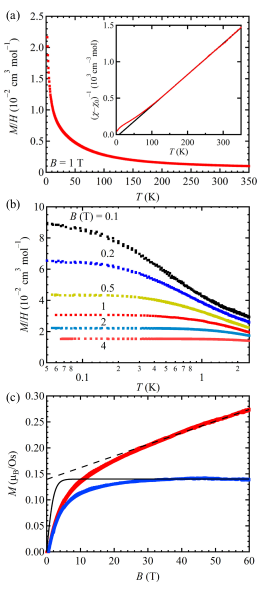}
\caption{\label{susmag6}(c) (2020) The Physical Society of Japan. In panel (a), temperature dependence of magnetization is shown for poly-crystalline sample of $\rm Os_xCl_3$. Magnetic field of 1 Tesla is applied and temperature was brought down up to 2 Kelvin. The inset 	shows the inverse susceptibilities which fits well with Curie-Weiss formula(shown by black line) at high and moderate temperatures. The
temperature-independent term $\chi_0=3.10(3) \times 10^{-4}\rm{cm^{3} mol^{-1}} $ has been subtracted to obtain the inverse susceptibilities. In panel (b), the low temperature susceptibilities have been shown for a range of magnetic field starting from 0.1 Tesla to 4 Tesla.  In panel (c) evolution of magnetization  at 1.4 K is shown against magnetic field by red line. It has linear and nonlinear part. The linear part is extracted and shown by black dashed line. The black line is a theoretical fit by Brillouin curve calculated for non-interacting spins yielding the limiting magnetization determined by dashed line. The  blue line shows the nonlinear part of the experimental data. {\it Permission from the authors to reuse the figure from  JPSJ {\bf 89}, 114709 (2020)\cite{kataoka-jpsj-2020} is gratefully acknowledged.}}
\end{figure}
A comparison with the theoretical work on Kitaev-nano clusters\cite{pervez2023decipheringI,kataoka-2022} shows that at low temperature the qualitative behavior of magnetization is similar so that it reaches a saturation depending on the magnetic field,  as we note from panel (b) in Fig. \ref{susmag6}. Interestingly we note that similar to the theoretical study the saturation magnetization decreases as one increases the magnetic field. This suggests that the periodic boundary condition taken in theoretical study is not relevant and the bond-dependent Kitaev interactions assumed in $\rm Os_xCl_3$ are able to induce significant Kitaev physics. \\
\indent
We note that for large magnetic field the temperature independent magnetization  suggests that Kitaev physics is non-existent and spins are behaving mostly paramagnetic. However at low magnetic field, the Kitaev physics is still relevant and induces a competing mechanism at shortest length scale governed by bond-dependent interactions. We also note that this particular study might be relevant for study of vacancies and disorder in Kitaev model. Figure \ref{susmag6}(c) shows the field dependence of magnetization at 1.4 K. 
The red curve represents the magnetization of the $\rm Os_xCl_3$ which  clearly has two regions with distinct field dependencies. At relatively large magnetic field i.e above 30K it shows a linear dependence and below 30K it shows a nonlinear dependence. When the linear part is extrapolated to lower temperature (represented by  dashed line) it yields 0.14 $\mu_{\rm B}$ per $\rm Os$. The black line shows the Brillouin curve estimated for such free spins. The Blue line then represents the subtraction from  red curve by the linear part.  A comparison between the solid curve and the blue line shows that the saturation field needed is much higher for the blue one. This suggests that  the  defect spins are not entirely free and interact with reduced magnetic interactions with neighboring spins. The gradual linear increase and not saturating even at 60 B indicates that  the  exchange interactions between the spins are still relevant. The  scenario could be compared with Fig. \ref{sus-mag1} panel (b) and (c) where a nonlinear dependency is shown for small magnetic field. However more analytical explanations are required to explain the reported nonlinear dependency and may be of future interest. \\
\indent
\subsection{{\Large Thermal Hall effect}}

{\bf{Search for  transport signature:}} Experiments dealing with various transport mechanism are essential probes to determine the possible phases of a given system. However unlike  magnetization, susceptibility or specific heat, the transport signature depends on existence of some mobile carriers whose response due to an appropriate external field could be detected. Unfortunately absence of such mobile carriers limits  scope of transport experiments on Kitaev materials. The materials with possible Kitaev  like exchange interaction are insulators which do not have the most common type of carriers such as electrons. Thus the  electrical transport measurement is not useful to detect the possible spin liquid phase in Kitaev materials. However there exist several other efficient methods to detect such insulating magnetic phases by appropriately coupling the excitations with  suitable external field and thermal Hall effect\cite{katsura-thermal-hall} emerged one among them.
In spin liquid phase (or other quantum paramagnetic phase) and other ordered magnetic phases spinons and magnons constitute the carriers of charge neutral heat current respectively. However we note that unlike charge current by electrons, the magnons can only carry heat transport provided the magnetically ordered state satisfies certain criteria which depends on the specific lattice geometry and underlying manetic structure such that magnons can  be effectively coupled with external magnetic field\cite{katsura-thermal-hall}.
On the other hand, for the spinons, there is no such restrictions and it gives rise to thermal Hall response due to an emergent Lorentz force \cite{gangchen-2020,moessner2023}.
For more theoretical understanding on different origin of thermal Hall effect due to magnons and  magnon-phonon interactions one may look at the following references \cite{murakami-the-2017,zhang2023thermal,zhang-magnon-phonons,Park2020,Takeda2024,matsumoto-2014}. \\
\indent
\begin{figure*}[ht!]
	\includegraphics[width=0.59\linewidth]{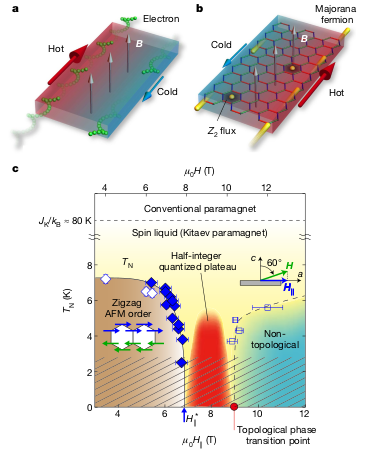}
	\caption{\label{thermal1}In panel (a) and (b) the schematic experimental set up of thermal Hall effect is shown for ordinary electron and Kitaev system respectively. In (c) a phase diagram is shown in $T-H_{\parallel}$ plane. For more detail see text.  {\it Permission from the authors to reuse the figure from   Nature {\bf 559}, 227-231 (2018)\cite{Kasahara2018} is gratefully acknowledged.}}
\end{figure*}
{\bf{The carriers of transport:}} Given the above discussion, the pertinent question is what are the possible source of carriers for candidate materials having Kitaev interactions.  Kitaev-Heisenberg system shows various competing phases such as zig-zag ordered magnetic ground state  at very low temperature and  at moderately high temperature Kitaev  spin liquid phase  is argued to be revived.
In the zig-zag ordered phase the possible excitations are magnons as well as complex classical vortices\cite{chern-2021}. On the other hand, in the spin liquid phase, the Majorana fermions appear as the excitations due to fractionalization of spins. There exists two kind of Majorana fermions, one dynamic and other static which gives rise to flux excitations $B_p$. We note that all these excitations (be it magnon or Majorana fermions) are charge neutral which prohibits the standard electrical  Hall experiments to be useful method to detect them. However the thermal Hall effect experiment is  potential alternative as magnons or Majorana fermions both can respond to temperature.  As far as the Kitaev model is concerned, due to thermal effect   magnetically ordered phase  gives rise low energy classical magnetic vortex liquid instead of ordinary magnons. Such vortex liquid is shown to yield thermal Hall effect\cite{chern-2021}. However this thermal Hall effect does not correspond any topological origin. On the other hand, in the spin liquid phase the  thermal Hall effect due to dynamic Majorana fermions should  have a distinctly different behavior due to topological reason.  Indeed the transverse thermal Hall conductivity in the spin liquid phase is shown to be quantized (at certain range of temperature and applied magnetic field) recently\cite{Kasahara2018} establishing the veracity of the spin liquid phase.  We also note that in a related study\cite{czajka-2021}, the longitudinal Hall conductivity is shown to exhibit interesting oscillations as magnetic field is varied. These oscillations are explained  due to existence of spin liquid phase.
Here we mainly discuss the experimental results obtained in Ref.\onlinecite{Kasahara2018}. In
Fig. \ref{thermal1} (a) and (b) a schematic set up of thermal Hall experiments are shown and a comparison between the conventional electronic thermal Hall effect and that due to Majorana fermions have been shown respectively.  Where in for electrons, the skipping orbits in quantum Hall regime is responsible for charge thermal Hall effect, for Majorana fermions, the ballistic motions (in the absence of scattering) are in play.
In the thermal Hall experiment set up a magnetic field is applied perpendicular to a two dimensional slab and a temperature gradient is maintained across the transverse  as well as longitudinal direction. The hot and cold regions are shown by red and blue sheds respectively in Fig. \ref{thermal1} (a) and (b). The thermal gradient  generates an effective electric field which gives rise to non-vanishing force to elementary particles such as electrons (in (a)) and Majorana fermions (in (b)). As the flux excitations are considered static, in the spin liquid regime, only the Majorana fermions can contribute to thermal Hall response and would correspond to fractionalization of spins  yielding a finite transverse Hall conductivity $\sigma_{xy}$.\\
\indent
\begin{figure*}[ht!]
\includegraphics[width=0.89\linewidth]{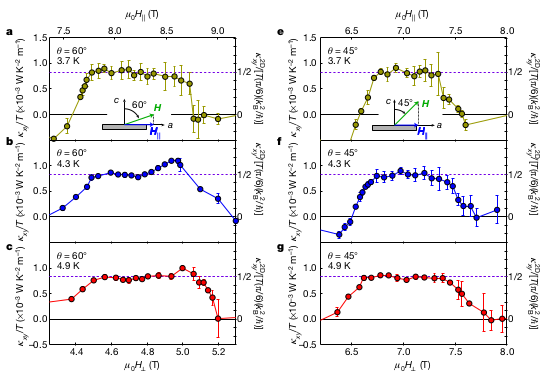}
\caption{\label{thermal2}Transverse Hall conductivity is plotted against $\mu_0 H_{\perp}$ for various temperature and  angle of applied external magnetic field for $\theta=60^{\circ}$ in (a,b,c) and for $\theta=45^{\circ}$ in (e,f,g). The temperature for upper, lower and middle panels are taken as 3.7 K, 4.3 K and 4.9 K respectively. The dashed line indicates the rescaled value corresponding to 1/2 quantization. {\it Permission from the authors to reuse the figure from   Nature {\bf 559}, 227-231 (2018)\cite{Kasahara2018} is gratefully acknowledged.} }
\end{figure*}
{\bf{Thermal Hall effect:}} We now move on to describe the intriguing results of thermal Hall effect obtained in Ref.\onlinecite{Kasahara2018}. The result  has been put in Fig. \ref{thermal1} (c) where a density plot of thermal Hall coefficient in $T-H_{\parallel}$ has been shown. The $T-H_{\parallel}$ plane can be divided into two regions, the low temperature regime which has been shown by dashed region and the high temperature region which is above that. In the low temperature regime, we observe that there exist three different regions as one increases the external magnetic field from zero to a high value as 12 T. These three regions are represented by brown, red and the light blue colors. The low temperature, low magnetic field region denotes a magnetically  ordered state with zig-zag order. This happens due to the presence of non-Kitaev terms and is the ground state at zero temperature. As the temperature or magnetic field is increased, this phase is destabilized and fractionalization of spins are revived. In this region the spins are fractionalized into a dynamic Majorana fermion and a pair of static fluxes. These dynamic Majorana fermions  do contribute to thermal Hall effect and it is supposed to be quantized at 1/2. The red region denotes the experimentally obtained one where the thermal Hall coefficient is half quantized. As the magnetic field is further increased, the fractionalization  ceases to exist due to the confinement of Majorana fermions and its binding to the fluxes leading to a non-topological phase. It is interesting that the same magnetic field which facilitates  the revival of fractionalization at low field, prohibits the same at high magnetic field. This indicates the subtle role of quantum fluctuations in the presence of competing interactions at low field value. Observation of such integer thermal Hall effect\cite{saurabh-2019} for fermions and half-intger thermal Hall effect for fractional quantum hall quasiparicles has been observed previously\cite{Banerjee2018}. As one increases the temperature and goes beyond the dashed region, one obtains a quantum paramagnetic phase (very close to Kitaev spin liquid phase)  denoted by light yellow shed where remnants of spin-fractionalization survives but the thermal Hall coefficient is not quantized to half. The possible explanation could be that due to thermal excitation fluxes  also contribute and this could change the magnitude of thermal Hall current. The physical reason of cessation of half-quantization of thermal Hall effect in the non-topological and Kitaev spin liquid phase is different. In non-topological phase the fractionalization ceases and the magnons are the contributors. On the other hand in the Kitaev spin liquid phase at high temperature fluxes do move and contribute changing the value from half quantization. At very high temperature  i.e above $J/H=80 $K one obtains the conventional paramagnetic phase.\\
\indent
\indent
{\bf{Dependency on temperature and field angle:}}

  We now proceed to discuss more on the intriguing dependence of thermal Hall conductivity on temperature and magnetic field as mentioned in the study\cite{Kasahara2018}. For this purpose two different tilting angles of magnetic field  have been taken and the results are plotted in Fig. \ref{thermal2}. Here transverse thermal Hall conductivity is plotted against $H_{\perp}$ for  $\theta= 60^{\circ}$ in (a,b,c) and for $\theta= 45^{\circ}$ in (e,f,g) at different temperatures as mentioned at the upper left corner of the panel. It is interesting to observe that for a large interval of magnetic field the half-quantization is observed. Secondly the departure of half-quantization and the width of magnetic field for which the half-quantization prevails largely depends on system parameters such as field angle, temperature. For example comparing (a) and (e) we observe that a change in $\theta$
from $60^{\circ}$ to $\theta=45^{\circ}$ reduces the width of magnetic field for which half-quantization has been observed. Third, the occasional departure from half-quantization
around $\mu_0 H_{\perp}=5$ at $\theta= 60^{\circ}$ (panel (b) and (c)) is brought back at half-quantization for $ \theta= 45^{\circ}$.  Similarly comparing (a,b,c) and (e,g,f)  
one notices that at low magnetic field, the thermal Hall conductivity reaches a lower value
for $\theta=45$ in comparison to $\theta=60^{\circ}$. We think this particular aspect should be investigate theoretically in more details as a future interest. \\
\indent
It may be noted that  in a significant study the half-quantization of thermal Hall effect due to fractionalization has been critically questioned as no phase transition has been observed \cite{bachus-2020}.  In a 2017 study\cite{Do2017}, the role of Majorana fermions and fractionalization have been established by a comparison of theoretical and experimental study. However most recent detailed experimental suggest that half-quantization is of Majorana fermion origin\cite{Bruin2022,imamura-2024}.\\
\indent
\subsection{Specific heat and entropy}
We note that in a number of theoretical studies  the specific heat of pure Kitaev model is shown to exhibit a two peak structure at different temperatures $T_1$ and $T_2$\cite{tathagata-2020,yamaji-2016,bachus-2021,feng-2020,nasu-2015}. As we explained earlier that the low temperature peak is associated with flux excitations (or localized Majorana fermions) and high temperature peak is associated with itinerant Majorana fermions. This happens due to the fractionalization of spins into localized fluxes (equivalent to localized Majorana fermions) and dynamical Majorana fermions. However in the presence of non-Kitaev interactions, the fractionalization is not realized perfectly and various long range magnetic orders set in at low temperature.  All these facts make it very intriguing to investigate the specific heat in this context.  Here we discuss  salient experiments done on various Kitaev candidate materials such as $\rm \alpha-RuCl_3, Na_2 IrO_3, Os_xCl_3$ etc.   For this review we mainly discuss the experimental work  discussed earlier\cite{Do2017,yogesh-2017,Widmann2019,kataoka-jpsj-2020,Laurell2020}.\\
\begin{figure}[ht!]
	\includegraphics[width=0.92\linewidth]{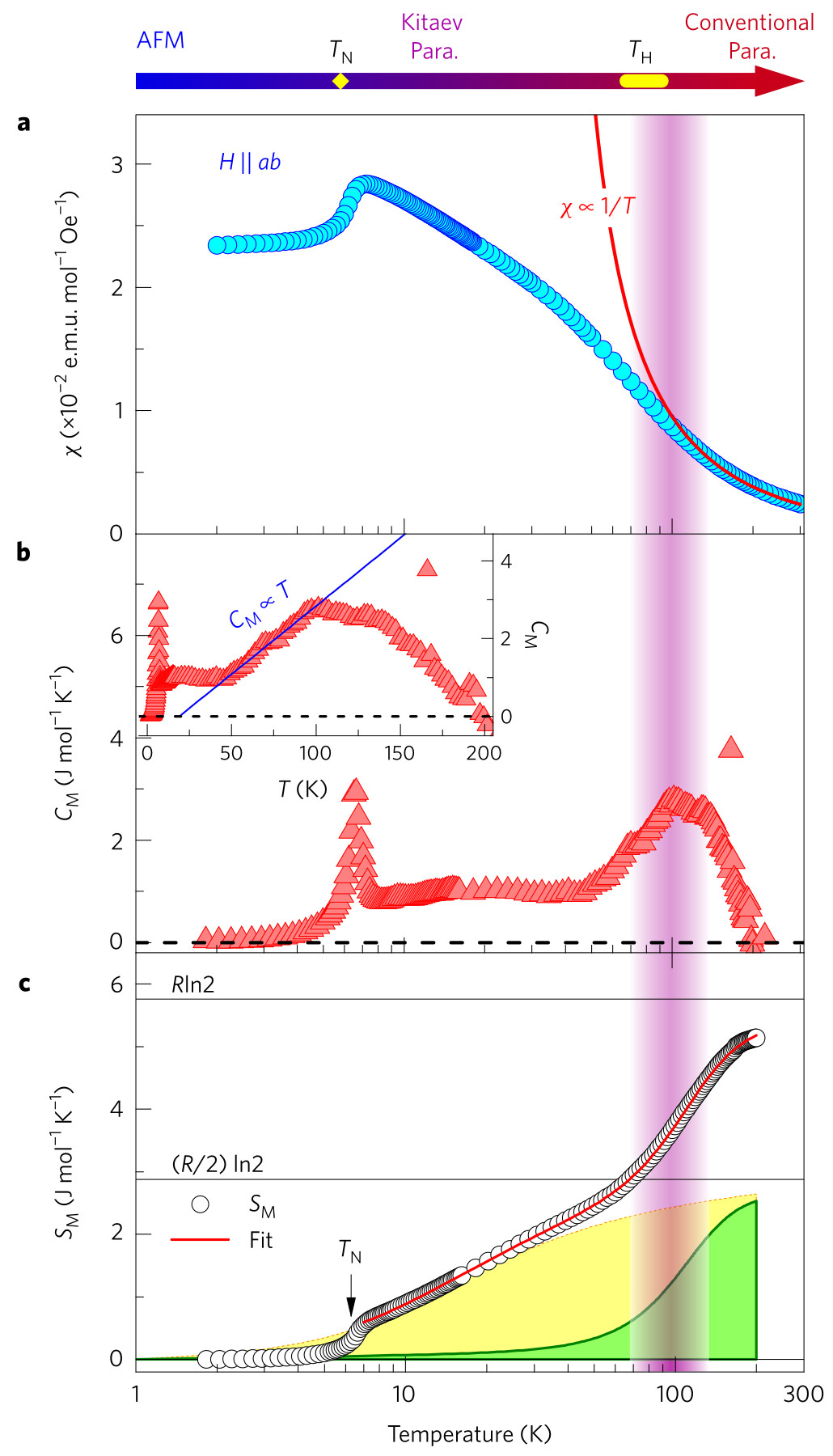}
	\caption{\label{seung1} In  all the panel above the horizontal axis represents temperature presented in semi-log scale. In panel (a), static magnetic susceptibility($\chi$) of $\rm \alpha-RuCl_3$ is plotted. The red line represents the normal Curie-Weiss dependence and a depurture from it below $T=140 $ K is obvious as shown by blue circles. The kink at $\rm T_N=6.5$K signifies onset of zig-zag-type AFM order.   In panel (b), magnetic specific heat $\rm C_M$ is plotted. It shows three features. Firstly, one observes a sharp peak at $\rm T_N$ (at the onset of AFM order), secondly there is a broad bump for temperature in between $T_N$ and 50K, and thirdly a pronounced bump near 100 K. The bumps around 50K and 100K  denote excitations of localized and itinerant Majorana fermions respectively.  One notes that specific heat has linear dependence on temperature in between 50K and 100K as shown in the inset of panel (b). In panel (c)  magnetic entropy is plotted. One notices that the entropy reaches $\rm (R/2)$ln2 around 100K establishing fractionalization of spins into itinerant Majorana fermions and static fluxes. The yellow line and green line denote a theoretical contribution to total entropy by the itinerant and flux excitations respectively. The red line is the sum of these two conributions. {\it Permission to reuse the figure from   Nature Physics {\bf 13}, 1079-1084 (2017)\cite{Do2017} is gratefully acknowledged.}}
\end{figure}
\indent
\begin{figure}[ht!]
	\includegraphics[width=0.99\linewidth]{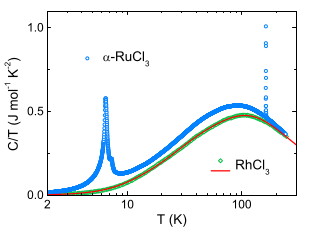}
	\caption{\label{widspheat1} The blue open circles denote the temperature dependence of heat capacities $\rm C/T$ of $\rm \alpha-RuCl_3$ while the green open circles denote the same for $\rm RhCl_3$, a nonmagnetic reference compound. The temperature is plotted on a semi-logarithmic plot. The solid red line denotes a fit according to  Debye-Einstein model.  To account for correct phonon background contribution to heat capacity of the rhodium compound was rescaled by a factor of $0.92$.  {\it Permission from the authors to reuse the figure from   Phys. Rev. B {\bf 99}, 094415 (2019)\cite{Widmann2019} is gratefully acknowledged.}}
\end{figure}
{\bf{Two peaks and T-linear specific heat}} We first  review  the 2017 study\cite{Do2017} where  role of Majorana fermions and fractionalization has been established by a comparison of theoretical and experimental study of specific heat and neutron diffraction. We refer Fig.\ref{seung1} (taken from  reference \onlinecite{Do2017}) where upper, middle and lower panel( (a), (b) and (c)) correspond the magnetic susceptibility ( $\chi(\rm T)$), magnetic specific heat ($\rm{C_M}$ ) and entropy ($\rm{S_M}$) respectively. One can easily notice that for temperature below 140 K, the susceptibility departs from Curie-Weiss curve denoting absence of long-range correlation.  However further lowering the temperature at 6.5 K, the system establishes a zig-zag type AFM order. This onset of zig-zag AFM order at characteristic temperature denoted by $\rm T_N$ is associated with anomalies in specific heat in the form of pronounced peak. Apart from this anomaly at $\rm T_N$, the specific heat shows two broad maxima. The prominent one is seen around 100 K. The other one is not very apparent and happens near $\rm T_N$. This two extra anomalies are understood as the contribution from itinerant and localized Majorana fermions. We note that low temperature  maxima is not prominent due to onset of competing presence of AFM order.
It is remarkable to notice that $\rm{C_M}$ shows a linear dependence on temperature in between these two peaks. This governs a metallic behavior of itinerant Majorana fermions. It is worth mentioning that such linear dependence of specific heat on temperature is also theoretically obtained \cite{nasu-2015}. \\
\indent
{\bf{Entropy:}} We now explains the features obtained in entropy study as shown Fig.\ref{seung1} (from reference \onlinecite{Do2017}) lower panel. While the low temperature zig-zag order corresponds to vanishing entropy, the emergence of localized and itinerant Majorana fermions at different temperature should correspond release of entropy. We see sudden increase of entropy near $\rm T_N$ and a gradual increment further as temperature is increased. As reported the entropy ($\rm{S_M(T)= \int C_M/T dT}$ ) reached close to the ideal value ( 90$\%$) R$\rm{ln}$2  at $T = 200$ K . Around $T = 100 K$ where itinerant Majorana fermions are activated fully the entropy obtained is 0.46R${\rm ln}$2 (very close to the expected value 0.5R$\rm{ln}$2). Near $T_N$ (in the vicinity ) the entropy released amounts to (40$\%$ of 0.5R$\rm{ln}2$ (ref. \cite{Park2024}) which indicates that 60$\%$ of the fluxes remains static does not contribute to entropy release from the AFM order. For more qualitative discussions on contributions of localized flux excitations and dynamic Majorana excitations into total entropy release, we refer the original work of reference \onlinecite{Do2017}.
\indent
\begin{figure}[h!]
\includegraphics[width=1.1\linewidth]{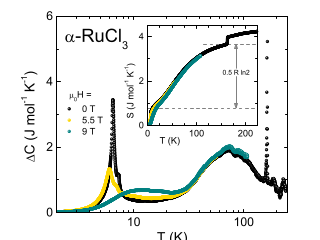}
\caption{\label{widspheat2}Excess heat capacity(after subtracting the background phonon contribution) $\Delta C$ is plotted for a series of in plane magnetic field. As before the temperature scale is on a semi-logarithmic one. Though the quantum criticality associated with AFM transition  vanishes  for field strength $> 7$, one observes a two peak structure. In the inset  entropy released is shown and it shows a jump  once it reaches $\rm 0.5 Rln2$. {\it Permission from the authors to reuse the figure from   Phys. Rev. B {\bf 99}, 094415 (2019)\cite{Widmann2019} is gratefully acknowledged.}}
\end{figure}
\begin{figure}[h!]
\includegraphics[width=0.99\linewidth]{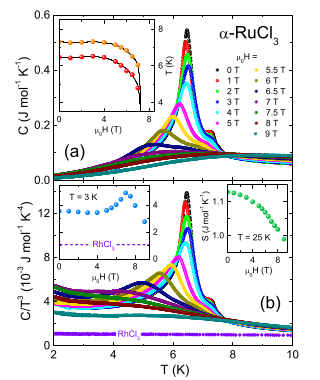}
\caption{\label{widspheat3}In continuation of Fig. \ref{widspheat2}, more results on the specific heat in the presence of in plane magnetic field is shown in (a) for a range of external magnetic fields. In (b) scaled specific heat $\rm C/T^3$ is shown where heat capacity of $\rm RhCl_3$ is also indicated by magenta line. In the inset of (a), the field dependence of two anomalies associated with two peaks(yellow represents the stronger peaks and red represents the weaker peaks) are plotted vs magnetic field. In (b) the left ( right) inset shows the field dependence of $\rm C/T^3$ (entropy measured) at temperature 3K(25K). {\it Permission from the authors to reuse the figure from   Phys. Rev. B {\bf 99}, 094415 (2019)\cite{Widmann2019} is gratefully acknowledged.}}
\end{figure}
\begin{figure*}[ht!]
	\includegraphics[width=0.8\linewidth]{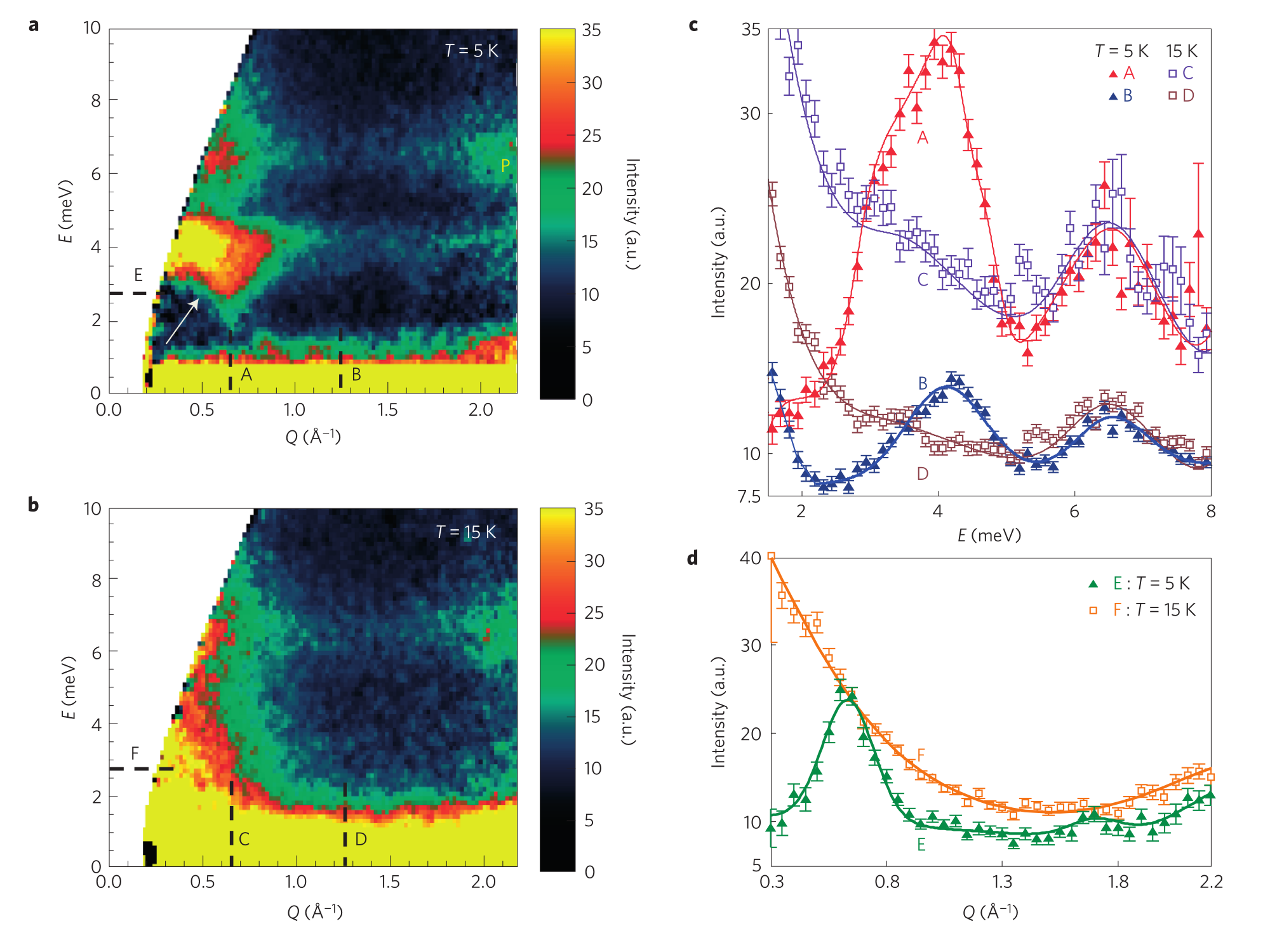}
	\caption{\label{neutron1} Here inelastic neutron  scattering data depicting  collective magnetic modes are presented. Neutrons with 25 meV are used. Panel (a) and (b) shows the false color plot of the data at 5K and 15 K respectively. A density plot of the intensity is plotted in E-$Q$ plane where E stands for energy and $Q$ for momentum. Panel (a) contains two distinct excitation modes $M_1$ and $M_2$ obtained around 4 and 6 meV. The associated $Q$ is near the M point of honeycomb lattice Brillouine zone.  The white arrow points at the conclave lower edge of the $M_1$ mode. The latter `P' in panel (a) signifies a phonon mode which is activated at higher momenta $Q > 2$A$^{\circ -1}$. In panel (c), a plot of intensity for $Q$ at points A and B (in panel (a)) as well as for $Q$ at C and D (in panel (b)) is plotted against E. We notice the characteristic peaks around $E=4$ meV at 5K has disappeared at 15K. The panel (b) shows that due to strong quasiparticle scattering the collective mode at $M_1$ has washed away. However the phonon mode and the collective modes at $M_2$ are still present. In panel (d), intensity is plotted against $Q$ for energy around E and F points as shown in panel (a) and (b) respectively. The green and orange lines show result for temperature 5K and 15K respectively. 
		{\it Permission from the authors to reuse the figure from   Nature Materials {\bf 15}, 733-740 (2016)\cite{abanerjee-2016} is gratefully acknowledged.} }
\end{figure*}
\\
{{\bf Absence of linear T dependence:}} 

In Fig. \ref{widspheat1}, we present $C/T$ against a semi-log T plot as obtained in \onlinecite{Widmann2019}. The blue circles denote the results for $\rm \alpha-RuCl_3$ and the green one for $\rm RhCl_3$, a nonmagnetic reference material. The plot for 
$\rm \alpha-RuCl_3$ shows multiple features. It contains a pronounced peak near 6 K, a small shoulder just after it, a broad maxima around 70 K and  a sharp peak at 163 K.
The absence of low temperature anomalies in $\rm RhCl_3$ conforms the magnetic origin of anomalies in $\rm \alpha-RuCl_3$. The broad peak around 70 K is understood due to itinerant Majorana fermions which are further confirmed by the associated entropy release close to $0.5{\rm Rln}2$  as shown in the inset of \ref{widspheat2}.  The sharp transition at 163 K is ascribed to occurrence of a first-order structural transition from the low-temperature rhombohedral to the high-temperature monoclinic phase. The shoulder after 6 K is explained due to possible stacking faults and not ascribed to generation of localized Majorana fermions. 
In Fig. \ref{widspheat2} \cite{Widmann2019}, we present the magnetic contribution to specific heat (after subtracting the temperature rescaled phonon contribution obtained from $\rm RhCl_3$ ) under different values of external magnetic field. It clearly shows the four different characteristic peaks or broad maxima as discussed before. At low temperature onset of AFM order can be of only usual spin degree  of freedom. However unlike previous experiment discussed\cite{Do2017}, we find absence of linear T dependence. Thus at low temperature it favors  a magnon like behavior which undergoes a crossover to spin-fractionalization of unknown origin at high temperature\cite{rouso-prb-2019}. 
Interestingly the entropy released at high temperature is close to 0.5$\rm Rln$2 and at low temperature it is less than  0.2$\rm Rln$2. This qualitatively  agrees with the experiment discussed before\cite{Do2017}. This 0.2R$\rm ln$2 also come from theoretical model prediction. Interestingly such a low value of entropy (less than 0.2R$\rm ln$2) is even lower than the entropy associated with a AFM transition. Thus we see there is a clear indication of need for further clarification. \\
\noindent
{\bf{Effect of magnetic field: }} To unravel the low temperature specific heat anomaly in more detail a series of magnetic field have been applied and the results are shown in Fig. \ref{widspheat3} (a) and (b) in which specific heat $\rm C$ and scaled specific heat $\rm C/T^3$ have been presented respectively. The low temperature peak signifying AFM transition gradually shifts to left as field is increased. The peak height also gradually decreases as the field is increased. This indicates that the external field is prohibiting onset of AFM order. When the field is above 7 T, the nature of peak changes and it resembles  a broad maxima signifying absence of AFM order after a quantum critical point. As the field is further increased, the broad maxima persists. However it is argued that in the absence of AFM order, the phase does not correspond to localized Majorana excitations as claimed theoretically \cite{anjana-2018,nasu-2015,udagawa-prl-2014,yoshitake-2016-prl,yamaji-2016,Catuneanu2018,Laurell2020}. Rather the it corresponds to field induced PM  AFM phase and no Kitaev-type  quantum spin liquid phase. We may note that this conclusion is little different than what has been argued in Fig. \ref{widspheat1}  or in  recent works\cite{wolter-2017}. \\
\indent
\begin{figure*}[ht!]
	\includegraphics[width=0.99\linewidth]{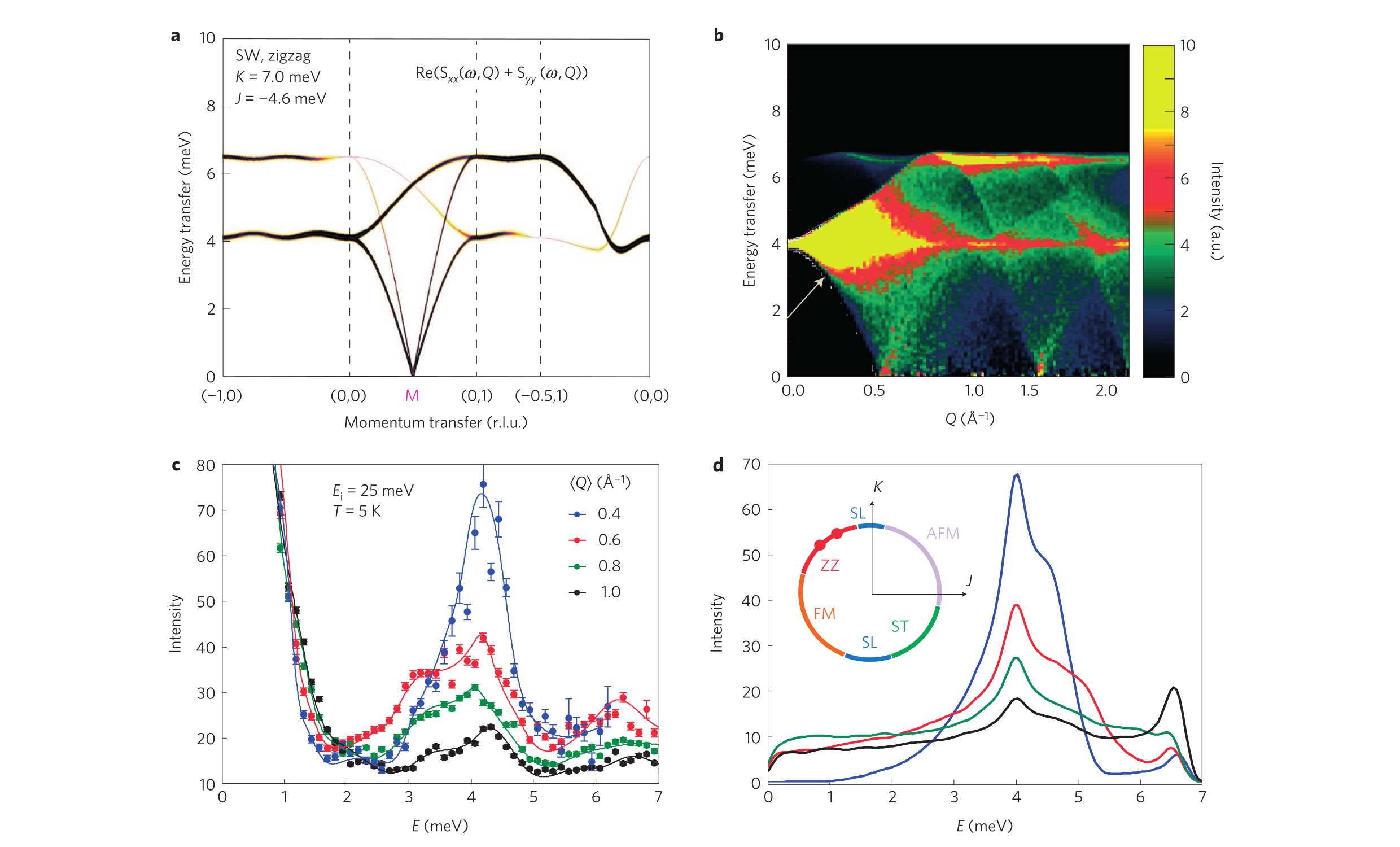}
	\caption{\label{neutron2} Spin wave  theory calculations.  In panel (a) spin wave simulation for the Kitaev-Heisenberg model wth ($K,J$)= (-4.6 meV,~7.0meV) is plotted with zig-zag ground state being considered. Panel (b) shows the calculated powder-averaged scattering including the magnetic form factor. The white arrow is used to indicate the similarity between the experimentally observed spin wave spectrum and theoretically obtained one near $M$ point. Panel (c) shows the cuts through the data of Fig. \ref{neutron1} for representative $Q$. It is interesting to note that actual data shows  large elastic response from Bragg and incoherent scattering at $E=0$ meV. Panel (d) depicts the calculated scattering shown in panel (b). Various colors correspond the same $Q$ considered in panel (c). 	{\it Permission from the authors to reuse the figure from   Nature Materials {\bf 15}, 733-740 (2016)\cite{abanerjee-2016} is gratefully acknowledged.}  }
\end{figure*}
  Now we pay attention to Fig. \ref{widspheat3}  (b) where $\rm{C/T^3}$ has been plotted vs T for many values of magnetic fields. Apart from the gradual shifts of peaks toward left in increasing fields, we note that the plots become flatter at low temperature. The  results for $\rm RhCl_3$ show a flat behavior which are explained as coming  from phonon or lattice contributions. In comparison to this, for $\rm \alpha-RuCl_3$, we see a little upturn in $\rm{C/T^3}$ which signifies that  at low temperature $\rm C \propto T^x$ with $x=2.5 <3$. This reduced power law is in agreement with previous study\cite{wolter-2017}. Also the entropy released are  very low in comparison to the expected  0.5Rln2. In brief, this particular study does not convey a strong results in support of localized Majorana excitations with signature Kitaev spin liquid  character. Though, the thermal Hall effect does show a contradictory fractionalized thermal quantum spin Hall effect \cite{Kasahara2018}. However we end with another related study \cite{Banerjee2018} which stress upon the existence of kitaev spin liquid in low temperature under magnetic field as found in neutron diffraction experiment. \\
  \indent 
Before moving to the inelastic neutron diffraction study of $\rm{\alpha-RuCl_3}$ we briefly mention few other important aspects of Kitaev materials.  Giant phonon anomalies in the proximate Kitaev quantum spin liquid $\rm{\alpha-RuCl_3}$ has been observed experimentally\cite{haoxiang-nature-2012}. Magnon bound states\cite{wulferding-2020} versus anyonic Majorana excitations in the Kitaev honeycomb magnet $\rm \alpha-RuCl_3$ has also been discussed. Effect of  Raman spectra and susceptibility to detect Kitaev quantum spin liquid is discussed\cite{wang-2020}. For further study  of effect of magnons, phonons, and thermal Hall effect in Kitaev magnet  namely $\rm \alpha-RuCl_3$ one may consult \onlinecite{shuyi-2023}. Magnon damping in the zig-zag phase of the Kitaev-Heisenberg-$\mathrm{\ensuremath{\Gamma}}$ model on a honeycomb lattice has also been studied\cite{smit-2020}.\\
\indent
\subsection{Neutron diffraction study on $\rm RuCl_3$}

We complete the survey of important experimental works done on Kitaev materials by reviewing the inelastic neutron diffraction study\cite{abanerjee-2016} performed on $\rm \alpha- RuCl_3$ and results are compared with possible connection to Kitaev-Heisenberg Hamiltonian given below,
\begin{eqnarray}
\label{arnabH}
H= \sum^{\gamma}_{\langle i,j \rangle_{\gamma}} K s^{\gamma}_i s^{\gamma}_j + J \vec{S}_i \cdot \vec{S}_j~.
\end{eqnarray}

\begin{figure*}[ht!]
	\includegraphics[width=0.99\linewidth]{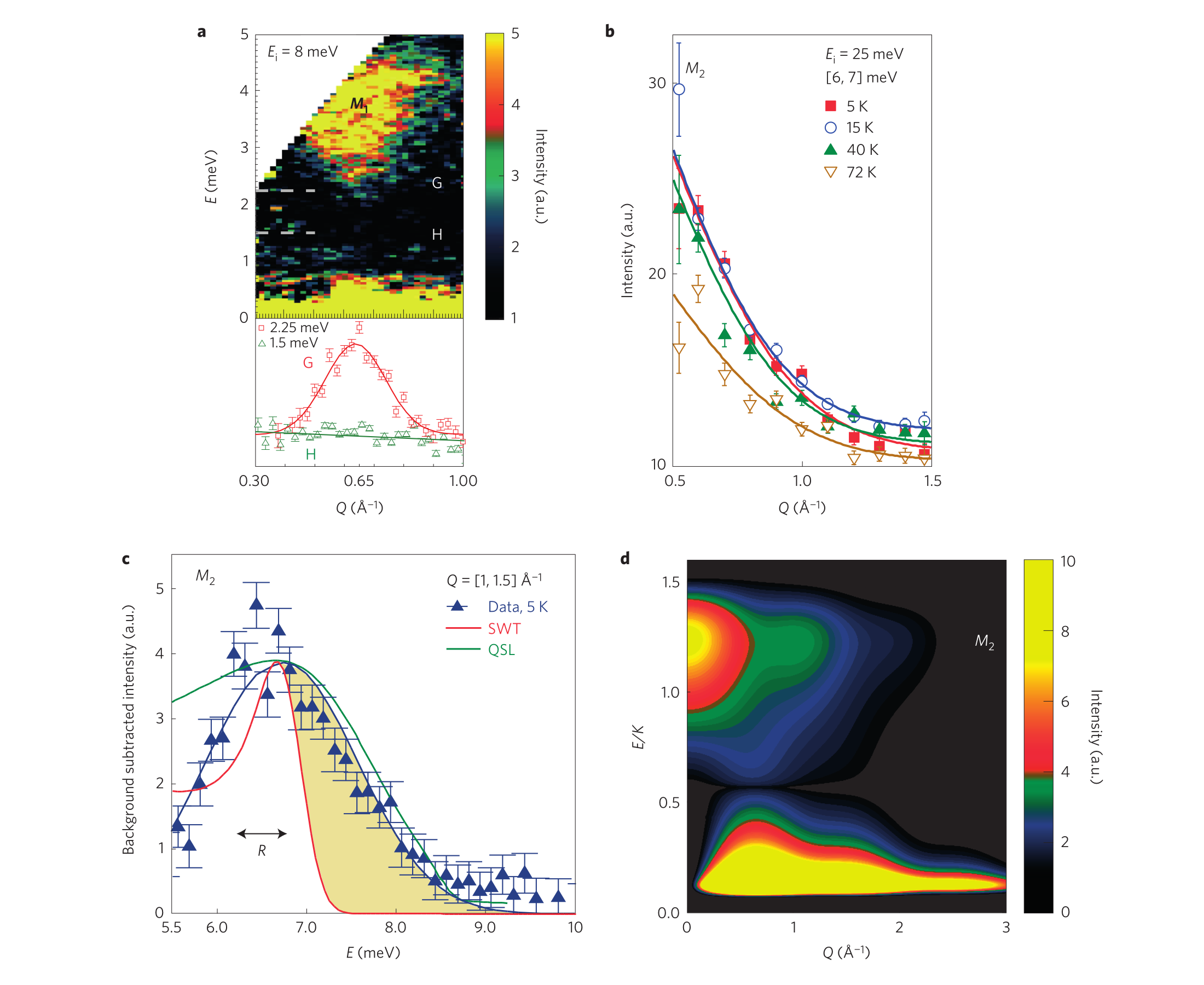}
	\caption{\label{neutron3}Better agreement with quantum spin liquid calculation in comparison to spin wave approximation: Panel (a) depicts inelastic neutron scattering  around mode $M_1$ at $T= 5$ K with $E_i= 8 \rm~ meV$. In the lower panel intensity is plotted for constant energy cut around point `G' and `H'. Absence of any well defined scattering below 2 meV, establishes the gap in magnonic excitation spectrum. Panel (b) shows the intensity around the mode $M_2$ at four different temperature using $E$. Panel (c) present the background subtracted intensity around $Q$(within range (1,1.5)) near $M_2$. The triangle represents the actual data and the blue line is fit to a Gaussian peak. The red and green line are calculated scattering from spin wave approximation and Kitaev spin liquid response function respectively. The shaded region shows the discrepancy between the spin wave and spin liquid calculation. Panel (d) presents powder-averaged scattering calculated from a model Hamiltonian after Ref \onlinecite{knolle-2014}.
	{\it Permission from the authors to reuse the figure from   Nature Materials {\bf 15}, 733-740 (2016)\cite{abanerjee-2016} is gratefully acknowledged.}}
\end{figure*}

Similar to the previous experimental works reviewed, at low temperature the elastic neutron diffraction study confirmed existence of zig-zag ordered magnetic state. The transition temperature is reported to be $\rm T_N \approx 14$K. Having established the magnetically ordered state, low-energy excitations are probed via inelastic neutron scattering (INS) study and  its dependence on energy, momentum and temperature are carefully presented and analyzed. Further the results were compared with analytical and numerical simulation of Hamiltonian in Eq. \ref{arnabH} for appropriate exchange coupling. This establishes the possible connection to the spin liquid physics that could be present in the low-energy excitations. In Figs. \ref{neutron1}, \ref{neutron2} and \ref{neutron3} we present the inelastic neutron scattering results and associated theoretical simulation.

{\bf Low energy vs high energy modes of excitation:} We first begin by discussing the density plot of INS intensity in energy(E)-vs-momentum (Q) plane at different temperatures. In Fig. \ref{neutron1} a and b this intensity is plotted for temperature 5K and 15K respectively. We note that powder of $\rm \alpha-RuCl_3$ was used and neutron energy was taken to be $\rm E_i=25 meV$. From Fig. \ref{neutron1}, panel a, we can easily note two region of excitations centered around $\rm E= 4 meV$ and $\rm 6 meV$ respectively. These two modes are referred as $M_1$ and $M_2$ respectively in \onlinecite{abanerjee-2016}.  The $M_1$ excitation has a minimum for $Q=0.62$ (denoted by letter A) and it conforms with theoretical spin wave calculation done for two dimensional system. The concave shape(represented by a white arrow) of the excitation at $M_1$  also matches with spin wave calculation done for zig-zag state. All these firmly points out the magnetic origin of $M_1$ associated with the zig-zag ordered state and exclude the possibility of any other kind of excitations. The other excitation mode $M_2$ is centered around $\rm E=6.5 meV$ and also of magnetic origin of different nature. The temperature dependence of these two modes show interesting behavior as shown in Fig. \ref{neutron1} panel (b). We observe that the mode $M_2$ remains unaffected  but the mode $M_1$  is softened and shifts toward $Q=0$. \\
\indent
In Fig. \ref{neutron1} c intensity is plotted with $\rm E$ for fixed values of $Q$ corresponding to the points $A$ and $B$ for $\rm T=5$K. $\rm C$ and $\rm D$ denote the plots for temperature 15K. The well defined peak structure confirms well defined excitations. We note that the peak at higher energy remains unaffected as the temperature is increased. Similarly in Fig. \ref{neutron1} d constant energy cut plot for intensity against $Q$ reveals the disappearance of well defined peaks. The stability of the $M_2$ mode and disappearances of $M_1$ mode possibly hint that the origin of $M_2$ mode is due to different reason and not connected to $M_1$ mode whose origin is explained due to zig-zag ordered ground state.

{\bf Spin wave calculation  vs spin liquid speculation:} To understand the physical origin of these two modes $M_1$ and $M_2$, theoretical calculations are done for model Hamiltonian(in Eq. \ref{arnabH})and compared with the experimental data presented in Fig. \ref{neutron2}. In Fig. \ref{neutron2} a, theoretical calculation based on spin wave approximation is presented and in (b) powder averaged neutron scattering data is presented. For the spin wave theory value of $K$ and $J$ are taken as $\rm 7.0 meV$ and $\rm -4.6 meV$ respectively which are obtained by equating the experimental peak at $E_1$ and $E_2$ with the theoretically obtained peak values. From Fig. \ref{neutron2} b, we find that the calculated spin wave data inferred from experiment matches well for $Q$ ranging from  0.0 to 0.5 as denoted by white arrow. In Fig. \ref{neutron2} (c) scattering intensity is plotted against $E$ for constant $Q$ and integrated over $0.2A^{o-1}$  and taken from Fig. \ref{neutron1} a. In Fig. \ref{neutron2} (d), we plot the same from Fig. \ref{neutron2} (b). We observe that at higher energy both the plots agree with each other but at lower energy there are significant differences. \\
In Fig. \ref{neutron3} a INS for $E_i=8 \rm meV$ is plotted in E-Q plane showing $M_1$ mode only. The bottom of panel a shows intensity plot for constant energy around `H' and `G' points. We note that below $E=2 \rm meV$ there is no peak structure which confirms the gap for $M_1$ mode. In Fig. \ref{neutron3} b, the INS intensity around mode $M_2$ is plotted for different temperature with incident neutron energy $E_i=25 \rm meV$. We note that for quite a large temperature range $M_2$ mode remains stable. Now in Fig. \ref{neutron3} c, a comparison of experiment, spin wave theory and quantum spin liquid(QSL) based calculation are presented. We note that QSL calculation matched quite well with the experimental data presented by blue triangles. In Fig. \ref{neutron3} d, a theoretical calculation of INS based on isotropic Kitaev interaction with antiferromagnetic Kitaev exchange coupling is shown and its close similarity with panel (a) is remarkable. All these discussions point out a strong indication that at higher temperature the spin liquid phase become relevant and fractionalization of spins are realized to great extent.

\section{Discussion and Conclusions}
\label{dis}
In this basic review, we have presented  some exact theoretical aspects of Kitaev model \cite{kitaev-2006} as well  recent important experiments related to materials which realize Kitaev Hamiltonian among other interactions being present.  We have explained in detail how the Majorana fermionization  is used to solve the Kitaev  model exactly. The exact solvability  of Kitaev model  is a distinct feature which makes it stand apart from other  exactly solvable models where in some cases we have only the ground state \cite{ghosh-majumder} available. It may be noted that Kitaev model can be defined in any lattices with coordination number  three and exact solution can be obtained \cite{saptarshi-2009,saptarshi-2014}. This particular aspect motivated many interesting works to explore the properties of Kitaev model in other lattices. Though most of such models use the Kitaev's Majorana fermionization, an alternative way of solving Kitaev model without using Majorana fermionization can also be obtained~\cite{nussinov-1} implementing algebra of bond-Hamiltonians. In all these, Kitaev model is reduced to a $Z_2$ gauge theory and in the ground state sector  all the $Z_2$ gauge fields have to conform the Lieb theorem. For honeycomb Kitaev model  the product of $Z_2$ gauge fields  over a plaquette becomes unity. Uniform choice of $Z_2$ gauge fields for every link is a convenient choice for such configuration. The spectrum, phase diagram and calculation of correlation function  have been shown explicitly. Notable feature of the phase diagram is that it consists of  gapless and gapped phase both. In the gapless phase, the spectrum  varies linearly with momentum around the gap closing points. On the hand where  spectrum is gapped, the energy varies quadratically with momentum.  \\
\indent
The correlation function calculation shows that two-spin correlation function is  short-range and bond dependent. This  property of correlation function is  true for both static and dynamic correlation function. The short-range nature of correlation function is different than what found in classical paramagnetic phase. Moreover this short range two-spin correlation function exists for any eigenstates of Kitaev model. This particular property is unique for Kitaev model. We have shown that the action of a spin operator on any eigenstate is two fold. Firstly it changes the occupation number of a $Z_2$ gauge field on  a particular bond by changing the value of $B_p$ on the adjacent plaquettes. Secondly it also adds a Majorana fermion at a given site the spin belongs to. However the time evolution of the state shows that the pair of fluxes created do not move but the Majorana fermion added can move and is not confined or attached to that site only. This phenomenon is  referred as fractionalization of spin into static fluxes and dynamic Majorana fermion. The itinerant Majorana fermion defines a deconfinement phase which is gapless with linear  spectrum at low energy. \\
\indent
Moving further we have shown how the long-range entanglement exists in an eigenfunction by showing  non-zero  multi-spin correlation function and explained how it is different than other magnetic phases with short-range two-spin correlation. Existence of topological degeneracy of any eigenfunction in thermodynamic limit has been  explained schematically.  Historically Kitaev model was conceived  mainly with the aim of possible implementation to quantum computation due to the existence of topological order and non-trivial topological excitations called anyons. It may be noted that the gapped phase of Kitaev model realizes Abelian anyons and  gapless phase realizes non-Abelian anyons. The braiding properties of these anyons in relation to realization of quantum gates have been discussed in detail in the original work \onlinecite{kitaev-2003, kitaev-2006}. It may be noted that the anisotropic limit of Kitaev model in different lattices constitute a class of model which is known as Toric code model. Depending on the magnitude of the spin considered and the details of lattices, the zero and finite temperature properties of the Toric code model shows different characteristics.  Kitaev model is an example of certain generalized model known as compass model where the spin-spin interactions are bond or direction dependent. A good account of  discussions of such compass models and its connection to Kitaev model can be found in Ref. \onlinecite{nussinov-2015}.\\ 

\indent
Kitaev model is  exactly solvable in the sense that all the eigenfunctions and eigenvalues can be obtained formally. The exact solvability is lost once other more conventional spin interactions such as Ising or Heisenberg interaction \cite{saptarshi-2011,subhro-1,subhro-2} are added to Kitaev model. In this case the $Z_2$ gauge fields defined on each bond  are no longer  conserved. The  flux operator $B_p$ also does not commute with the Hamiltonian any more. Physically this means that the fractionalization of spins into Majorana fermion and static fluxes  are no longer possible. Though  the flux excitations become dynamic  along with the Majorana fermions  their individual dynamics may be at different time and length scale. Interestingly,  the Kitaev model was proposed with a theoretical foundation for quantum computation, however, it was rather surprising that this unusual interaction happens to exist in certain materials such as Iridiate  and $\rm RuCl_3$ etc \cite{jackeli,abanerjee-2016}.  However due to the presence of other non-Kitaev interactions(such as Heisenberg interaction etc) along with Kitaev interaction, the exact aspects of Kitav model are difficult to be realized experimentally. Presently intense research is underway to explore various aspects of Kitaev model such as fractionalization, deconfinement to confinement transition, detection of spin liquid phase. Interested readers are requested to go through some of the recent developments \cite{Takagi2019} and the references there in.   \\
\indent
 To make this review a self content one we have included basic material aspect of Kitaev model and some overview of important experiments in relevant materials.  The specific mechanism that is needed to realize the anisotropic interactions of Kitaev model requires strong spin-orbit interaction. In the presence of strong spin-orbit interactions and accompanying tetragonal surrounding in certain TM oxides, the isotropic part of Heisenberg interaction are nearly zero and yields  a large Kitaev component. The complete spin Hamiltonian thus obtained includes anisotropic $\Gamma$-term as well. The presence of non-Kitaev terms in related materials makes it challenging to realize some of the remarkable characteristics of  Kitaev model. At zero temperature, theoretical works predict that the presence of other non-Kitaev term  causes various long-range ordered classical ground states. These have been confirmed experimentally  at low temperature. However at moderate temperature the Kitaev interaction induced mechanism of spin-fractionalization  revives and some aspect of Kitaev spin liquid phase have been confirmed experimentally. We have included some experimental thermodynamic and transport results elucidating these. These include specific heat, susceptibility, magnetization, thermal Hall effect and neutron diffraction. All of these experiments seem to confirm that there is a range of temperature where Kitaev spin liquid physics is dominantly present. This Kitaev spin liquid phase finally goes to usual paramagnetic phase through a certain quantum-para magnetic state. The main difference between the Kitaev spin liquid phase and other quantum paramagnet phase is that  in Kitaev spin liquid phase the flux excitations are  significantly suppressed  in numbers than in the quantum paramagnet phase.  We end this brief review on Kitaev model with some overview of few relevant experimental and theoretical works. \\
\indent
{{\bf Effect of magnetic field:}} Magnetic field has been found to induce a variety of interesting effects in Kitaev model as well as in Kitaev-Heisenberg model. Along this development the first one was by  Tikhonov, Feigel'man and Kitaev in 2011\cite{tikhonov2011} where they  considered the effect of Zeeman field on Kitaev model.  It shows that in the presence of Zeeman field the gapless phase opens up a gap and spin liquid phase becomes critical by developing a power-law spin-spin correlation function. Effect of magnetic field for Kitaev-Heisenberg model has been studied in \onlinecite{lunkin2019}. Magnetic field causes various interesting effect in Kitaev related model such as i) oscillations in the thermal conductivity in the presence of perpendicular magnetic field \cite{czajka-2021}, ii) existence of an intermediate quantum  spin liquid phase as observed  theoretically using DMRG\cite{niravkumar-2019}. In another recent DMRG study\cite{tanmoy-2023} of Kitaev model under external magnetic field it is shown to produce an emergent glassy phase instead of QSL phase before going to spin polarized phase. It appears that effect of external magnetic field is similar to thermal effect. Thermal effect and magnetic field both acts as destabilizing the magnetically ordered phase.  Whether due to increased temperature  the spin liquid state is revived, the magnetic field acts as stabilizing it when temperature is further increased to prohibit  the onset of quantum paramagnetic state. However at much increased temperature it eventually leads to an ordinary paramagnetic phase.  Field induced intermediate ordered state and anisotropic  inter layer interaction has also been  discussed\cite{balz-prb-2021,berke-prb-2020}. In depth study using multiple many-body methods can be found here \onlinecite{li-2021-ncom} which explains many major observations in $\rm \alpha-RuCl_3$, including the zig-zag order, two peak structure in specific heat, magnetic anisotropy, and the characteristic M-star dynamical spin structure. A comprehensive study of Kitaev model in the presence of magnetic field in (111)  direction \cite{Zhang2022} shows non-Abelian phase with Chern number $C = \pm 1$ and Abelian phase with Chern number $C=\pm 4$ belonging to Kitaev's 16 fold phases. Anomalous thermal Hall effect in the presence of magnetic field has been shown theoretically\cite{yokoi-2021-anomalous-hall}. The stability of Kitaev spin liquid state with fractionalized Majorana excitations considering AFM Kitaev model has been studied as well \cite{berke-prb-2020}.\\
\indent
{\bf Ferromagnetic vs Antiferromagnetic  Kitaev model:} 
In the beginning we have remarked that physics of Kitaev model is independent of the sign of the coupling parameters namely $J_x, J_y, J_z$. Indeed for pure Kitaev model various formal aspects discussed so far such as spectrum, spin-spin correlation function, fractionalization and topological order are independent of the signs of $J_{\alpha}$s. However it does not imply that there are no differences that could depend on the signs of coupling parameter. We now outline some of the recent important works\cite{Hickey-2021,zhu-2018,Hickey-2019,yilmaz-2022,pervez2023decipheringI,pervez2023decipheringII} which have explicitly demonstrated that response of Kitaev model in the presence of other perturbation such as external magnetic field, Heisenberg interaction or $\Gamma$ interactions does depend on the signs of exchange coupling in Kitaev interaction, namely $J_{\alpha}$. In \onlinecite{zhu-2018}, a remarkable difference between FM and AFM Kitaev model has been found in the presence of magnetic field in $[1,1,1]$ direction. While the FM Kitaev spin liquid is robust up to a critical magnetic field of the order of $0.02$ of Kitaev exchange coupling, the AFM Kitaev spin liquid phase is robust up to $0.2$. Beside it, the AFM Kitaev model shows a gapless phase before going to high field polarized phase. In a similar study\cite{Hickey-2019}, identical results were discovered and  the intermediate gapless phase is recognized as $\rm U1$ gapless spin liquid phase. Here\cite{Hickey-2019}, magnetic field considered has a out of plane component and phase diagram as an angle between the two in-plane axis has been charted out. In a significant study\cite{yilmaz-2022} the topological phase diagram was charted out in such rotating magnetic field. There it was found that AFM Kitaev model yields much richer phases with Chern number $C=0, \pm 1, \pm 2$ where as FM Kitaev model only yields $C=0, \pm 1$. The above difference in regard to FM and AFM Kitaev model was also found in other tricoordinated lattices such as decorated honeycomb lattice and square octagon lattice\cite{Hickey-2021}. For a recent review on the effect of magnetic field on Kitaev physics one may consult Ref. \onlinecite{tripathy2024}  \\ 
\indent
{\bf{Many-body aspects of Kitaev model:}} After soon Kitaev model has been introduced, many other models which are extended version of Kitaev model have been proposed by several authors and  presently there are still efforts to explore further along these lines.  All such extensions are unique in their own way and offer a potential opportunity to investigate various aspects of many-body physics in general. Though it is impossible to enlist and discuss all such extensions, we briefly mention few such studies for minimal completeness.  Extended Kitaev models can in general be categorized into different classes. The first class mainly extended the Kitaev model  into other two\cite{kivelson-2007,philip,kekule-kitaev} and three dimensional\cite{saptarshi-2009,saptarshi-2014} lattices with three coordination number. Depending on the specific lattice, the free Majorana fermion hopping problem shows contrasting aspects such as gapless contour, gapped contour  or Majorana Fermi surface. Apart from these, diverse phases such as Abelian or non-Abelian spin liquid, chiral spin liquid phases have been found. Recent extension in hyperbolic space\cite{dusel2024chiralgaplessspinliquid}  is also studied. In the second class of extension  effect of various perturbations\cite{dung-hai-lee-2007,wang-extended-KM} which keeps the exact solvability of Kitaev model intact have been included. These studies show additional phase transitions as the model parameters are tuned. These phase transitions could be of different types such as topological to non-topological phase transitions, or uniform flux phase to different type of vison crystal phases\cite{batista-vison} where the plaquette conserved quantity $B_p$ assumes regular pattern.  Thirdly there are theoretical works which consider perturbations that destroy the exact solvability of Kitaev model. There have been some very useful approaches such as variational wave function or quantum Monte-Carlo simulations. Within such approaches remarkable new insights into the Kitaev model have been obtained. For example the study\cite{batista-variation} found that for Kitaev-Heisenberg-$\Gamma$ model, the fractionalized excitations form bound states in specific regions of parameter space and these bound states cause phase transition.  Further work on many-body aspects using dynamical meanfield study or functional renormalization study, perturbative continuous unitary transformation  study\cite{vidal-pcut-2008,vidal-pcut-2008-1} have also been followed in exploring specific aspects\cite{Zhang2022,subhro-2,subhro-1,feng-2022,yoshitake-2016-prl,tensorkitaev,kimfrgkitaev}. Recently various dynamics under time dependent perturbation has been investigated. For recent works on dynamics of Kitaev model following references are useful\cite{Kumar2022,saikat-faraday,shreyoshi-prl,shreyoshi-prb,shreyoshi-prb-ext,Sasidharan_2024,saptarshi-quench,saptarshi-entanglement} and all of these contributed to fundamental aspect of quantum mechanics. There is a lot of scope to investigate further along  these lines.

{\bf{Effect of disorder and vacancies in Kitaev model:}} Before we conclude we briefly discuss the effect of disorder and vacancies in Kitaev model. The study of effect of disorder and vacancies are not of only theoretical interest but of practical as well. Because in material realization of Kitaev model we expect that some amount of disorder or vacancies be invariably present. Understanding the effect of disorder and vacancies may also  lead us to control the physical state for better future application.  Various kind of disorder have been studied such as bond disorder where $x$, $y$ and $z$-type bonds can arbitrarily appear maintaining the local rule of meeting three different type of bonds at a site.  Another disorder that has been studied is  the absence of some site which is known as vacancy disorder.  At the non-interacting level 
(say in the absence of any other non-Kitaev interaction), these two different kind of irregularity of otherwise translational invariant Kitaev model yields different aspect of flux conserve quantities namely $B_P$. In the absence of vacancies when the coupling strength $J_{\alpha}$ is assigned any random values, $B_p$ can still be defined for each plaquette and it is conserved, but its values can not be straightforwardly determined as Lieb's theorem can no longer be applied. Thus the ground state sector may belong to certain flux distribution which is not uniform. In second class of irregularities, namely, vacancies, absence of any site makes certain bond disappear  and all the $B_p$ previously defined sharing these bonds become absent. Instead a larger loop conserved quantity appears  whose
value is undefined and needs to be numerically determined. In addition to these consequences, the other interesting aspect of the vacancies, is that the certain Majorana fermions which were previously defined on bonds sharing that particular site become absent from the Hamiltonian. This corresponds to zero energy immobile Majorana mode.  Thus at zeroth order the irregular version of Kitaev model in the absence of other non-Kitaev interaction is a non-trivial problem itself. \\
\indent
In an interesting  study along this line, it is found that the vacancies cause local moment formation\cite{willans2010,willans2011} by binding a Majorana fermion with a flux excitation. This local moment can be polarized in the presence of an external magnetic field. On the other hand it is claimed that the randomness in the exchange interaction does not qualitatively change the susceptibility compared to the clean case. In a series of studies\cite{kao-2021,kao2021prx}, the authors investigated some important consequences of disorder and vacancies in thermodynamic and transport responses. Apart from the original Kitaev interaction, these studies included an additional time reversal breaking three spin terms and its effect has been found to alter the effect of disorder significantly. Among various intriguing effects of disorder, it is found that in the presence of large magnetic field the flux free sector becomes the ground state. The effect of disorder and vacancies  on low-energy Majorana thermal transport reveals that the half-quantization of the thermal Hall current\cite{nasu-2020,yamada-2020} is very fragile against vacancies. For the bond-disorder case, the thermal Hall current remains half-quantized up to a small disorder before going to a immobile phase named as Anderson-Kitaev spin liquid. Other aspect of disorder in two and three dimensional Kitaev model we refer\cite{Brennan2016,santosh-2012,shaozhi2023,sreejith-2016} for the interested readers. \\
\indent
{ \bf Acknowledgment :} The author acknowledges late Prof. A. Jayannavar for encouragement to write a pedagogical article on Kitaev model which finally took the form of this review. The author also happily acknowledges the mentors and collaborators on the journey of Kitaev model namely Prof. G. Baskaran, Prof. R. Shankar, Prof. K. Sengupta, Dr. Naveen Surendran, Dr. Subhro Bhattacharjee and Prof. Julien Vidal. Financial assistance from SERB with sanction number CRG/2021/006934 is gratefully acknowledged. \\
\indent
{\bf Comment:}  For any clarifications  or explanations of physical concepts or figures used, the readers are requested to contact the author at saptarshi@iopb.res.in. 

\end{document}